\documentclass[useAMS,usenatbib]{mnras}
\usepackage{natbib}
\usepackage{aas_macros}
\usepackage{graphicx}
\usepackage{savesym}
\usepackage{listings}
\usepackage[intlimits]{amsmath}
\savesymbol{iint}
\usepackage{txfonts}
\usepackage{bm}
\restoresymbol{TXF}{iint}
\usepackage{amssymb}
\usepackage{hyperref}
\usepackage{amsmath}
\usepackage{tabularx} 
\usepackage{color}
\usepackage{soul}
\usepackage{ulem}
\newcommand{\cE}{{\cal E}}
\newcommand{\cL}{{\cal L}}
\newcommand{\cT}{{\cal T}}
\newcommand{\cR}{{\cal R}}
\newcommand{\err}{\mbox{Er}}

\newcommand{\vv}[1]{\bmath{#1} }

\newcommand{\etext}[1]{\quad\mbox{#1}\quad}
\newcommand{\spr}[2]{\bmath{#1} \!\cdot\! \bmath{#2}}
\newcommand{\vpr}[2]{\bmath{#1} \!\times\! \bmath{#2}}
\newcommand{\vdiv}{\bmath\nabla\!\cdot\!}
\newcommand{\vgrad}[1]{\nabla{#1}}
\newcommand{\vcurl}[1]{\vpr{\nabla}{#1}}

\newcommand{\oder}[2]{\frac{d #1}{d #2}}
\newcommand{\soder}[2]{\frac{d^2 #1}{d #2^2}}
\newcommand{\pder}[2]{\frac{\partial #1}{\partial #2}}
\newcommand{\spder}[2]{\frac{\partial^2 #1}{\partial #2^2}}
\newcommand{\Pd}[1]{\partial_{#1}}

\newcommand{\ort}[1]{ \bmath{e}_{#1} }

\newcommand{\sub}[1]{_{\mbox{\tiny #1}}}
\newcommand{\beq}{\begin{equation}}
\newcommand{\eeq}{\end{equation}}

\newcommand{\fracp}[2]{\left(\frac{#1}{#2}\right)}
\newcommand{\0}{{(0)}}
\newcommand{\1}{{(1)}}
\newcommand\T{\rule{0pt}{2.6ex}}       
\newcommand\B{\rule[-1.2ex]{0pt}{0pt}} 

\def\apjl{Astrophys. J. Lett.}
\def\apjs{Astrophys. J. Supp. Ser. }

\def\aap{Astron. Astrophys. }

\title[Splitting method for  RMHD]{A splitting method for numerical relativistic magnetohydrodynamics }
\author[Komissarov \& Phillips] 
{
Serguei S. Komissarov \thanks{E-mail: s.s.komissarov@leeds.ac.uk (SSK)},
David Phillips \thanks{E-mail: mini1000@btinternet.com (DP)}\\
Department of Applied Mathematics, The University of Leeds, Leeds, LS2 9JT, UK}

\begin{document}
\date{Received/Accepted}
\maketitle

\begin{abstract} 
We describe a novel splitting approach to numerical relativistic magnetohydrodynamics (RMHD) designed to expand its applicability to the domain of  ultra-high magnetisation (high-$\sigma$). In this approach, the electromagnetic field is split into the force-free component and its perturbation due to the plasma inertia. Accordingly, the system of RMHD equations is extended to include the subsystem of force-free degenerate electrodynamics and the subsystem governing the plasma dynamics and the perturbation of the force-free field. The combined system of conservation laws is integrated simultaneously, to which aim various numerical techniques can be used, and the force-free field is recombined with its perturbation at the end of every timestep.  To explore the potential of this splitting approach, we combined it with a 3rd-order WENO method, and carried out a variety of 1D and 2D test simulations.  The simulations confirm the robustness of the splitting method in the high-$\sigma$ regime, and also show that it remains accurate in the low-$\sigma$ regime, all the way down to $\sigma=0$.  Thus, the method can be used for simulating complex astrophysical flows involving a wide range of physical parameters. The numerical resistivity of the code obeys a simple ansatz and allows fast magnetic reconnection in the plasmoid-dominated regime.  The results of simulations involving thin and long current sheets agree very well with the theory of resistive magnetic reconnection.        
\end{abstract}   
                                                                                       

\section{Introduction}
\label{introduction}

The strong gravity of astrophysical black holes and neutron stars creates some of the most extreme physical conditions in the Universe which cannot be achieved in research laboratories. In particular, they naturally develop magnetospheres with extremely high plasma magnetisation.  Highly-relativistic winds and jets emerging from these magnetospheres create spectacular structures on enormous scales, from parsecs (pulsar wind nebulae) to hundreds of kiloparsecs (jets of active galactic nuclei).  These flows transport huge amounts of energy in the form of Poynting flux and the kinetic energy of the bulk motion, and drive the observed phenomena via magnetic reconnection and shock interaction with the external plasma.  For plasma flows on such huge scales, relativistic magnetohydrodynamics (RMHD) is the most  suitable framework.  

The starting point of modern numerical schemes for compressible hydrodynamics (HD) and magnetohydrodynamics (MHD) is their differential equations written in the form of conservation laws, with the aim of developing a numerical analog of these laws which provides conservation of the scheme-specific numerical approximation for the conserved quantities integrated over the volume of individual computational cells and the whole computational domain down to the processor rounding (machine) error. This is mainly because of the superior ability of such schemes to accurately capture shock waves.  Both finite-volume and finite-difference schemes can be developed along these lines.  The schemes may differ in many other aspects, like the type of Riemann solvers, the order of accuracy, etc.  In terms of the integration of the Faraday equation, the modern numerical schemes for MHD and RMHD split into two major groups, those which use the generalised method of Lagrange multiplier \citep[GLM,][]{Munz00,Dedner02} and those that use the constrained transport method \citep[CT,][]{CT}.            

In form, the Faraday equation of the MHD (and RMHD) system can be written as a law describing the conservation of the magnetic field integrated over the volume. This approach allows to integrate the Faraday equation in the same fashion as the other conservation laws, which is very convenient. However, it makes impossible to ensure that the magnetic field remains divergence-free, whatever numerical approximation of the divergence is chosen, because this condition involves not the volume integral of the magnetic field but its integral (magnetic flux) over a close surface.  The uncontrolled deviation from the divergence-free state is not just an error for the magnetic field alone but may have severe implication for other flow parameters \citep{BB80}. For this reason, additional algorithms were developed  to keep the magnetic field close to a divergence-free state.

In the CT method, the Faraday equation is treated as the law describing the conservation of magnetic flux over a surface.    As the result, the normal components of the magnetic field have to be defined at the faces of computational  cells and the electric field over their edges. This leads to the so-called staggered grid, where different physical parameters are defined at different points of the grid making this approach rather involved \citep[for the recent analysis and comparison of various CT schemes see ][]{MD21}. Moreover, since the energy (in Newtonian MHD) and the whole energy-momentum vector (in RMHD) include the electromagnetic contribution, integration of the corresponding conservation laws requires to define the magnetic field inside the computational cells as well. As a result, there are two numerical solutions for the magnetic fields, one defined at the cell faces and advanced using the surface form of the integral Faraday equation, and one defined inside the cells and advanced using the volume form of this equation. They may diverge over time, resulting in a rather uncomfortable state where the volume-based conserved variables like energy and momentum are inconsistent with the face-based magnetic field. In highly magnetised regions, this may even lead to an unphysical state with negative gas pressure. To prevent this, one may readjust the volume-based magnetic field and hence the conserved variables, thus making the numerical scheme not fully conservative. In Newtonian MHD, where out of all other conserved quantities the magnetic field is present only in the total energy, this can be done by simply removing the energy contribution due to the volume-based magnetic field and replacing it with the contribution due to the surface-base magnetic field \citep{BS99}.  This algorithm keeps the plasma contribution to the energy and hence the corresponding primitive variables (velocity, density and thermal pressure of plasma) unchanged, thus making the deviation from the energy conservation superficial.  In RMHD, both the magnetic and the electric field contribute to all the components of the total energy-momentum vector. Even if the volume-based magnetic field is a conserved variable,  the volume-based electric field is not and can be found only via conversion of the conserved variables into the primitive ones (velocity, density and thermal pressure of plasma) first, and then applying the perfect conductivity condition $\vv{E}=-\vpr{v}{B}$.  As a result, the direct replacement of the electromagnetic contribution to the energy-momentum like in \citet{BS99} is impossible. Instead, one is forced to carry out the ``harmonisation'' only after the variable conversion \citep{ssk-godun99}.  The same algorithm can be applied in Newtonian MHD, where it is equivalent to that of \citet{BS99}.  

In the GLM method, the system of differential equations is extended by 1) introducing new scalar dependent variable and hence one more evolution equation, and 2) modifying the Faraday equation to couple the new variable and the magnetic field. As a result, $\vdiv{B}$ can be both transported away from the regions where it is generated, normally regions with large computational errors, and diffused over the computational domain effectively dumping its spacial oscillation.  However, some residual divergence of the magnetic field remains.  This method is very easy to implement, as it does not require any modification of the computational grid, and all the evolution equations are treated in the same way.  At least in ideal Newtonian MHD, the GLM and CT methods provide similar accuracy \citep{MTB10}.          

In contrast to Newtonian MHD, where it is sufficient to describe the plasma magnetisation by the ratio of thermal and magnetic  pressures ($\beta=p/p\sub{m}$), relativistic plasma requires a different parameter $\sigma=b^2/4\pi w$, where $w=\epsilon+p$ is the relativistic enthalpy of plasma (where the internal energy density $\epsilon$ includes the contribution due to the rest mass energy of plasma particles), and $b^2=B^2-E^2$ is the Lorentz invariant equal to the squared  strength of magnetic field as measured in the rest frame of plasma.  Direct extension of the numerical methods developed for Newtonian MHD to RMHD has been quite successful in the low-$\sigma$ regime.  However, the high-$\sigma$ regime has turned out to be very problematic, as these codes tend to crash, resulting in conserved variables which cannot be converted into physically meaningful set of primitive variables.   In multi-dimensional simulations, these schemes begin to fail when $\sigma\sim 1$. On the one hand, this is a very high magnetisation, never achieved in laboratory plasmas. On the other hand, it can be much higher in many problems of relativistic astrophysics.   For example, in the pulsar magnetospheres, $\sigma$ can be as high as $10^3-10^6$.   

It has been suggested that the origin of this issue is the stiffness of the conservation laws of RMHD in the high-$\sigma$ regime  \citep{ssk-michigan}.   Basically, when $\sigma$ is high, the electromagnetic field dominates in the total energy and momentum. In this case, it is reasonable to expect that even small errors in the magnetic field, associated with the numerical  integration of the Faraday equation, may lead to large errors in the plasma parameters, when they are computed from the conserved quantities.  The quantitative analysis of the errors is rather complicated, however.   To strengthen the argument, one may approach this problem from a different angle.  The dynamics of electromagnetic field in highly rarefied plasma can be described using the approximation of the Force-Free Degenerate Electrodynamics \citep[FFDE, e.g.][]{MT82,Uchida97,Gruzinov99,ssk-ffde}).  Normally, it is formulated as the Maxwell equations complimented with a constraint on the electric 4-current, which ensures vanishing Lorentz-force.  The density of electric charges required to satisfy this constraint is quite small and the corresponding energy-momentum density of plasma can be negligibly small compared to that of the electromagnetic field.   Alternatively, one may consider FFDE as RMHD in the limit  $\sigma\to\infty$ \citep{ssk-ffde}. In this limit, the set of the differential equations of RMHD reduces to the Faraday equation and the energy-momentum conservation laws for the electromagnetic field, complemented with the two perfect conductivity conditions.   However, this system is overdetermined, with only two out of the four components of the energy-momentum equation being independent.  For the numerical integration, this implies that any error in the computed magnetic field makes it inconsistent with the computed energy-momentum density.  

For adiabatic flows, one can eliminate the energy equation from the set of numerically integrated equations of RMHD and this helps to extend the range of manageable magnetisation up to $\sigma\approx 100$ \citep{kbvk-07,Noble09}.  The conversion of remaining conserved variables to the primitive variables may still fail from time to time, becoming increasingly more severe for higher $\sigma$ and requiring emergency fixes just to keep the simulations going.  Nonetheless, these results support the stiffness of RMHD equations as the reason for the high-$\sigma$ failures, as the omission of the energy equation reduces this stiffness.   

In their CT scheme for ideal RMHD, \citet{MB06} applied exactly the same energy correction as \citet{BS99}, before converting the conserved variables into the primitive ones and replacing the cell-based magnetic field with the face-based one.   This allowed them to succeed with the cylindrical explosion test for $B_0=1$ (see section \ref{sec:cyl-expl}), which would failed otherwise.  \citet{Marti15} went further and proposed an iterative algorithm for correcting the total energy-momentum vector, using the correction of \citet{MB06} as a first step.  This allowed them to avoid the conversion failure in the explosion test with $B_0=100$. The corresponding plasma magnetisation in this problem is extremely high, with $\sigma > 2.5\times 10^3$!  This is a remarkable achievement, but some caution is in order. First, these corrections of conserved variables are different to that of \citet{BS99} as they change not only the energy-momentum of the electromagnetic field but the energy-momentum of plasma as well. So, this is more than just resetting the cell-based magnetic field using the face-centred magnetic field. Second, there is no analysis to justify these corrections. The mathematical problem supposed to be solved by the iterative algorithm of \citet{Marti15} is actually not posed.  Even a qualitative understanding of why these corrections help to keep the conserved variables in the physical part of their domain is missing.  As the result, it is not clear how robust these fixes are.  A proper analytical and numerical investigation of this approach is required.       

In addition to the conversion failures, RMHD simulations of highly magnetised flows suffer from excessive artificial plasma heating due to numerical resistivity. For this reason, the high-$\sigma$ region is normally excluded when modelling black hole emission in simulations \citep[e.g.,][]{EHTVIII}. 

In this paper, we propose an alternative to emergency fixes in the form of a radically new approach to computational RMHD which allows to mitigate the stiffness of its equations in the high-$\sigma$ limit.  In part, it was motivated by the method used by \citet{Tanaka94} for MHD simulations of the collision between the Earth's magnetosphere and the solar wind. In this problem, the Earth's magnetic field increases by many orders of magnitude from the collision site to the troposphere, where it is largely stationary and dipolar, whereas the perturbation of this field remains of about the same magnitude and hence increasingly small relative to the undisturbed Earth's field on approach to the troposphere. If the total magnetic field is evolved numerically, the truncation errors become large in comparison to the perturbation amplitude and hence the numerical solution for the perturbation becomes corrupted.   To overcome this problem, \citet{Tanaka94} proposed to separate the strong stationary dipolar field from its perturbation and hence to integrate only the nonlinear equations governing the perturbation. This approach has been proved to be very effective, and nowadays it is widely used in numerical modelling of planetary magnetospheres \citep[e.g.][]{Eggington20}.  

Our problem is more complicated, however, as in the most interesting astrophysical applications, the strong background magnetic field is highly dynamic and cannot be considered as a known stationary component.  At the first sight, this could be handled by allowing it to evolve according to the evolution equations of FFDE \citep{ssk-bz01,ssk-mns,McK06,Mahl21}. However, over time, the RMHD solution for the electromagnetic field could significantly deviate from the FFDE solution, with the force-free component and its perturbation having similar amplitudes.  To keep the electromagnetic perturbation small, one could reset the division of the electromagnetic field into the strong force-free and perturbation components. The simplest way of doing this is to recombine the force-free field and its perturbation into the 'refreshed' force-free field, and to nullify the   perturbation at the same time.  In sufficiently simple problems, this could be done only so often. However, in some other problems, the perturbation may grow very rapidly. For example, consider a stationary fast magnetosonic shock. Since the fast modes of FFDE propagate with the speed of light, the FFDE solution will strongly deviate from the RMHD solution already after one time step of numerical integration.  This shows that to make the scheme robust, one has to invoke the resetting every time step.  

Numerical integration of FFDE equations, either in the form of the Maxwell equations with force-free current \citep{Gruzinov99} or in the form of reduced RMHD equations \citep{ssk-ffde}, does not conserve the electromagnetic energy-momentum and hence the splitting approach cannot ensure the conservation of the total energy-momentum down to the processor rounding error.  However, a departure from this conservation seems inevitable for high-$\sigma$ RMHD anyway, because it is the attempt to ensure the full conservation that leads to crashes. 
      
In this paper, we describe a successful attempt to develop the splitting method based on these ideas.  In section \ref{sec:principles}, we detail the key principles of the splitting method. Section \ref{sec:numerics} describes the specifics of its numerical implementation. The 1D test simulations are presented in section \ref{sec:1Dtests}. In addition to the standard tests involving hyperbolic waves of RMHD, this section also describes the investigation of the scheme's numerical resistivity and the possibility to control the plasma heating via numerical magnetic dissipation. Section \ref{sec:2Dtests} describes the test simulations for inherently 2D problems. These include the investigation of the anisotropy of numerical resistivity, and a number of problems involving current sheets. The latter constitute a study focusing on the ability of ideal MHD codes to capture fast magnetic reconnection, apparently the first study of this kind. The whole study is summarised in section \ref{sec:summary} and the key conclusions are stated in section \ref{sec:conclusions}.  Appendix \ref{sec:app-weno}  describes the novel 3rd-order WENO interpolation employed by our code, and  Appendix \ref{app:vc} gives the derivation of the key equations involved in the variable conversion algorithm.

\section{The key principles}
\label{sec:principles}

\subsection{Ideal Relativistic Magnetohydrodynamics}
\label{sec:IRMHD}

For an inertial frame of Minkowski spacetime, the system of ideal RMHD in consists of the Faraday equation
\beq
 \Pd{t}\vv{B} + \vcurl{E} =0 \,, 
\eeq
the energy equation
\beq
 \Pd{t} \left(\frac{E^2 +B^2}{2}+w\gamma^2 -p\right)  + \vdiv (\vpr{E}{B} + w\gamma^2\vv{v}) =0 \,,
\label{eq:en} 
\eeq
the momentum equation 
\begin{align}
\nonumber
&\Pd{t}\left( \vpr{E}{B}  + w\gamma^2 \vv{v}\right)  + \\
&\vdiv \left[ -\vv{E}\otimes\vv{E} - \vv{B}\otimes\vv{B} +w\gamma^2 \vv{v}\otimes\vv{v} 
+\mbox{\bf {g}} \left( \frac{E^2+B^2}{2} +p\right) \right]  =0 \,,
\label{eq:mom}
\end{align}
 the continuity equation 
 \beq
\Pd{t} (\rho\gamma) + \vdiv (\rho\gamma\vv{v}) =0 \,,
\label{eq:cont}
\eeq
 the divergence-free condition for the magnetic field
\beq
   \vdiv\vv{B}=0 \,,
\label{eq:divB}
\eeq
and the perfect conductivity condition 
\beq
    \vv{E} =  -\vv{v}\times\vv{B} \,.  
\label{eq:pcond}
\eeq
Here $p$ is the thermodynamic pressure, $\rho$ is the density of plasma particles rest mass,   ${\mathbf g}$ is the metric tensor of space, $\vv{v}$ is the fluid velocity, $\gamma$ is the corresponding Lorentz factor, $\vv{B}$ and $\vv{E}$ are the vectors of electric and magnetic field respectively as measured in the aforementioned inertial frame.  $w(p,\rho)$ is the relativistic enthalpy per unit volume. In what follows, we use the equation of state 
\beq
   w=\rho + \kappa p \etext{with} \kappa=\frac{\Gamma}{\Gamma-1}\,,
   \label{eq:eos}
\eeq
where $\Gamma$ is the ratio of specific heats. Here we utilise the relativistic units where neither the speed of light no the geometric factor $1/4\pi$ appear in the equations (for example $\sigma=b^2/w$). We also agree that for any 3-vector of the space $\vv{a}$, $a^2=a_ia^i$ and $a=\sqrt{a^2}$, and for any 4-vector $a^\nu$ of the spacetime, $a^2=a_\nu a^\nu$, and use $-+++$ signature for the spacetime.   
 
Let us now discuss how the errors in numerical integration of the energy-momentum equations can result in an unphysical state.  Consider the 4-vector of energy-momentum density, $\Pi^\mu =-T^{\mu\nu}n_\nu$, where $T^{\mu\nu}$ is stress-energy-momentum tensor, and $n_\nu$ is the 4-velocity of the fiducial observer who measures the energy and momentum. When the observer is at rest in the space, $n_\nu=-\delta^t_\nu$ and $\Pi^\nu=(\cE,\vv{S})$, where $\cE$ and $\vv{S}$ are the energy and momentum densities respectively.  For the electromagnetic field, this is 
\beq
\Pi\sub{em}^\nu= \left(\frac{E^2+B^2}{2},\vpr{E}{B}\right)\,, \etext{where} \vv{E}=-\vpr{v}{B}\,,  
\eeq 
and 
\beq
\Pi\sub{em}^2 =-\frac{1}{4}\left( B_\parallel^2 + \frac{B_\perp^2}{\gamma^2}\right)^2 <0\,,
\eeq 
where $\vv{B}_\parallel$ and $\vv{B}_\perp$ are the components of the magnetic field parallel and orthogonal to the velocity, respectively. 
Hence $\Pi^\nu\sub{em}$ is a time-like 4-vector.  For the plasma (fluid), 
\beq
\Pi^\nu\sub{pl} = (w\gamma^2-p,w\gamma^2\vv{v})\,,
\eeq
and
\beq
\Pi\sub{pl}^2=\gamma^2w(2p-w)-p^2 \,.
\eeq 
For the physical range of specific heats, $1\!<\!\Gamma\!<2\!$, and the combination $2p-w=p\dfrac{\Gamma-2}{\Gamma-1}-\rho$ is strictly negative. Hence  $\Pi^\nu\sub{pl}$ is also a time-like 4-vector.   For the total energy-momentum vector $\Pi\sub{t}=\Pi\sub{em}+\Pi\sub{pl}$,
\beq
\Pi\sub{t}^2 = \Pi\sub{em}^2+\Pi\sub{pl}^2 +2(-\cE\sub{em}\cE\sub{pl} +(\vv{S}\sub{em}\!\cdot\!\vv{S}\sub{pl}) ) <0 
\eeq
and hence $\Pi^\nu\sub{t}$ is also time-like (This is a particular case of the general result on the sum of future-directed time-like vectors of spacetime.).  

Obviously, if the numerical integration results in a space-like $\Pi\sub{t}^\nu$, the  conversion of conserved quantities into primitive ones will fail.  However, this is the same as in the numerical relativistic hydrodynamics,  and hence the truncation errors arising in the numerical integration of the energy-momentum equations  are unlikely to be the cause of the problems specific to the high-sigma regime of RMHD. The source of errors specific only to RMHD is the Faraday equation.   In the rest of this section, we demonstrate that sufficiently large errors in magnetic field can render the set of conserved variables unphysical by pushing the energy-momentum vector of plasma $\Pi\sub{pl}^\nu$ into the space-like domain.       

We start with a physical state with the magnetic field $\vv{B}_0$ and the total energy-momentum $\Pi^\nu\sub{t,0}$. Then we analyse other states with $\Pi\sub{t}^\nu =\Pi\sub{t,0}^\nu$ but  $\vv{B}=\vv{B}_0+\delta\vv{B}$, where the small perturbation $\delta\vv{B}$ plays the role of the computational error. The aim is to determine how large this error can be before the energy-momentum vector of plasma  $\Pi\sub{pl}^\nu$ turns space-like.  In general, the impact of this error is hard to analyse.  To simplify the analysis, we assume that $\delta\vv{B}\parallel \vv{B}\sub{0}$ and consider only the cases where $\vv{v}\sub{0}$ is either parallel or perpendicular to  $\vv{B}\sub{0}$.  Since the total momentum vector 
$$
\vv{S}\sub{t}=(B^2+w\gamma^2)\vv{v} -(\spr{v}{B})\vv{B} \,,
$$
$\vv{v}$ will remain either parallel or perpendicular to $\vv{B}$ in the perturbed state as well.   We will also assume that the magnetic field strength increases, $B=B\sub{0}+\delta B$ where $\delta B>0$, as only this case is constraining. 

When  $\vv{v}\sub{0}\parallel \vv{B}\sub{0}$, the electromagnetic momentum $\vv{S}\sub{em,0}=\vv{0}$, and hence
\beq
\Pi\sub{t,0}^2 = \Pi\sub{em,0}^2+\Pi\sub{pl,0}^2 -2\cE\sub{em,0}\cE\sub{pl,0} \,,
\eeq
and
\beq
\Pi\sub{t}^2 = \Pi\sub{em}^2+\Pi\sub{pl}^2 -2\cE\sub{em}\cE\sub{pl} \,. 
\eeq
Since $\Pi\sub{t}^2=\Pi\sub{t,0}^2$, this implies that
\beq
\Pi\sub{pl}^2=\Pi\sub{pl,0}^2  -\delta\Pi\sub{em}^2 + 2 (\cE\sub{em}\cE\sub{pl}- \cE\sub{em,0}\cE\sub{pl,0})  \,,
\label{eq:Pi-pl1}
\eeq
where $-\delta\Pi\sub{em}^2 =-(\Pi\sub{em}^2-\Pi\sub{em,0}^2) =B_0^3\delta B>0$. For  $\Pi\sub{pl}^\nu$ to be time-like, the whole expression on the right-hand side of this equation must be negative, which is the condition for the perturbed state to be physical. When $\cE\sub{pl}\ll\cE\sub{em}$, the term $\Pi\sub{pl,0}^2\ll\cE\sub{em,0}\cE\sub{pl}$ can be ignored. One may also ignore the small difference $\delta\cE\sub{em}$ between $\cE\sub{em}$ and $\cE\sub{em,0}$. With these simplifications,  the physicality condition reads
\beq
2\cE\sub{em,0} (\cE\sub{pl}-\cE\sub{pl,0}) - \delta\Pi\sub{em}^2  < 0 \,.
\label{eq:cond1}
\eeq
Since the last term in this expression is positive, this implies that the plasma energy in the perturbed state must be lower than the one in the original state. Moreover, this condition will not be satisfied by any $\cE\sub{pl}>0$ unless 
$$
- \delta\Pi\sub{em}^2 < 2\cE\sub{em,0}  \cE\sub{pl,0}  \,. 
$$
Using $\cE\sub{em,0}=B_0^2/2$, $\cE\sub{pl,0}=w\sub{0}\gamma\sub{0}^2-p\sub{0} \simeq w\sub{0}\gamma\sub{0}^2$, and utilising the fact that in this case $b^2=B^2$, we finally arrive to the upper limit on the maximum error in magnetic field   
\beq
\frac{\delta B}{B_0} \lesssim 2\frac{\gamma\sub{0}^2}{\sigma\sub{0}} \,.
\label{eq:Berr-para}
\eeq

When  $\vv{v}\sub{0}\perp \vv{B}\sub{0}$, the electromagnetic momentum does not vanish. Hence equation \eqref{eq:Pi-pl1} is replaced with 
\begin{align}
\nonumber
\Pi\sub{pl}^2&=\Pi\sub{pl,0}^2  -\delta\Pi\sub{em}^2 +\\
&+2 (\cE\sub{em}\cE\sub{pl}- \cE\sub{em,0}\cE\sub{pl,0})  
+ 2((\vv{S}\sub{em,0}\!\cdot\!\vv{S}\sub{pl,0})-(\vv{S}\sub{em}\!\cdot\!\vv{S}\sub{pl}))\,,
\label{eq:Pi-pl2}
\end{align}
where $-\delta\Pi\sub{em}^2=B\sub{0}^3\delta B/\gamma\sub{0}^4>0$, and equation \eqref{eq:cond1} with
\beq
2\cE\sub{em,0} \delta\cE\sub{pl} - 2(\vv{S}\sub{em,0}\!\cdot\!\delta\vv{S}\sub{pl}) - \delta\Pi\sub{em}^2  < 0 \,,
\label{eq:cond2}
\eeq
where $\cE\sub{em,0}=B\sub{0}^2(1+v\sub{0}^2)/2$, $\vv{S}\sub{em,0}= B\sub{0}^2 \vv{v}\sub{0}$, 
$\delta\cE\sub{pl}=\cE\sub{pl}-\cE\sub{pl,0}$, and $\delta\vv{S}\sub{pl}=\vv{S}\sub{pl}-\vv{S}\sub{pl,0}$. 
When $v\sub{0},v \ll 1$, the second term in \eqref{eq:cond2} can be ignored, leading to the condition \eqref{eq:Berr-para} with $\gamma\sub{0}=1$.  When $v\sub{0},v \simeq 1$,  one may use the approximation $\delta\vv{S}\sub{pl}\simeq \delta\cE\sub{pl} \vv{v}\sub{0}$, which yields  
$2(\vv{S}\sub{em,0}\!\cdot\!\delta\vv{S}\sub{pl}) \simeq 2B\sub{0}^2v\sub{0}^2\delta\cE\sub{pl}$.
Substituting these into \eqref{eq:cond2}, one obtains the simplified physicality condition 
$$
\frac{B\sub{0}^2\delta\cE\sub{pl}}{\gamma\sub{0}^2} - \delta\Pi\sub{em}^2  < 0 \,.
$$
Thus, like in the parallel case, $ \delta\cE\sub{pl}$ must be negative. Moreover, this condition will not be satisfied by any $\cE\sub{pl}>0$ unless 
$$
 - \delta\Pi\sub{em}^2  < \frac{B\sub{0}^2\cE\sub{pl,0}}{\gamma\sub{0}^2} \,,
$$
which leads to the upper limit on the error in magnetic field 
\beq
\frac{\delta B}{B_0} \lesssim \frac{\gamma\sub{0}^2}{\sigma\sub{0}} \,,
\label{eq:Berr-perp}
\eeq
where we applied $B^2=\gamma^2 b^2$. Interestingly, this limit differs from the one obtained for the parallel case only by the factor of two, suggesting that there is no strong dependence on the angle between the velocity and magnetic field.  The accuracy constraints may be even more restrictive than those we have derived. For example, we did not take into account that the plasma energy of the perturbed  state cannot fall below $D=\rho\sub{0}\gamma\sub{0}$, the conserved variable that has to be preserved in the perturbed state as well. 

In the standard approach, one may try to tackle the issue of conversion failures via increasing the accuracy of the magnetic field integration.   In the splitting approach, we seek to reduce the impact of the errors in the magnetic field on the energy-momentum of plasma.

\subsection{Splitting equations of ideal RMHD into the electromagnetic field and plasma components}

Let us split the electromagnetic field into two components

$$
   \vv{B} = \vv{B}_\0+\vv{B}_\1 \, ,\quad 
\vv{E} = \vv{E}_\0+\vv{E}_\1 \, ,
$$
where the component with the suffix 0 satisfies the equations of FFDE.  Here we use the formulation of FFDE due to \citet{ssk-ffde}.
\beq
 \Pd{t}\vv{B}_\0 + \vcurl{E}_\0 =0 \,, 
 \label{eq:Faraday0}
\eeq
\beq
 \Pd{t} \fracp{E_\0^2 +B_\0^2}{2}  + \vdiv (\vv{E}_\0 \!\times\!\vv{B}_\0) =0 \,,
\label{eq:en0} 
\eeq
\begin{align}
\nonumber
&\Pd{t}\left( \vpr{E_\0}{B_\0} \right)  + \\
&+\vdiv \left( -\vv{E}_\0\otimes\vv{E}_\0 - \vv{B}_\0\otimes\vv{B}_\0 
+\vv{g} \fracp{E_\0^2+B_\0^2}{2} \right)  =0 \,,
\label{eq:mom0}
\end{align}
\beq
   \vdiv\vv{B}_\0=0 \,,
 \label{eq:divb0}
\eeq
and
\beq
    (\vv{E}_\0\!\cdot\!\vv{B}_\0) = 0 \,.
\label{eq:EB}
\eeq
The last equation is one of the constraints imposed by the perfect conductivity. The second one is 
\beq
    B_\0> E_\0\,.
\label{eq:BgtE}
\eeq
The energy-momentum equations can be considered as the corresponding equations of RMHD in the limit of vanishing plasma inertia.  Equations \eqref{eq:EB} and \eqref{eq:BgtE} follow from the perfect conductivity condition $\vv{E}=-\vpr{v}{B}$.  
Conversely, equations \eqref{eq:EB} and \eqref{eq:BgtE}  ensure the existence of inertial frames where the electric field vanishes. One of these frames is the rest frame of plasma, the others move relative to it along the magnetic field.  
When conditions \eqref{eq:EB} and \eqref{eq:BgtE}  are satisfied, the FFDE system is hyperbolic, with a pair of fast magnetosonic modes and a pair of Alfv\'en modes.  There are seven evolution equations in the FFDE system, \eqref{eq:Faraday0}-\eqref{eq:mom0}. Together with the algebraic constraint \eqref{eq:EB} imposed by the prefect conductivity condition, this gives eight equations in total\footnote{The divergence-free state of the magnetic field is preserved by the Faraday condition and hence does not need to be counted. The condition $E>B$ does not affect the evolution, until it gets broken, and for this reason it is not counted too.}. 
This exceeds by two the number of dependent variables (components of $\vv{B}_\0$ and $\vv{E}_\0$).  This is because only two components of the energy-momentum equations are independent \citep{ssk-ffde}.  For numerical integration, however, this means that the system of equations is overdetermined, and in order to convert the energy-momentum density into the electric field, some of the components of the energy-momentum have to be ignored, which can be done in many different ways. Our algorithm will be described later in Sec.\ref{sec:variableconv}.

The component with suffix 1 describes the correction (perturbation) of the force-free field due to the plasma inertia. The equations governing this component of the electromagnetic field, and at the same time the motion of plasma, are obtained from the full system of RMHD by removing the terms cancelling each other via equations  \eqref{eq:Faraday0}-\eqref{eq:divb0}. This yields 
\beq
 \Pd{t}\vv{B}_\1 + \vcurl{E}_\1 =0 \, , 
 \label{eq:Faraday1}
\eeq
\begin{align}
\nonumber
 & \Pd{t}\left( \vv{E}_\0\!\cdot\!\vv{E}_\1 + \vv{B}_\0\!\cdot\!\vv{B}_\1 + \dfrac{(E_\1^2+B_\1^2 )}{2}+ w\gamma^2 -p \right)  +\\  
 & +\vdiv\left( \vv{E}_\0\!\times\!\vv{B}_\1 + \vv{E}_\1\!\times\!\vv{B}_\0 + \vv{E}_\1\!\times\!\vv{B}_\1 +
 w \gamma^2 \vv{v} \right) =0 \,, 
  \label{eq:en1}
\end{align}
\begin{align}
\nonumber
& \Pd{t} \Big( \vv{E}_\0\!\times\!\vv{B}_\1  +  \vv{E}_\1\!\times\!\vv{B}_\0 + \vv{E}_\1\!\times\!\vv{B}_\1  + w\gamma^2 \vv{v}  \Big) + \\
\nonumber
 & +\vdiv \Big(-\vv{E}_\1\otimes\vv{E}_\0 -\vv{E}_\0\otimes\vv{E}_\1 
 -  \vv{E}_\1\otimes\vv{E}_\1  \\
 \nonumber
 & \quad\quad\,\,\, - \vv{B}_\1\otimes\vv{B}_\0 - \vv{B}_\0\otimes\vv{B}_\1 - \vv{B}_\1\otimes\vv{B}_\1 \\ 
\nonumber
 &  \quad\quad\,\,\, +\vv{g} \Big[  \vv{E}_\0\!\cdot\!\vv{E}_\1  + \vv{B}_\0\!\cdot\!\vv{B}_\1 + \dfrac{E_\1^2 +B_\1^2}{2}\Big]   \\
 &\quad\quad\,\,\, + w\gamma^2 \vv{v}\otimes\vv{v} +\vv{g}p  \Big)  =0 \,, \\
  \label{eq:mom1}
\end{align}
\beq
\Pd{t} \rho\gamma + \vdiv \rho\gamma\vv{v} =0 \, ,
\label{eq:cont1}
\eeq
\beq
   \vdiv\vv{B}_\1=0 \,,
     \label{eq:divb1}
\eeq
\beq
    \vv{E}_\1 =  -\vv{v}\times\vv{B}_\1  -(\vv{E}_\0  +\vv{v}\times\vv{B}_\0) \,.  
\label{eq:pcond1}
\eeq
The energy-momentum equations \eqref{eq:en1}-\eqref{eq:mom1} do not involve the terms quadratic in $\vv{B}_\0$ and $\vv{E}_\0$, which are dominant in problems with high $\sigma$. As a result, the effect of the truncation error in calculations of  $\vv{B}_\0$  on plasma parameters is reduced.  The FFDE field still enters the plasma equations via linear terms. These interaction terms describe both the effect of the electromagnetic field on the plasma motion, and the effect plasma inertia on the evolution of the electromagnetic field.  The two components of the electromagnetic field are also coupled via the perfect conductivity equation \eqref{eq:pcond1}.

\subsection{Numerical splitting} 
\label{sec:NumSplitting}

Each time-step of numerical integration consists of following three sub-steps: 
\begin{enumerate}
\item Given the solution at the time $t^n$, including $\vv{B}^n$ and $\vv{E}^n$, one introduces 

\begin{align}
\vv{B}_\0^n&=\vv{B}^n,\\
\vv{E}_\0^n&=\vv{E}^n,\\
\vv{B}_\1^n&=0,\\
\vv{E}_\1^n&=0.
\end{align}
  
\item
The combined equations of the FFDE and perturbation subsystems are integrated simultaneously to obtain the solution at the time $t^{n+1}=t^n+\Delta t$. The evolution equations of the combined system are conservation laws and can be written as a single vector equation
\beq
\Pd{t}\vv{q}+\vdiv\vv{f}=0,  
\eeq
where 
$$
\vv{q}=\begin{pmatrix}
\vv{q}_\0 \\ \vv{q}_\1
\end{pmatrix}
\etext{and}  
\vv{f}=\begin{pmatrix}
\vv{f}_\0 \\ \vv{f}_\1
\end{pmatrix} \,,
$$ 
are  the vectors of conserved variables and their fluxes respectively. 
The sub-vector of  conserved FFDE variables is
\beq
\vv{q}_\0=\begin{pmatrix}
\vv{B}_\0 \\ \cE_\0 \\ \vv{S}_\0
\end{pmatrix}  \,,
\eeq
where 
\beq
\cE_\0=\dfrac{E_\0^2+B_\0^2}{2}\,,\quad \vv{S}_\0= \vv{E}_\0\times\vv{B}_\0\,,
\eeq
and the sub-vector of conserved perturbation variables is
\beq
\vv{q}_\1=\begin{pmatrix}
\vv{B}_\1 \\ \cE_\1 \\ \vv{S}_\1\\D
\end{pmatrix} \,,
\eeq 
where
\beq
\cE_\1=(\vv{E}_\0\!\cdot\!\vv{E}_\1) + (\vv{B}_\0\!\cdot\!\vv{B}_\1) + \dfrac{(E_\1^2+B_\1^2 )}{2}+ w\gamma^2 -p
\label{eq:energy1}
\eeq
\beq
\vv{S}_\1=\vv{E}_\0\!\times\!\vv{B}_\1  +  \vv{E}_\1\!\times\!\vv{B}_\0 + \vv{E}_\1\!\times\!\vv{B}_\1 
+ w\gamma^2 \vv{v} 
\eeq
\beq
D=\rho\gamma \,.
\eeq

\item 
The total electromagnetic field vectors at the time $t^{n+1}$ are computed  via 
\begin{align}
\vv{B}^{n+1} &=\vv{B}_\0^{n+1}+\vv{B}_\1^{n+1},\\
\vv{E}^{n+1} &=\vv{E}_\0^{n+1}+\vv{E}_\1^{n+1}.
\end{align}
 
\end{enumerate}

Using simple conditional switches, the computer code based on this splitting scheme can be turned into a FFDE code and a standard ( unsplit ) RMHD code. To run it in the FFDE mode, one simply has to integrate only the FFDE equations and keep $\vv{B}_\1=\vv{E}_\1=0$. To run it in the standard RMHD mode, one has bypass the splitting step (i),  integrate only the perturbation equations, and keep  $\vv{B}_\0=\vv{E}_\0=0$.  This will be used later for testing the splitting approach against the standard one in the low-$\sigma$ regime.

\subsection{Controlled energy transfer}
\label{sec:controllednr}

In the splitting method, the energy-momentum of the force-free component of the electromagnetic field is separated from the energy-momentum of plasma.  Thanks to this separation, the errors arising in the numerical integration of the Faraday equation for the this field can no longer directly impact the state of plasma and result in a conversion failure. This is the main advantage of the separation approach.   On the other hand, this separation also prohibits the plasma heating via numerical resistivity. In some cases, this can be considered as beneficial. However, this may be detrimental in problems involving current sheets, where the magnetic dissipation and plasma heating are paramount.   

In ideal MHD simulations, the numerical resistivity  arises via the rounding errors emerging in numerical integration of the Faraday equation. In 'good' schemes, it leads mostly to diffusion of the magnetic field through plasma and reduction of its spatial gradients. This smoothing out of the magnetic field is accompanied by reduction (dissipation)  of the magnetic energy.  In standard conservative schemes for MHD, the total energy is conserved, which implies that this reduction of magnetic energy is fully compensated by the increase of plasma energy.  The rate of this numerical plasma heating is fixed implicitly by the algorithm for integration of the Faraday equation.  This lack of control may lead to undesirable numerical heating of plasma.  For example, the highly magnetised plasma in the accretion disk funnel emerging in numerical simulations of the black hole accretion gets heated to extremely high temperature for this very reason  \citep[e.g.,][]{EHTVIII}.   
The splitting approach, allows us to introduce control over the energy transfer between the electromagnetic field and plasma associated with the rounding errors.          

At some point during the integration step (ii), the conserved quantities are converted into the primitive ones.  In particular,  $\cE_\0^{n+1}$ and $\vv{S}_\0^{n+1}$ are converted into $\vv{E}_\0^{n+1}$. Given the nominal over-determinacy of the FFDE subsystem, this can be done only if we reduce the number of equations used for the conversion. There are many ways of doing this, each time departing from the computed conserved variables in one way or another.   Here we follow \citet{ssk-ffde} and compute the electric field via 
\beq
\vv{E}_\0^{n+1}=\frac{1}{{B_\0^{n+1}}^2}\vv{S}_\0^{n+1} \times\vv{B}_\0^{n+1} 
\label{eq:S0E0}
\eeq
if $\vv{B}_\0^{n+1}\ne\vv{0}$, otherwise $\vv{E}_\0^{n+1}=\vv{0}$.  Computing the electric field this way ensures the perfect conductivity condition $\vv{E}_\0^{n+1}\cdot\vv{B}_\0^{n+1}=0$.  However, the obtained electric field may exceed, especially at current sheets, the magnetic one, breaking the second perfect conductivity condition  \eqref{eq:BgtE}. Whenever this takes place, the electric field $E_\0$ is reduced somewhat below $B_\0$ ( in the test simulations to the level $0.9999B_\0$ ), or  to zero if $B_\0=0$.  This amounts to dissipation of the FFDE electromagnetic energy \citep[e.g. ][]{ssk-ebh,ssk-mns}. 
Even without this rescaling of the electric field, the electromagnetic energy density based on the obtained $E_\0^{n+1}$ and $B_\0^{n+1}$,
\beq
\tilde{\cE}_\0^{n+1}=({{E}_\0^{n+1}}^2+{{B}_\0^{n+1}}^2)/2\,,
\eeq
will be different from $\cE_\0^{n+1}$ obtained via integration of the energy equation \eqref{eq:en0}, giving rise to the energy difference    
\beq
\delta \cE_\0^{n+1} =\cE_\0^{n+1}-\tilde{\cE}_\0^{n+1}\,. 
\eeq
When $\delta \cE_\0^{n+1}>0$, the electromagnetic energy dissipates. Transferring the dissipated energy to the perturbation subsystem can only decrease $\Pi\sub{pl}^2$ and should not result in conversion failure.     

To further support this conclusion, consider unmagnetised fluid with conserved variables $D=\rho\gamma$,  $S\sub{pl}=w\gamma^2v$,  $\cE\sub{pl}=w\gamma^2-p$, and determine the response of the  gas pressure $\delta p$ to  the energy variation $\delta\cE\sub{pl}$ under constant $D$ and $S$.   Straightforward calculations show that 
$$
\delta p= {\cal A}\,\delta \cE\sub{pl} \,,
$$
where
$$
    {\cal A}=\frac{w\gamma^2 + \kappa p(\gamma^2-1)}{\rho\gamma^2(\kappa-1) +\kappa p (\gamma^2(\kappa-2) +1)} \,,
$$
$\kappa=\Gamma/(\Gamma-1)$, and $\Gamma$ is the ratio of specific heats. For $1<\Gamma<2$, the proportionality coefficient ${\cal A}$ is positive, and hence  $\delta \cE$ and $\delta p$ have the same sign. This suggests that the transfer of $\delta \cE_\0^{n+1}>0$ to the perturbation system 
\beq
\cE_\1^{n+1} \to \cE_\1^{n+1}+ \delta \cE_\0^{n+1} \,.
\eeq
will result in plasma heating. 

When $\delta \cE_\0^{n+1}<0$, its transfer to the perturbation subsystem may increase $\Pi\sub{pl}^2$ and even make it positive, thus leading to the variable conversion failure. To avert the danger, in this case the energy transfer is turned off. Our test simulations show that this allows to almost completely eliminate the conversion failures even in problems with extremely high $\sigma$.       

In smooth regions away from current sheets, the numerical heating of plasma can be undesirable. Thus, one may opt not to transfer the numerically dissipated energy of the FFDE system to the perturbation system even if $\delta \cE_\0^{n+1}>0$. In such smooth regions, $\delta \cE_\0^{n+1}$ is significantly smaller than in current sheets and this can be used to design a suitable switch-off criterion. In our code, we implemented the energy transfer condition   
\beq
     \delta \cE_\0^{n+1}>\alpha\sub{e} \cE_\0^{n+1} \,,
     \label{eq:etrans}
\eeq 
where the switch-off parameter $\alpha\sub{e}\ge 0$. When $\alpha\sub{e}=0$, the transfer takes place whenever  $\delta \cE_\0^{n+1}>0$, and  when $\alpha\sub{e}=1$, it is turned off completely.  In most of the test simulations, we used  $\alpha\sub{e}=10^{-3}$.

Finally, equation \eqref{eq:S0E0} ignores the component of  $\vv{S}_\0^{n+1}$ aligned with $\vv{B}_\0^{n+1}$,  emerging because of the computational errors. As the result, the momentum density corresponding to  $\vv{B}_\0^{n+1}$ and $\vv{E}_\0^{n+1}$, obtained via the variables conversion algorithm,   
$$
\tilde{\vv{S}}_\0^{n+1}= \vv{E}_\0^{n+1}\times\vv{B}_\0^{n+1}\,,
$$
also differs from the conserved variable $\vv{S}_\0^{n+1}$,   
\beq
\delta \vv{S}_\0^{n+1}=\vv{S}_\0^{n+1}-\tilde{\vv{S}}_\0^{n+1} \ne 0\,.
\eeq
Thus, one may consider transferring not only energy but the momentum as well. We have not been able to find an suitable algorithm for this transfer, though.

\section{Numerical implementation}
\label{sec:numerics}

To integrate the conservation laws of the split RMHD, we used a third-order finite-difference scheme. In this, we closely followed the scheme ECHO developed by \citep{DZ07} for unsplit RMHD equations. There are, however, few significant differences: 1) Use of the GLM approach  instead of the CT method to enforce the differential constraints \eqref{eq:divb0} and \eqref{eq:divb1}. GLM delivers similar accuracy to CT \citep{MTB10} for ideal MHD, but it is much easier to implement;   2) Use of a novel 3rd-order WENO reconstruction algorithm; 3) Switching the DER operator \citep{DZ07} off at shock waves to reduce numerical oscillations; 
4) New variables conversion algorithm adjusted to the peculiarities of the split RMHD equations.              

\subsection{GLM approach} 

To keep the magnetic field approximately divergence-free, we follow the method called Generalised Lagrange Multiplier \citep[GLM, ][]{Munz00,Dedner02}. Hence, we introduce two additional dependent variables $\Phi_\0$ and $\Phi_\1$, one per each subsystems, and replace the Faraday equations (\ref{eq:Faraday0},\ref{eq:Faraday1}) and the divergence-free conditions  (\ref{eq:divb0},\ref{eq:divb1}) with 
\beq
 \Pd{t}\vv{B}\sub{(s)} + \vcurl{E}\sub{(s)}  +\vgrad\Phi\sub{(s)} =0 \,, 
 \label{eq:Faradays}
\eeq
\beq
   \Pd{t}\Phi\sub{(s)} +\vdiv\vv{B}\sub{(s)} =-\kappa\Phi\sub{(s)}  \,,
 \label{eq:divbs}
\eeq
where $s=0,1$. In the test simulation, we use $\kappa=0.2/\Delta t$, making the $e$-folding time for $\Phi\sub{(s)} $ (in the case of vanishing $\vdiv\vv{B}\sub{(s)} $) equal to 5 integration time-steps $\Delta t$.  

\subsection{Time integration} 
\label{sec:TI}

Since this is a finite-difference scheme, the numerical solution $\vv{q}^n_{i,j,k}$ describes the values of variables $\vv{q}$ at the grid-points with coordinates $(x_i,y_j,z_k)$ at the discrete time $t^n$.  Here we utilise Cartesian coordinates and uniform spatial grid with $x_i=x_1 +(i-1) \Delta x$, $y_j=y_1 + (j-1) \Delta y $,  $z_k=z_1 + (k-1)\Delta z $, where $\Delta x=\Delta y=\Delta z=h$. These grid-points can considered as central points of rectangular computational cells. The interfaces of these cells are located at $x_{i\pm1/2} =x_i \pm h/2$, $y_{j\pm1/2} =y_j \pm h/2$, and $z_{k\pm1/2} =z_k \pm h/2$. The time grid is also uniform, $t^n=t_0 +\Delta t n$ with $\Delta t = \mbox{C} h$, where C is the Courant number.     
 
The finite-difference equations have the form
\beq
\oder{\cal Q}{t}={\cal F}({\cal Q}) \,,
\label{eq:FDS}
\eeq
where ${\cal Q}$ is a one, two, or three dimensional  array, depending on the dimensionality of the problem. Each entry of this array is the vector $\vv{q}$  at the corresponding grid point. ${\cal F}$ is an array of the same dimension and size as ${\cal Q}$. Each entry of this array is the numerical finite-difference approximation for $-\vdiv{\vv{f}}+{\cal S}\sub{Q}$ at the corresponding grid point, where ${\cal S}\sub{Q}$ is the vector of source terms. In the case of Cartesian coordinates, the source terms emerge only in the GLM equations.   
The system of ODEs \eqref{eq:FDS} is integrated using 3rd order Strong Stability Preserving (SSP) version of the Runge-Kutta method \citep{ShOs88}. Hence, 
\beq
{\cal Q}^{n+1} = {\cal Q}^n + \frac{\Delta t}{6} (k_1+k_2+4k_3) \,,
\eeq
where
\begin{align*}
k_1=&{\cal F}({\cal Q}^n)\,,\\
k_2=&{\cal F}({\cal Q}^n+\Delta t\, k_1) \,,\\
k_3=&{\cal F}\left({\cal Q}^n+\frac{\Delta t}{4} (k_1+k_2)\right) \,.
\end{align*}

The finite-difference approximation for $\vdiv{\vv{f}}$ is computed in the following steps: 

\begin{enumerate}
\item Conserved variables are converted into the primitive variables. This is needed because interpolating conserved variables may yield an unphysical  state.    

\item A 3rd order WENO interpolation is used to setup Riemann problems at the cell interfaces. 

\item HLL Riemann solver \citep{HLL} is used to find upwind flux densities $\vv{f}$ at the interfaces.  

\item  Central quartic polynomial interpolation is used to reconstruct the distribution of $\vv{f}$ in each coordinate direction and hence to find  the 3rd-order approximation for  $\vdiv{\vv{f}}$  \citep[DER operation of ][]{DZ07}.  This works fine for smooth solutions, but may introduce oscillations at shocks, often leading to crashes in high-$\sigma$ regime.  To avoid this, the computational domain is scanned for shock fronts and a 'safety zone' is set around them. Within the safety zone, a second-order TVD interpolation is used instead of the WENO interpolation.  
\end{enumerate}

\subsection{3rd order WENO interpolation}
\label{sec:WENO}

Weighted Essentially Non-Oscillatory (WENO) interpolation invokes linear combination of lower order sub-stencil polynomials to achieve a higher-order accuracy in smooth sections of numerical solution  and lower-order almost-non-oscillatory interpolation in rough sections  \citep[shocks ][]{LOC94,Shu20}. This is achieved by making the weights of the polynomials dependent on some quantitative roughness indicators. WENO approach have enjoyed great success over the years, especially after its efficient implementation by  \citet{JS96}. 
Later, however, it was found that their nonlinear weights have a drawback, resulting in significant reduction of accuracy in smooth regions with critical points. Since realistic numerical models often involve local extrema in numerous locations, especially in the case of turbulent flows, this is a major disadvantage.  \citet{HKY20} proposed new weights for 3rd-order WENO interpolation. Their test results look impressive, but the approach is not intuitive and hard to comprehend.    \citet{HEN05} derived new weights for 5th-order WENO interpolation via mapping the original weights of \citet{JS96} to the improved set. Here, we adopt a similar strategy to derive an improved set of weights for a 3rd-order scheme.  In particular,  we start with the weights of the second order Total Variation Diminishing (TVD) scheme \citep{F-91},  modify it to address the issue of critical points, and then use these TVD weights to produce 3rd-order WENO weights (see Appendix \ref{sec:app-weno}).

Below,  only the interpolation in the $x$ direction is considered, and all other spatial indices are dropped for brevity. In the other directions, the procedure is the same.

\subsection{Hyperbolic fluxes}

Given the left $\vv{u}^l$ and right $\vv{u}^r$ states at the interface, the flux density normal to the interface is computed using the approximate Riemann solver by \citet{HLL}. Namely, 

\beq
    \vv{f}_n = \frac{a^{+}\vv{f}_n^l+a^{-}\vv{f}_n^r}{a^{+}+a^{-}} - 
    a^{+}a^{-}\frac{\vv{q}^r-\vv{q}^l}{a^{+}+a^{-}} \,,
\eeq
where $\vv{f}_n^{l,r}=\vv{f}_n(\vv{u}^{l,r})$, $\vv{q}^{l,r}=\vv{q}(\vv{u}^{l,r})$,  and  
\beq
a^{\pm}=\max(0,\pm\lambda_n^{\pm}(\vv{u}^l), \pm\lambda_n^{\pm}(\vv{u}^r),   
\eeq
where $\lambda_n^\pm$ are the speeds of fastest hyperbolic modes moving relative to plasma in the positive and negative directions along the normal to the interface.  We use separate wave speeds for the FFDE and perturbation subsystems. For the FFDE subsystem,  $\lambda_n^\pm=\pm 1$. For the perturbation subsystem, we use the speeds of fast magnetosonic waves (as in unsplit  RMHD equations).   These are computed using the computationally-cheap approximation 
\beq
\lambda_n^\pm = \frac{(1-a^2)v_n \pm\sqrt{a^2(1-v^2)\left[ (1-v^2a^2)-(1-a^2)v_n^2) \right]}}{1-v^2a^2} \,,
\eeq
where 
\beq
a^2=c_s^2 +c\sub{A}^2 - c_s^2 c\sub{A}^2 \,, 
\eeq
$c_s$ is the sound speed, $c\sub{A}$ is the Alfv\'en speed, and $v_n$ is the velocity component normal to the interface \citep{GMT03}.
The HLL solver is stable and diffusive. Its diffusivity can be a drawback, but it is also a strength. It helps to smooth-out the numerical solution in complex regions with non-monotonic spatial variations of large amplitude, where large truncation errors may lead to an unphysical set  of conserved variables.   

\subsection{Finite-difference approximation for the flux divergence}
\label{sec:DER}

Given the array of upwind fluxes at cell interfaces,  we look for a 3rd-order accurate approximation for $\vdiv\vv{f}$ at the cell centres (grid points).  To simplify the presentation, consider a gridline aligned with the $x$ direction, choose a particular grid point on this line, reset its index to zero, and measure the position of other points relative to this one, so that $x_0=0$. Then introduce the 4-point stencil $S=\{x_{-3/2},x_{-1/2},x_{1/2},x_{3/2}\}$ centred on this grid point, denote the corresponding upwind fluxes in the direction of the gridline as $\{ \vv{f}_{-3/2},\vv{f}_{-1/2},\vv{f}_{1/2},\vv{f}_{3/2} \}$, and use the 3rd-order interpolating polynomial  $\vv{p}(x)=\vv{a}_3x^3+\vv{a}_2x^2+\vv{a}_1x+\vv{a}_0$ to reconstruct the distribution of $\vv{f}$ around $x=0$. Its derivative $d\vv{p}/dx(0)=\vv{a_1}$ gives us the require 3rd-order approximation for $\Pd{x}\vv{f}_0$. It is easy to verify that the final results is
\beq
\Pd{x}\vv{f}_{_0} = \frac{9}{8}\frac{\vv{f}_{1/2}-\vv{f}_{-1/2}}{\Delta x} -\frac{1}{8} \frac{\vv{f}_{3/2}-\vv{f}_{-3/2}}{3\Delta x} \,.
\label{eq:fda}
\eeq    

Using a somewhat different approach, \citet{DZ07} derived this result (where it is called the DER step) in a different form. Restoring the normal cell indexation,  it reads
\beq
\Pd{x}\vv{f}_{_i} = \frac{(\vv{F}_{i+1/2}-\vv{F}_{i-1/2})}{\Delta x}\,,
\label{eq:fda1} 
\eeq
where 
\beq
  \vv{F}_{i+1/2} =-\frac{1}{24} \vv{f}_{i+3/2} +\frac{26}{24} \vv{f}_{i+1/2} -\frac{1}{24} \vv{f}_{i-1/2} \,.
\eeq
Equation \eqref{eq:fda1} is the same form as in finite volume schemes for conservation laws, where the place of $\vv{F}$ is taken by the interface  flux $\vv{f}$ at the cell interface.  This tells us that this finite-difference scheme provides an exact conservation to the integral quantities computed via the second order accurate approximation 
\beq
\int_{v} \vv{q}dV \approx \sum_{i,j,k} \vv{q}_{i,j,k}\Delta V_{i,j,k} \,.
\label{eq:conserved-value}
\eeq
This approximation is neither upwind nor ENO/WENO, and hence may, and does, introduce oscillations at strong shocks.  In the high-$\sigma$ regime, these oscillations can be fatal, resulting in a failure of the variable conversion.  For this reason, we implemented a strong-shock-finder algorithm and, in a safety zone around them,  replace \eqref{eq:fda1} with  
\beq
(\Pd{x}\vv{f})_{_i} = \frac{\vv{f}_{i+1/2}-\vv{f}_{i-1/2}}{\Delta x}\,. 
\eeq
This is a step towards the 2nd-order TVD scheme, like in \citep{ssk-godun99}, which allows to prevent the shock oscillations almost completely.  

The strong-shock identification algorithm is currently based on these two criteria. 

1) The central difference approximation is used to estimate the 3-divergence of $\vv{u}=\gamma \vv{v}$ at the tested grid point.  It is required to be negative with 
$$
|\vdiv{\vv{u}}|>\alpha_u u\,,
$$
where $\alpha_u>0$ is a strength factor, and $u$ is the amplitude of $\vv{u}$ at this point. 

2) The same approximation is used to estimate the gradient of total pressure $p\sub{tot}=p+(B^2+E^2)/2$.  It it required to satisfy the condition
$$
|\vgrad{p}\sub{tot}|>\alpha_p p\,,
$$   
where $\alpha_p>0$ is another strength factor, and $p$ is the value of gas pressure at this point. The $p\sub{tot}$ variation pressure is compared against the gas pressure $p$, because in the high-$\sigma$ regime the relative variation of magnetic pressure can remain low even at strong shocks, where other flow parameters change significantly.  In the test simulations, we use, $\alpha_u=\alpha_p=0.5$.   

One can make one more step and replace even the WENO interpolation with the TVD interpolation in the safety zone.  

\subsection{Variables conversion}
\label{sec:variableconv}

For the FFDE subsystem, the conversion is relatively straightforward and already described in Sec.\ref{sec:controllednr}.
For the perturbation subsystem, $B_\1$ and $\Phi_\1$ are both the primitive and conservative at the same time and do not need converting. Thus, we need to compute the primitive variables $p$, $\rho$, $w$, $\vv{v}$, and $\vv{E}_\1$ given the conservative variables $\cE_\1$, $\vv{S}_\1$, $D$,  the values of $B_\1$, $B_\0$ and $E_\0$,  the equation of state $w=w(p,\rho)$, and the perfect conductivity equation \eqref{eq:pcond1}.  Using the conductivity equation one can easily eliminate $\vv{E}_\1$ from the set of unknowns. It is also relatively easy to eliminate one of the thermodynamic variables using the equation of state \eqref{eq:eos}.  Then one can use the Newton-Raphson method to solve the remaining system of five equations for the five unknowns,  but it is rather slow due to the high dimensionality of the problem. However, we have found a way to reduce the number of equations. The key first step of this algorithm is the recombination of the conserved variables of the FFDE and perturbation system. This yields the conserved variables of the unsplit RMHD system and hence allows to use any of the existing methods for the conversion of its variables.      Here we adapt the approach described by \citep{DZ07}.  

The recombination of conserved variables may have an adverse effect on the accuracy of the conversion, as in the high-$\sigma$ regime this involves the mixing of very large and very small terms.  However, the induction equation \eqref{eq:pcond1} alone already reintroduces the terms quadratic in $B_\0^2$ and $E_\0^2$ into the expressions for $\cE_\1$ and $\vv{S}_\1$, and so the mixing issue exists in any case.  So anyway, extra care has to be taken in order to avoid unnecessary loss accuracy in conversion calculations.   After lengthy calculations described in Appendix \ref{app:vc}, the conversion problem is reduced to finding the root of  
the transcendental equation
\beq
F(X,W(X)) = 0 \,,
\label{eq:F}
\eeq
where $X=u^2-u_0^2$, $u^2=v^2\gamma(v)^2$, $u_0^2=v_0^2\gamma(v_0)^2$, where $\vv{v}_\0=\vv{E}_\0\!\times\!\vv{B}_\0/B^2_\0$ is the drift velocity of the FFDE subsystem, and $W=w\gamma^2$. The function $F(X,W)$ is defined by the equation 
\beq
F(X,W)=W^2v^2+4\bar{\cE}_1 W +4(P(W,X)-W)\left(W+\frac{B^2}{2}\right) - A \,,
\label{eq:fF}
\eeq
where 
\beq
A=S_\1^2 +2(\vv{S}_\1\!\cdot\!\vv{S}_\0) - 2\hat{\cE}_1 B^2 - B_\0^2 v_0^2 (B_\1^2+2(\vv{B}_\0\!\cdot\!\vv{B}_\1))\,,
\eeq
\beq
\bar{\cE}_1=\cE_\1 +\frac{E_\0^2}{2}-\frac{B_\1^2}{2}-(\vv{B}_\0\!\cdot\!\vv{B}_\1) \,,
\eeq
and
\beq
\hat{\cE}_1=\cE_\1-\frac{B_\1^2}{2}-(\vv{B}_\0\!\cdot\!\vv{B}_\1) \,,
\eeq	
are constants, and  
\beq
P(W,X)=\frac{1}{\kappa}(W/\gamma^2-D/\gamma)\,,
\eeq
is the function describing the gas pressure as a function of the enthalpy and flow speed. In these equations, $B^2$ is the squared magnitude of the total magnetic field $\vv{B}=\vv{B}_\0+\vv{B}_\1$.

The function $W(X)$ is defined as the positive root of the cubic equation
\beq
W^3+a_2(X) W^2 + a_0 =0 \,,
\label{eq:W}
\eeq
where
\beq
a_2(X) = \frac{A_1(X) +A_2}{A_3(X)}
\eeq
where
$$
A_1(X)=\frac{B_\0^2 X}{2(1+u^2)(1+u_\0^2)}+ 
\frac{v^2}{2}(B_\1^2+\vv{B}_\0\!\cdot\!\vv{B}_\1) + \frac{D}{\gamma\kappa} \,,
$$
$$
A_2= - \cE_\1+\frac{B_\1^2}{2}+\vv{B}_\0\!\cdot\!\vv{B}_\1 \,,
$$
$$
A_3(X)=1-\frac{1}{\gamma^2\kappa}\,,
$$
and 
\beq
a_0=-\frac{1}{2}\frac{(\vv{S}_\0\!\cdot\!\vv{B}_\1+\vv{S}_\1\!\cdot\!\vv{B})^2}{A_3(X)} \,.
\eeq
\citet{DZ07} used $W$ and $v^2$ as their iteration variables. We opted for $X=u^2-u_\0^2\in[-u_\0^2,+\infty)$ instead of $v^2$, and hence $\gamma^2=\gamma^2(X)$ and $u^2=u^2(X)$, to increase the accuracy in computation of the cubic coefficient $a_2$. If we used $v^2$, the calculations of $A_1(v^2)$ would involve the subtraction $v^2-v_\0^2$, resulting in a significant loss of accuracy when $v^2 \approx v_\0^2\approx 1$. In the high-$\sigma$ regime, this error would further increase due to the multiplication by the large factor $B_\0^2$.  

Since $a_0\le0$, this cubic equation always has one non-negative real root. This root vanishes only when $a_2>0$ and $a_0=0$. When $a_2<0$, this is the only real root of the cubic.  Obviously, finding accurate numerical value for the root is important for the accuracy of the whole conversion algorithm. If $a_2<0$, it is sufficient to use the modified  Cardano's method as described in \citep{Press92}, though one has to avoid numerical subtraction of almost equal large terms when computing the discriminant of the reduced cubic when $|a_0/a_2^3|\ll 1$.   The first step of this method involves depression of the cubic via introduction of the new variable $Y=W+a_2/3$.   When $a_2>0$ and $|a_0/a_2^3|\ll 1$, the positive root  $W \ll a_2$, and computing it via $W=Y-a_2/3$ involves significant loss of accuracy.  In this case, we follow \citet{Blinn06} and introduce another variable $\bar{Y}=1/W$, which also reduces the cubic equation to the depressed form, but no shift is involved. After this, the standard prescription is used again.         

Equation \eqref{eq:F} is solved numerically via either the secant or the Brent-Dekker method \citep{Dekker69,Brent71}. 
The secant method is tried first, using the value of $X$ in the solution at the previous timestep as the initial guess. When $\sigma$ is not extremely large, this method finds the root  $X\ge -u_0^2$, provided it exists,  down to the rounding error (machine precision) after no more than 10 iterations. When $\sigma$ is very high, it may fail to converge, getting trapped in an oscillation about the root. Whenever the secant method fails, the Brent-Dekker method is tried instead. To start the method, one has to find an interval $[a,b]$, with $a\ge -u_0^2$, which includes the root, and hence  $F(a)F(b)<0$.  We start with a reasonably narrow interval containing the initial guess first, and then, if it does not contain the root, exponentially decrease the distance between $a$ and $-u_0^2$ and exponentially increase the distance between $a$ and $b$. When such interval is found, the method always converges to the root, though in extreme cases this may take up to 60 iterations to reach the rounding-error level.  

\begin{figure*}
\centering
 \includegraphics[width=0.5\textwidth]{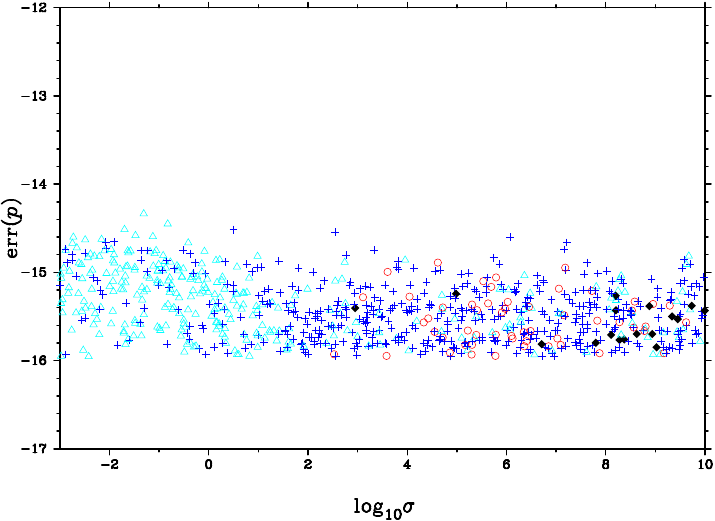}
 \caption{Relative error of pressure  in the variables conversion algorithm. The type of plotting marker describes the number of iterations $n\sub{it}$ required: green triangles when $n\sub{it}\le 5$, blue crosses when $5<n\sub{it}\le 10$, red circles when $10<n\sub{it}\le 20$, and black diamonds when  $n\sub{it}> 20$.}  
 \label{fig:conv-errors}
\end{figure*}

To test the conversion algorithm, we used the Monte-Carlo method, first to set up the exact parameter state within the parameter space, and then to produce the initial guess. Figure \ref{fig:conv-errors} shows the relative error in the gas pressure against the magnetisation $\sigma$, for one of such tests.  Given the extreme values of $\sigma$ used in the test and not a single incident of convergence failure, we are almost 100 percent certain that when the variables conversion fails in real simulations, this is not due to some deficiencies of the conversion algorithm, but because the root $X \ge -u_0^2$ does not exist.

Once the root of (\ref{eq:F}) is found, the primitive variables are computed via
\beq
u^2=X+u_0^2, \quad v^2=u^2/(1+u^2), \quad w=W(X)/\gamma^2
\eeq
\beq
  \rho=\frac{D}{\gamma} \,,
\eeq    
\beq
p=\frac{1}{\kappa}(w-\rho) \,,
\eeq
\beq
 \vv{v}= \frac{\vv{S}+(\spr{S}{B})\vv{B}/W}{B^2+W} \,,
\eeq
and 
\beq
\vv{E}_\1=-\vpr{v}{B} -\vv{E}_\0 \,.
\eeq

\section{1D test simulations}
\label{sec:1Dtests}

In the simulations we use the EoS of ideal gas with the ratio of specific heats $\Gamma=4/3$, even when the sound speed is well below the speed of light.  In all the simulations, the Courant number C=0.5, with the exception of the Alfv\'en wave test where C=0.4. 

\subsection{Alfv\'en wave. Convergency study}  

In addition to being a fundamental wave in RMHD, the Alfv\'en wave is a great option for testing the scheme convergency rate. It is quite complex in structure due to the rotation of electromagnetic and velocity fields, quite simple to be describe analytically even without the assumption of small amplitude \citep{ssk-aw},  and allows solutions with continuous higher-order derivatives.  In the Hoffmann-Teller frame \citep{HT50}, the wave is stationary, with $B^2\,, \gamma\,, p \,,\rho\,= \mbox{const}$, $\vv{E}=0$, and 
\beq
v^i=\pm\frac{1}{\sqrt{\cal E}} B^i \,,
\eeq
where ${\cal E}=w+B^2$, and the sign decides the direction of the wave vector.  

For the test simulation we set $p=\rho=1$, and  
\beq
B^x=0.3 B_0\,, \quad B^y=B_0 \cos\phi\,, \quad B^z=B_0 \sin\phi\,, 
\eeq
where the phase variable 
$$
\phi = \arcsin(a\sin(k x))\,.
$$
To set the wave in motion, we use the Lorentz transformation to the lab frame moving with the speed $v=0.5$ in the positive x direction.  The wavenumber $k$  in the Hoffmann-Teller frame is set to yield the wavelength $\lambda=2$ in the lab frame. We set the phase variation amplitude to $a=0.3$, to ensure that the Lorentz factor does not become excessively high even for the model with the highest explored magnetisation.  The simulations run from $t=0$ to $t=2$, by which time the wave shifts to the left by exactly one half of its wavelength, and in the exact solution the profile of $B^y$ coincides with the initial one.  The Courant number is set to C=0.4 to ensure that in all runs of the convergency study the final time $t=2$ is a whole number of timesteps.  

\begin{figure*}
\centering
 \includegraphics[width=0.3\textwidth]{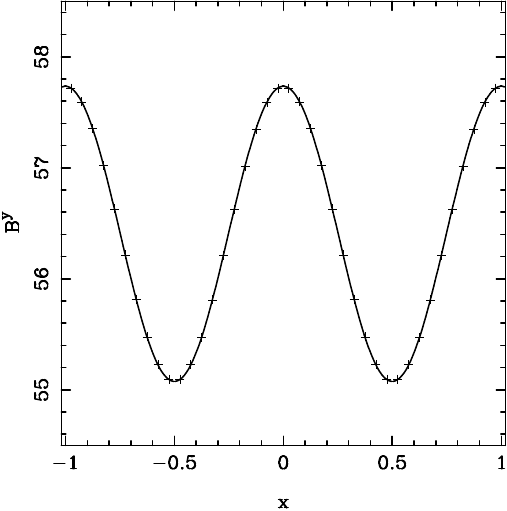}
\includegraphics[width=0.3\textwidth]{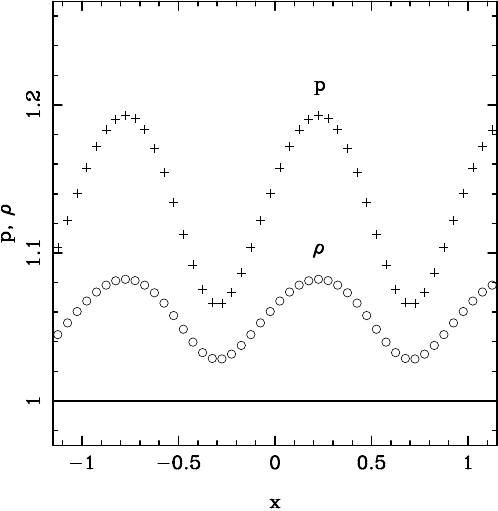} 
 \includegraphics[width=0.3\textwidth]{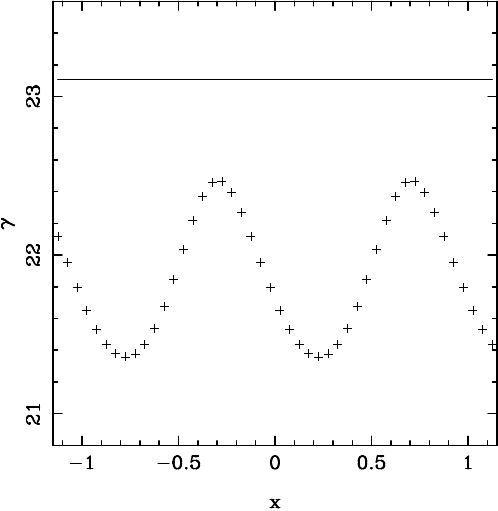} 
 \caption{Alfwen wave test.  The solid lines show the exact solution and markers show the numerical solution for the model with $B_0=50$ ($\sigma=545$) with the resolution $n_x=40$ at $t=2$. }  
 \label{fig:AW}
\end{figure*}

Figure \ref{fig:AW}  shows the results for the model with $B_0=50$, $p=\rho=1$, with the corresponding magnetisation $\sigma=545$. Table \ref{tab:AW1}  shows the results of convergency study based on this model.  One can see that the scheme shows 3rd-order behaviour already at the very low resolution. For the resolution $n_x=20$, the characteristic variation length scale for $B^y$ is only five cells.    

By varying the value of $B_0$,  it s found that $L_1(B^y)\propto \sqrt{\sigma}$ and  $L_1(\rho), L_1(p) \propto \sigma$ when $\sigma \gg 1$.       

\begin{table}
\caption{Convergency test with Alfv\'en wave simulations. Here, $n_x$ is the number of grid points (the resolution), $n_t$ is the number of time steps from the start of the run, $L_1(A)$ is the $L1$-error of the variable $A$, $r$ is the two-point estimate of the order of accuracy based on the errors for the current and previous resolutions.} 
 \label{tab:AW1}
\begin{center}
\begin{tabular}{ llcccccc } 
 \hline
 $n_x$    &    $n_t$    &    $L_1(B^y)$  & $r$ &   $L_1(\rho)$  & $r$ &   $L_1(p)$ & $r$  \T\B\\ 
 \hline
20     &  50   &  0.358e-1   & -     &  0.388e+0   &   -    &  0.107e+1  & -  \T\\
40     &  100 &  0.307e-2   & 3.5 &  0.572e-1    &  2.8 & 0.133e+0   &  3.0 \\
80     &  200 &  0.355e-3   & 3.1 &  0.755e-2    &  2.9 &  0.171e-1   & 2.9 \\
160   &  400 &  0.447e-4   & 3.0 &  0.950e-3    &  3.0 &  0.214e-2   & 3.0 \\
320   &  800 &  0.536e-5   & 3.1 &  0.119e-3    &   3.0 &  0.268e-3  & 3.0 \B\\
\hline
\end{tabular}
\end{center}
\end{table}

\subsection{Harris current sheet. Mechanisms of numerical plasma heating} 
\label{sec:heating}

The numerical resistivity determines the evolution of current sheets in ideal RMHD simulations, which makes this test particularly important for studying the possibility to control the numerical plasma heating associated with the resistivity as described in Section \ref{sec:controllednr}.   

In the initial solution,  the magnetic field $\vv{B}=(0,B^y(x),0)$ has no guide component, and  
\beq
B^y(x)=B_0 \tanh(x/a) \,,
\eeq
where $a$ is the characteristic width of the sheet and $B_0$ is the asymptotic field strength.   The electric field $\vv{E}=0$ 
and the magnetic pressure is balanced by the gas pressure   
\beq
p(x)= p_0+ \frac{B_0^2}{2} (1-\tanh^2(x/a) )\,.
\eeq
In the test problem,  $B_0 =500$, $p_0=1$ and $a=0.02$. The plasma mass density is uniform $\rho(x)=\rho_0$, with $\rho_0=1$.  The corresponding asymptotic (as $x\to\infty$) magnetisation $\sigma=54500$.  The computational domain is $(-5,5)$ with 500 grid points. This makes the current sheet  approximately 4 computational cells wide, so it is resolved but only just. Such thin current sheets do emerge in the 2D simulations described in Sec.\ref{sec:2Dtests}. To explore the impact of the energy transfer on the solution we made few runs with different values of the energy-transfer parameter $\alpha\sub{e}$. Here, the results for $\alpha\sub{e}=0$, 0.001, and 1 are presented.  In many respects, they are surprisingly similar. However, there are some revealing differences concerning the energy balance. 

\begin{figure*}
\centering
 \includegraphics[width=0.3\textwidth]{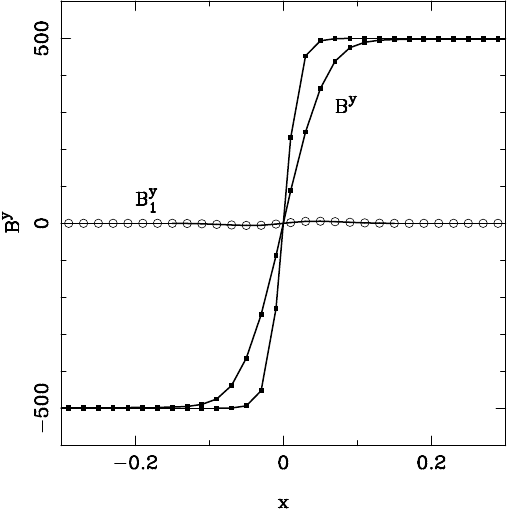} 
 \includegraphics[width=0.3\textwidth]{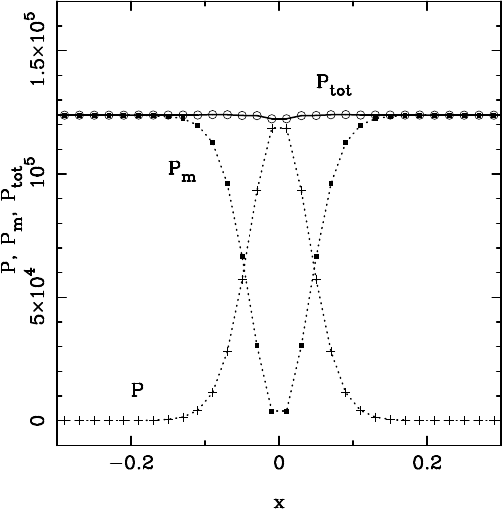}
 \includegraphics[width=0.3\textwidth]{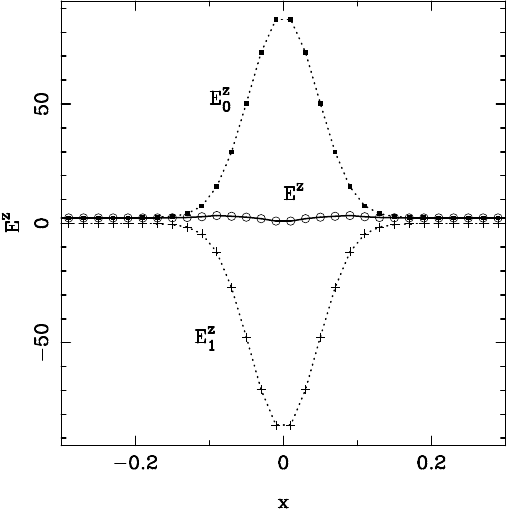} 
 \caption{1D current sheet test.  Left panel: the total magnetic field $B^y$ at $t=$0 and 5 and its perturbation component $B_\1^y$ at $t=5$. Middle panel: the gas pressure $p$, the magnetic pressure $p\sub{m}$ and the total pressure $p\sub{tot}=p+p\sub{m}$ at $t=5$. Right panel:  the total electric field $E^z$, its FFDE component $E^z_\0$ and its perturbation $E^z_\1$ at $t=5$. The energy transfer parameter is $\alpha\sub{e}=1$. In the models with $\alpha\sub{e}=0$ and 0.001, the results are very similar.}  
 \label{fig:hcsh}
\end{figure*}

Initially, the numerical resistivity is too high for the solution to maintain the pressure balance. Both the magnetic and total pressures in the middle of sheet reduce, and this tiggers fast rarefaction waves moving out at almost the speed of light. 
These waves initiate plasma flow into the current sheet. Inside the current sheet, the plasma gets heated to very high temperatures, and soon the total pressure balance across the current sheet is restored.  This active phase last up to $t=0.15$, by which time the current sheet thickness increases to about six cells.  This phase is followed by the phase of slow diffusive spreading, and by the end of the simulations, at $t=5$, the current sheet thickness is still only about ten cells ( see figure \ref{fig:hcsh}).   

The right panel of figure \ref{fig:hcsh} shows the total electric field $E^z=E^z_\0+E^z_\1$ and its force-free and perturbation components at $t=5$.  The force-free component has the sign consistent with the flow of electromagnetic energy into  the current sheet. In the pure FFDE numerical solution to this problem, $E_\0 \approx B_\0$, and  the electromagnetic energy flows into the current sheet at the speed of light. Inside the current sheet, it disappears at the central discontinuity via enforcement of the condition $B>E$,  In the split-RMHD simulations, the FFDE electric field $E^z_\0$  is checked by the perturbation field $E^z_\1$, and the total electric field $E^z$ almost vanishes.   

\begin{figure*}
\centering
 \includegraphics[width=0.3\textwidth]{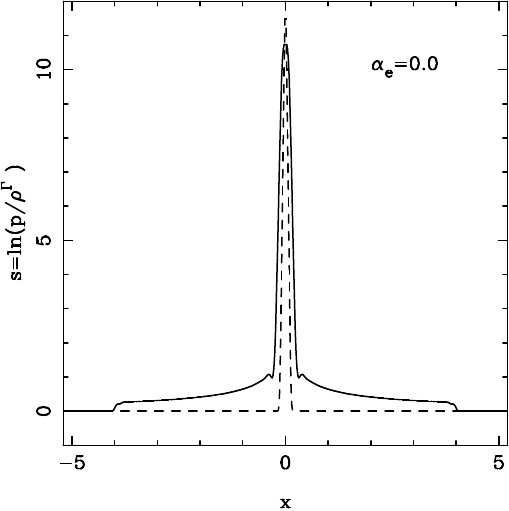}
 \includegraphics[width=0.3\textwidth]{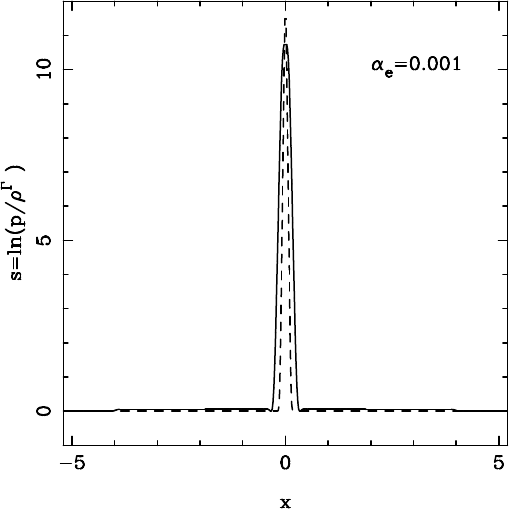} 
 \includegraphics[width=0.3\textwidth]{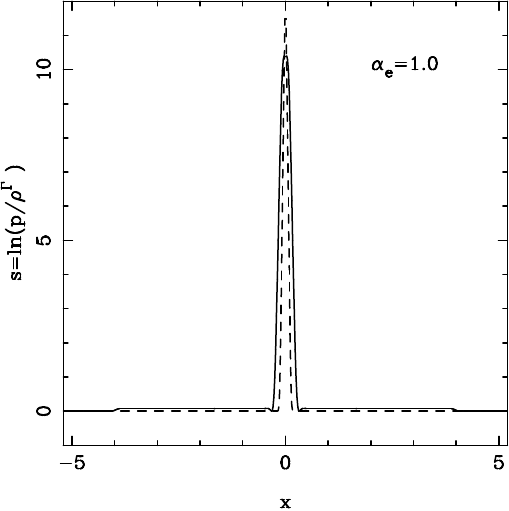}
 \caption{1D current sheet test.  The plasma entropy in the runs the energy transfer parameter $\alpha\sub{e}=0$, $10^{-3}$ and 1. The dashed lines show the initial solution and the solid lines show the solution at  $t=4$.}  
 \label{fig:csh-entr}
\end{figure*}

\begin{figure*}
\centering
 \includegraphics[width=0.3\textwidth]{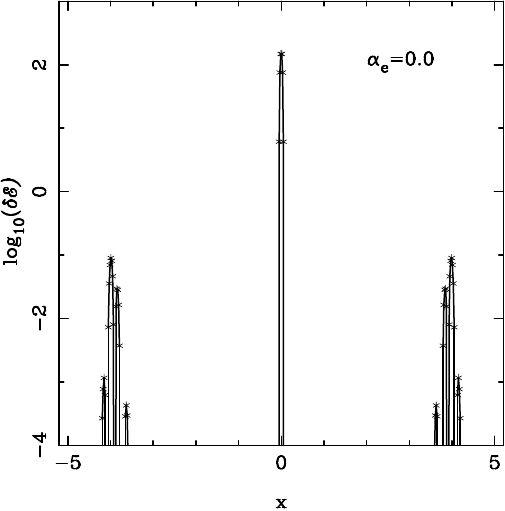}
 \includegraphics[width=0.3\textwidth]{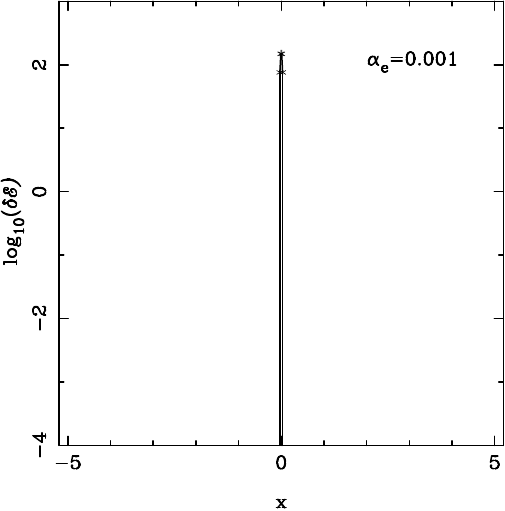} 
  \includegraphics[width=0.31\textwidth]{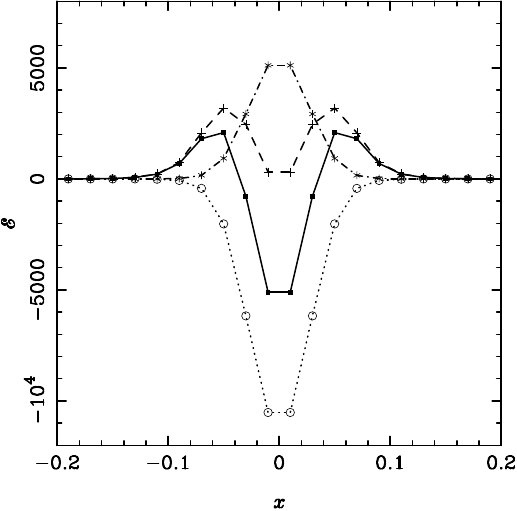} 
 \caption{1D current sheet test. {\it Left and middle panels}: Plasma heating per one integration time-step at $t=4$ for the runs with $\alpha\sub{e}=0$ (left panel) and $\alpha\sub{e}=0.01$ (middle panel). {\it Right panel}:  $\cE_{int}+\cE_{per}$ (solid line and filled squares), $\cE_{per}$ (dashed line and crosses), $(\vv{B}_\0\!\cdot\!\vv{B}_\1)$ (the dash-dotted line and stars), $(\vv{E}_\0\!\cdot\!\vv{E}_\1)$ (dotted line and circles)  at $t=2$ for the run with $\alpha\sub{e}=1$.     }
 \label{fig:csh-de}
\end{figure*}

Figure \ref{fig:csh-entr} shows the entropy $s=\ln(p/\rho^\Gamma)$, for the models with $\alpha\sub{e}=0$, 0.001 and 1 at $t=4$. The most conspicuous feature of these plot is the central peak. It manifests the plasma heating in the current sheet itself. In all three models, the peak has almost the same height and width.  The plots also show weak 'wings', most pronounced in the model with $\alpha\sub{e}=0$, which spread out by $\Delta x =4$ in the both directions. This is the wake left by the fast rarefaction wave emitted by the current sheet at the start of the simulations.   The left panel of figure \ref{fig:csh-de}  shows the energy transfer rate per time-step for the run with $\alpha\sub{e}=0$ at $t=4.5$. The central peak is the current sheet, where the numerical plasma heating continues in the central six cells.  Moreover, there are additional regions of numerical heating, which are clustered around the rarefaction waves.  They are responsible for the entropy wings in figure  \ref{fig:csh-entr}.      The irregular structure of plasma heating in the rarefaction waves shows that the sign of $\delta{\cal E}_\0^{n+1}$ fluctuates there.  One may argue that like at shock waves, the numerical heating in current sheets imitates the proper physical  processes known to operate there.   On the contrary, its is hard to see how the numerical heating at rarefaction waves can be anything but an unwelcome numerical artefact.  Fortunately, it can be suppressed by setting $\alpha\sub{e}$ slightly above zero.  The middle panel of figure \ref{fig:csh-de} shows that in the run with $\alpha\sub{e}=0.001$ the energy transfer operates only in the current sheet.           

\begin{table}
\caption{1D current sheet test. Integral energy variation by $t=4.5$ for runs with different energy transfer parameter $\alpha\sub{e}$.  ${\cal E}\sub{em}$, ${\cal E}\sub{pl}$, ${\cal E}\sub{tot}$ are the electromagnetic energy, the plasma energy and the total energy, respectively.  $\delta \tilde\cE\sub{pl}$ is the contribution of the interaction terms to the plasma energy variation. The total energy at the start is ${\cal E}\sub{tot,0}=6.30\times10^7$. } 
\label{tab:CSH}
\begin{center}
\begin{tabular}{ ccccc } 
 \hline
 $\alpha\sub{e}$    &    $\delta {\cal E}\sub{tot}$    &    $\delta {\cal E}\sub{em}$  & $\delta {\cal E}\sub{pl}$  & $\delta \tilde{\cE}\sub{pl}$  \T\B \\ 
 \hline
 
1.0      &  $-7.81\times10^5$    &  $-1.874\times10^6$     & $1.098\times10^6$  & $1.098\times10^6$ \T \\
$10^{-2}$    &  $-9.72\times10^4$    &  $-1.315\times10^6$     & $1.223\times10^6 $ & $0.748\times 10^6$ \\
$10^{-3}$  &  $4.34\times10^4$     &  $-1.224\times10^6$   & $1.273\times10^6$  & $0.690\times 10^6$ \\
0.0      &  $5.04\times10^4$     &  $-1.222\times10^6$   & $1.294\times10^6$  & $0.689 \times 10^6$ \B\\
\hline
\end{tabular}
\end{center}

\end{table}
 
Table \ref{tab:CSH} shows the variation of the total energy $\cE\sub{tot}=\cE\sub{em}+\cE\sub{pl}$ and its components for the whole system over the whole run (up to $t=4.5$).  The integrals are computed using the conservative approximation \eqref{eq:conserved-value}, 
\beq
\cE = \sum_{i=1}^{n_x} \cE_i \,,
\label{eq:tot-energy}
\eeq
where $\cE_i$ is the energy density at the $i$th grid point (the cell-length factor is ignored). In the standard conservative RMHD mode of the code, the total energy of the system would remain unchanged, $\delta {\cal E}\sub{tot}=0$ down to the rounding error, because by $t=4.5$ the rarefaction waves have not reached the domain boundaries.  The splitting scheme is not fully conservative, however, and a non-vanishing   $\delta {\cal E}\sub{tot}$ is expected.  

In the run with fully suppressed energy transfer ($\alpha\sub{e}=1$), the total energy of the system decreases by about 1\%. Some decrease is expected because the numerical resistivity reduces the energy of the FFDE system, and this reduction is not compensated via the energy transfer algorithm.  Interestingly, the plasma energy of the system still increases. Because in these simulations the bulk motion energy of plasma is very small compared to its thermal energy, this increase indicates the existence of numerical heating mechanism unrelated to the energy transfer algorithm.   To understand this mechanism, recall that  the conserved energy of the perturbation system $\cE_\1$ contains not only the plasma energy $\cE\sub{pl}=w\gamma^2 -p$, but also the interaction energy $\cE\sub{int}=(\vv{E}_\0\!\cdot\!\vv{E}_\1) +(\vv{B}_\0\!\cdot\!\vv{B}_\1)$ and the energy of the electromagnetic perturbation $\cE\sub{per}={(E_\1^2+B_\1^2 )}/{2}$ (see eq.\ref{eq:energy1}).  Hence the plasma energy itself is not conserved. 
At the start of $(n+1)$th time step, $\vv{E}_{(1),i}^n=\vv{B}_{(1),i}^n = 0$, and hence $(\cE\sub{int})_i^n+(\cE\sub{per})_i^n=0$, $(\cE_\1)_i^n=(\cE\sub{pl})_i^n$. By the end of the time step,  $(\vv{E}_\1)_i^{n+1}, (\vv{B}_\1)_i^{n+1}\ne 0$, $(\cE\sub{int})_i^{n+1} + (\cE\sub{per})_i^{n+1}\ne0$, and as a result, the plasma energy changes by  $-(\cE\sub{int})_i^{n+1}-(\cE\sub{per})_i^{n+1}$. The corresponding change of the plasma energy for the whole system during the time step is
\beq
     \delta \tilde{\cE}\sub{pl}^{n+1} = -\sum_{i=1}^{n_x} \left((\cE\sub{int})_i^{n+1}+(\cE\sub{per})_i^{n+1}\right) \,,
\eeq 
where the summation is taken over the whole grid. Over the whole run, this yields
\beq
\delta \tilde{\cE}\sub{pl} = \sum_{n=2}^{n_t} \delta \tilde{\cE}\sub{pl}^{n} \,. 
\eeq
The value of $\delta \tilde{\cE}\sub{pl} $ is shown in the last column of table \ref{tab:CSH}. For the run with $\alpha\sub{e}=1$, $\delta \tilde{\cE}\sub{pl}=\delta\cE\sub{pl}$, confirming that in this run the plasma heating is entirely via this mechanism. 

In the run with full energy transfer ($\alpha\sub{e}=0$), the solution is closer to the perfect energy conservation. Now $\delta\cE\sub{tot}$ varies by about 0.09\% only, and, in contrast to the run with $\alpha\sub{e}=1$, the total energy of the system increases.   The increase of $\cE\sub{tot}$  in this run is expected because any deficit of $\cE_\0$ is fully compensated via increase of $\cE_\1$, but the occasional surplus of $\cE_\0$  is not compensated via decrease of $\cE_\1$.  The energy transfer accounts for about  47\% of the plasma heating. For the run with $\alpha\sub{e}=0.001$, the numbers are similar, with a slight improvement of the total energy conservation. For $\alpha\sub{e}=0.01$, $\cE\sub{tot}$ decreases again and its variation grows in amplitude.   
   
In summary, the energy transfer is not the only channel of plasma heating in the splitting scheme.  However, it helps to improve the energy conservation and then accounts for up to 50\% of plasma heating in current sheets. To suppress the low-level parasitic heating away from current sheets, it helps to introduce a threshold on the transferred energy, and in the rest of the test simulations we use the threshold parameter $\alpha\sub{e}=10^{-3}$ as a default value.

\subsection{Degenerate Alfv\'en wave. The study of numerical resistivity}
\label{sec:daw-1d}

In the MHD approximation, basic theories of magnetic reconnection introduce diffusion of magnetic field lines through plasma using the model of scalar (isotropic) resistivity $\eta$, which is properly justified only for collisional plasma. It yields a relatively simple relation between the electric field and the electric current. In the 3+1 framework of resistive RMHD, this relation reads 
\beq
\vv{j} =\frac{\gamma}{\eta}\left(\vv{E}+\vpr{v}{B} - (\spr{E}{v})\vv{v}\right)+q\vv{v} \,,
\label{eq:relat-resist}
\eeq
where $q$ is the electric charge density of plasma \citep[e.g.][]{ssk-rrmhd}. For electrically-neutral plasma with the flow speed $v\ll 1$, this reduces to 
$$
\vv{j} =\frac{1}{\eta}\left(\vv{E}+\vpr{v}{B}\right) \,,  
$$  
which further reduces to $\vv{E}=\eta \vv{j}$ when $v=0$.  
When $\eta$ is constant, the magnetic field evolves according to the equation  
\beq 
\eta\left(\spder{\vv{B}}{t}-\nabla^2 \vv{B}\right) +\left(\pder{\vv{B}}{t} - \nabla\!\times\!(\vpr{v}{B})\right)=0\,.
\label{eq:telegraph0}
\eeq 
When $\cL/\cT \ll1$, where $\cL$ and $\cT$ the characteristic length and time scales of the problem, the second derivative term can be ignored and we obtain the equation of Newtonian MHD 
\beq 
\pder{\vv{B}}{t} - \nabla\!\times\!(\vpr{v}{B})-\eta\nabla^2 \vv{B} =0\,.
\label{eq:newtonB}
\eeq 
Denote as $\cT_\eta $ the time scale introduced by the resistivity. Then from \eqref{eq:newtonB} it follows that 

\beq
\cT_\eta= \eta^{-1} \cL^2 \,.
\eeq

Since we solve equations of ideal RMHD, the only kind of resistivity available in our simulations and controlling the magnetic reconnection is the numerical one. The numerical resistivity, like the numerical diffusion and viscosity, emerges from the truncation errors of the numerical scheme.  For a Runge-Kutta scheme with temporal and spatial accuracies of the same order $r$, the rounding error $\cR_1$ after one time step scales with the resolution $n_x=L/\Delta x$, where $L$ is the domain size, as 
$$
\cR^1 = \mbox{O}(n_x)^{-(r+1)} \etext{as} n_x\to\infty \,,
$$    
assuming a smooth solution \citep{ShOs88}. However, this error is local, and for a feature of the characteristic length scale $\cL\ll L$, the size of the domain does not matter.  What matters is $n_\cL=\cL/\Delta x$, the number of grid points per $\cL$.  Hence, the local error 
$$
\cR_\cL^1 \propto n_\cL^{-(r+1)} \etext{for} n_\cL\gg 1 \,. 
$$    
The number of timesteps required to reach the resistive timescale $\cT_\eta$ is $n_\eta=\cT_\eta/\Delta t$ and the total error accumulated by this time is   
$$
\cR_\cL^{n_\eta} \propto n_\eta (n_\cL)^{-(r+1)} = 
\frac{\cL^2}{\eta\Delta t} (n_\cL)^{-(r+1)}  \etext{for} n_\cL\gg 1 \,. 
$$
This accumulated error is the overall $\delta B/B$ on the resistive timescale, and hence a constant which does not depend of the particular values of $\Delta x$, $\Delta t$, and $\cL$. Hence,    

\beq
\eta\sub{num} = A_\eta \frac{\cL^2}{\Delta t}\fracp{\Delta x}{\cL}^{r+1} = A_\eta \frac{\Delta x}{\Delta t} \cL \fracp{\Delta x}{\cL}^r \,,
\label{eq:num-resist}
\eeq
where $A_\eta$ is the normalisation factor, and we replaced $\eta$ with $\eta\sub{num}$ to stress the fact this is the expression for the numerical resistivity. In this derivation, we assumed that the rounding error emerging in the numerical integration of the Faraday equation has the effect similar to that of the discretised diffusion term $\eta\nabla^2\vv{B}$. A proper analytical study of this issue is beyond the scope of this paper, and here we only check this via computer simulations.  For our scheme $r=3$, and, given the maximum wave speed being equal to the speed of light, ${\Delta x}/{\Delta t}=C^{-1}$.     

The result \eqref{eq:num-resist} is almost identical to the special case of the ansatz proposed by \citet{Remb17}, who based it on a mixture of physical and numerical reasons.  \citet{Remb17} tried to determine the normalisation factors of their ansatz by studying the decay of Alfv\'en and magnetosonic waves.  The decay of these waves depends both on the numerical viscosity and resistivity, which makes the computations rather involved. Curiously, they reported negative resistivity for their numerical scheme.  

Here we simplify the procedure by studying the problem which involves only the numerical resistivity and hence no decoupling is needed. Namely, we consider the 1D initial value problem, where in the initial solution $\vv{v}=0$, $p=p_0$, $\rho=\rho_0$, and the magnetic field $\vv{B}=B_0(0,\cos kx,\sin kx)$ rotates with $x$ at a constant rate. In ideal RMHD, this configuration is magnetostatic due to uniform magnetic and total pressures. It may be described as a degenerate limit of the Alfv\'en wave, when the wave vector $\vv{k}$ is orthogonal to the magnetic field.  In resistive RMHD with constant scalar resistivity, the magnetic field decays and this decay is accompanied by plasma heating.  However, because of the translational symmetry of the problem, the rate of decay and heating is independent on $x$ and the configuration remains magnetostatic. 

When $\vv{v}=0$, the magnetic field evolves according to the telegraph equation 
\beq 
\eta \spder{\vv{B}}{t}+\pder{\vv{B}}{t}-\eta\spder{\vv{B}}{x}=0\,.
\label{eq:telegraph}
\eeq 
When $\eta k\ll 1$, it allows the separable solution 
\beq
\vv{B}(t)=B_0 (0,\cos kx,\sin kx) \exp(-\omega t) \,,
\label{eq:wave-decay}
\eeq
where $\omega=\eta k^2$ is the decay rate of the magnetic field (This is the same as in the Newtonian limit, where the first term in \eqref{eq:telegraph} drops out.).  Thus, if the rounding errors of our scheme do indeed amount to numerical resistivity, one expects the magnetic field to decay exponentially, in which case the value of numerical resistivity can be found as $\eta\sub{num}=\omega/k^2$.   In the test simulations, the initial solution has $p_0=\rho_0=1$, and $B_\0=50$. The domain is $(0,1)$ with periodic boundary conditions and $C=0.5$. 

\begin{figure*}
\centering
\includegraphics[width=0.3\textwidth]{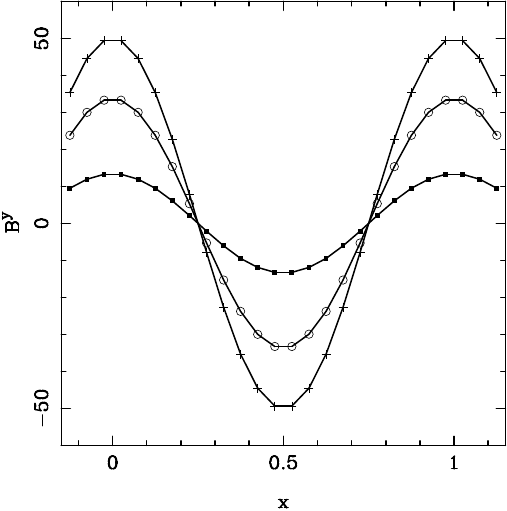}
 \includegraphics[width=0.3\textwidth]{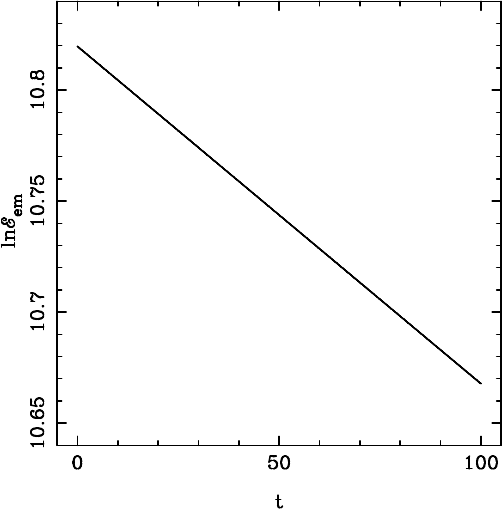}
 \includegraphics[width=0.3\textwidth]{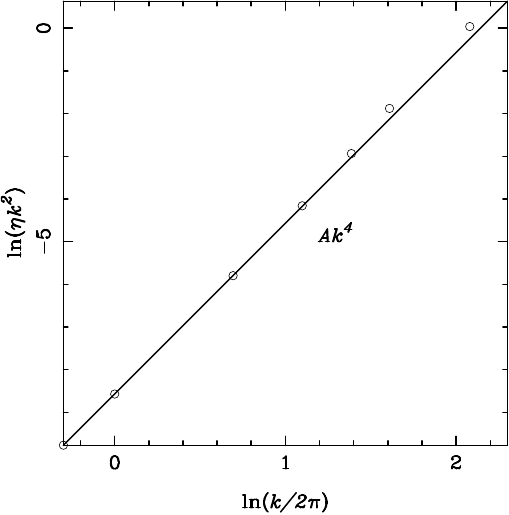}
 
  \caption{Degenerate Alfv\'en wave. Left panel: $B^y$ at $t=0$ (crosses), $t=30$ (circles), and $t=100$ (filled squares) in the run with $n_x=20$ and $k=2\pi$. Middle panel:  Evolution of the total electromagnetic energy $\cE\sub{em}$ in the run with $n_x=40$ and $k=2\pi$. Right panel: The wave decay rate $\omega = \eta k^2$ against $k$ for the models with the  resolution $n_x=80$.}
 \label{fig:resist}
\end{figure*}

The left panel of figure \ref{fig:resist} shows the evolution of the magnetic field for the model with $k=2\pi$. As expected, the wave decays keeping its shape intact. To measure the decay rate, we use the total magnetic energy of the system, computed via equation \eqref{eq:tot-energy}, which is expected to decay exponentially at the rate $2\omega=2\eta\sub{num}k^2$. It is indeed exponential, as illustrated in the middle panel of figure \ref{fig:resist}. Table \ref{tab:AW2} shows that the decay rate, and the value of $\eta\sub{num}$, decrease with the numerical resolution as $n_x^{-3}$ for sufficiently large $n_x$, in agreement with \eqref{eq:num-resist}. 

The characteristic length scale $\cL$ is based on the equation 
\beq
\soder{\vv{B}}{x} = \frac{\vv{B}}{\cL^2} \,,
\label{eq:lengthscale}
\eeq
and for this problem it yields $\cL=1/k$, independent of the location. Then equation  \eqref{eq:num-resist} predicts $\omega\propto k^4$, which is indeed the case as illustrated in the right panel of figure \ref{fig:resist}.  

Table \ref{tab:AW2} also shows the values of the normalisation constant $A_\eta$ obtained in the simulations with $k=2\pi$. One can see that for $n_x\gtrsim20$,  $A_\eta\approx 0.031$ independently of the resolution as expected.  For $n=10$, $A_\eta$ it is almost twice as high.  However, in this case the number of grid points per the length scale $n_\cL$ is only about 1.6 and a strong deviation from \eqref{eq:num-resist} is expected.  
The numerical magnetic Reynolds number of the wave problem,
\beq
\mbox{Re}_m = \frac{c\cL}{\eta\sub{num}}\,.
\label{eq:Rem}
\eeq

\begin{table}
\caption{Degenerate Alfv\'en wave simulations.  $n_x$ is the number of grid points, $r$ is the two-point estimate of the scheme order of accuracy, $\eta$ is the numerical resistivity, $Re_m$ is the magnetic Reynolds number based on the numerical resistivity, $A_\eta$ is the coefficient in the ansatz \eqref{eq:num-resist}. The wave used for the simulations has $k=2\pi$.} 
 \label{tab:AW2}
\begin{center}
\begin{tabular}{ |c|llll|} 
 \hline
 $n_x$    &    10    &    20  & 40 &  80 \T\B\\ 
 \hline
$2\omega$     &  0.39   &  $0.26\times10^{-1}$  &   $0.30\times10^{-2}$   &  $0.38\times10^{-3}$  \T \\
$\eta\sub{num}$            & $0.49\times10^{-2}$ & $0.33\times10^{-3}$ & $0.38\times10^{-4}$ & $0.48\times10^{-5}$\\
$r$ & - & 3.75 & 3.1 & 3 \\
$A_\eta$        &  0.063   & 0.034 &  0.031  &  0.031 \\
Re${_m}$ & $0.32\times10^2$  & $0.48\times10^3$ & $0.42\times10^4$ & $0.33\times10^5$ \B\\
\hline
\end{tabular}
\end{center}
\end{table}

\subsection{Self-similar rarefaction waves} 

Self-similar (simple) rarefaction waves provide very useful non-linear test problems.  Although no analytic solutions for these waves exist, the problem of finding exact numerical solutions is reduced to solving numerically a system of first order ordinary differential equations \citep[e.g.][]{ssk-godun99}. These wave are not particularly suitable for the convergency testing because of the loss of smoothness in the exact solutions at the leading (trailing) wavefronts, where already the first spatial derivative is discontinuous.  Since we have already verified the order of accuracy of our code, this is no longer required and just a visual comparison with the exact numerical solution is sufficient.  Here we present the results for a switch-off fast rarefaction and a slow rarefaction waves propagating through the same high-$\sigma$ state.

\begin{figure*}
\centering
 \includegraphics[width=0.3\textwidth]{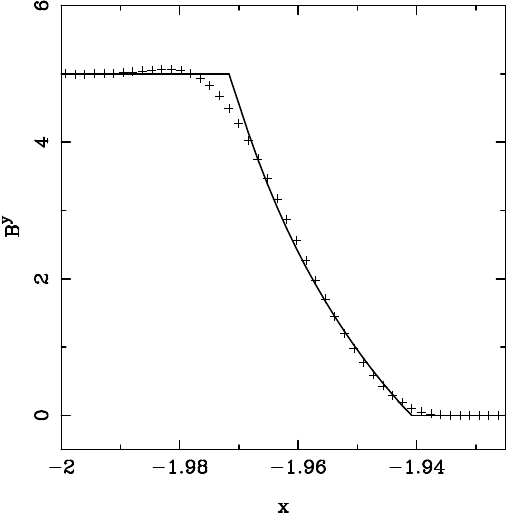}
\includegraphics[width=0.3\textwidth]{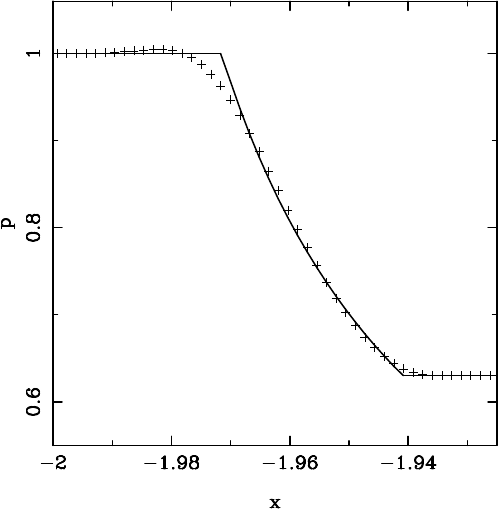} 
 \includegraphics[width=0.3\textwidth]{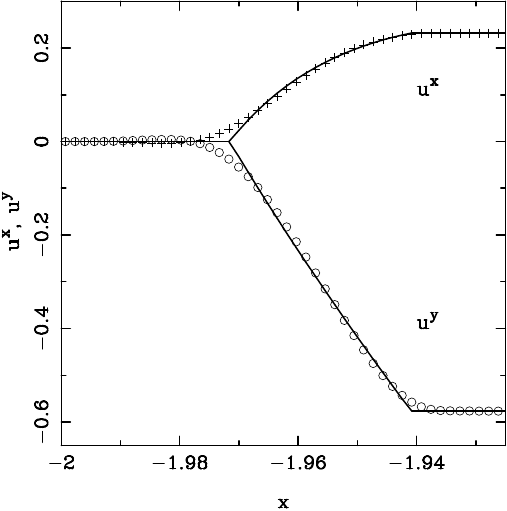} 
 \caption{Fast switch-off rarefaction wave test. The continuous lines show the exact solution, and the markers show the numerical solution at the integration time $t=1$, corresponding to the time $t\sub{R}=2$ since the resolution of the initial discontinuity.}  
 \label{fig:frw-off}
\end{figure*}

\vskip 0.2cm
\noindent
{\it Fast switch-off wave.}  This wave connects two uniforms states with the parameters $p=1$, $\rho=0.01$, $\vv{u}=(0,0,0)$, $\vv{B}=(10,5,0)$ for the left state, and $p=0.630$, $\rho=0.7076\times10^{-2}$, $\vv{u}=(0.232,-0.577,0)$, $\vv{B}=(10,0,0)$ for the right state.  The magnetisation $\sigma \approx 30$ in both the left and the right states. The wave moves to the left, with the wave speeds of the leading and trailing fronts being $v_l=-0.9856$ and $v_t=-0.9705$ respectively. These are so close because in high-$\sigma$ plasma the fast speed is very close to the speed of light, and the reduction of the tangential component of the magnetic field has little effect on the magnetisation when there is a strong normal component. Another interesting property of the wave is its limited strength in terms of the gas pressure variation. This is partly due to the fact that the fast rarefaction terminates as soon as the tangential magnetic field vanishes.    
The initial ($t\sub{R}=0$) discontinuity of the associated Riemann problem is set at $x=0$, whereas the initial ($t=t\sub{R}-1=0$) solution for the computer simulations is the exact solution to this Riemann problem at $t\sub{R}=1$.  The domain is $(-2.20, -0.90)$ with 800 grid points. Figure \ref{fig:frw-off} shows the exact numerical solution (solid lines) versus the results of computer simulations (markers) at the time $t=1$ ($t\sub{R}=2$). One can see that agreement between the solutions is quite good, apart from the vicinity of the leading and trailing fronts. The loss of accuracy near the fronts is expected due to the lack of continuity in the first spatial derivatives there.

\vskip 0.2cm
\noindent
{\it Slow switch-on wave.} This wave connects two uniforms states with the parameters  $p=1$, $\rho=0.01$, $\vv{u}=(0,0,0)$, $\vv{B}=(10, 5, 0)$ for the left state and $p=0.001$, $\rho=0.562\times10^{-5}$, $\vv{u}=(8.856,4.479,0)$, $\vv{B}=(10, 5.048, 0)$ for the right state. The magnetisation $\sigma \approx 30$ in the left state and $\sigma \approx 3\times10^{4}$ in the right state. The wave moves to the left, with the wave speeds of the leading and trailing fronts being $v_l=-0.516$ and $v_t=0.876$ respectively. Thus, relative to the computational grid, the trailing front now moves to the right.   The great contrast with 
the fast rarefaction in this regard is due to the fact that the sound speed, $c_s\approx 1/\sqrt{3}$ everywhere, is much lower than the speed of light, and so the speed of the slow mode is strongly influenced by the value of $v_x$.  Another contrasting feature is the large decrease of the gas pressure as the solution can be continued towards $p=0$ without limit. 

\begin{figure*}
\centering
 \includegraphics[width=0.3\textwidth]{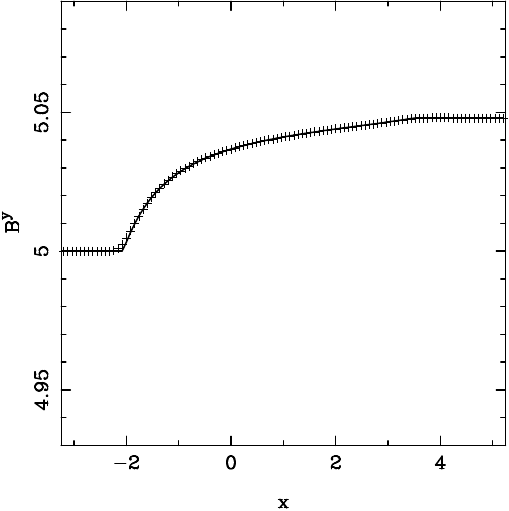}
\includegraphics[width=0.3\textwidth]{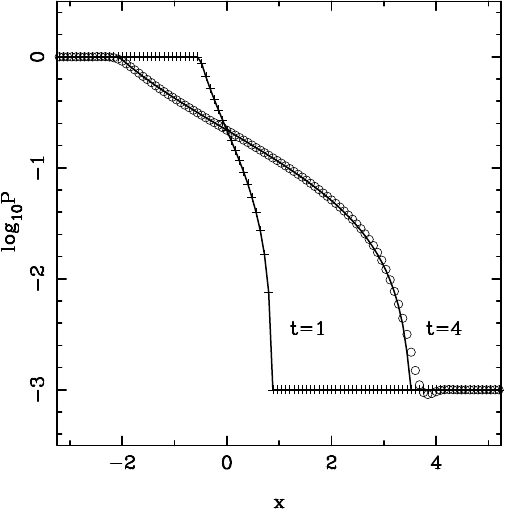} 
 \includegraphics[width=0.3\textwidth]{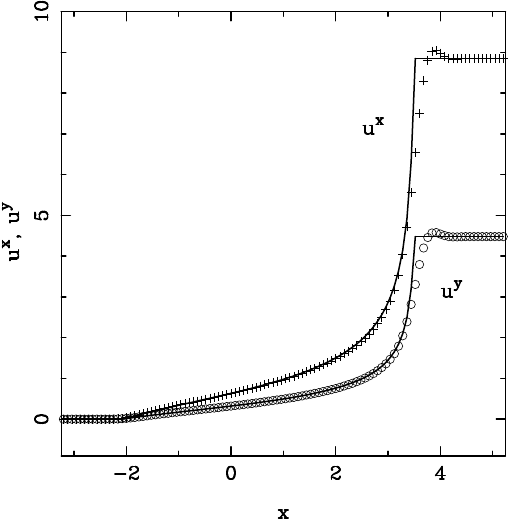} 
 \caption{Slow  rarefaction  wave test. The continuous lines show the exact solution, and the markers show the numerical solution at the integration time $t=3$, corresponding to $t=4$ since the resolution of the associated Riemann discontinuity. The middle panel also shows the exact solution at the Riemann time $t=1$, which served as an initial solution for this test. }  
 \label{fig:srw}
\end{figure*}

The initial ($t\sub{R}=0$) discontinuity of the associated Riemann problem is set at $x=0$, whereas the initial ($t=t\sub{R}-1=0$) solution for the computer simulations is the exact solution to this Riemann problem at $t\sub{R}=1$.  The domain is $(-3,5)$ with 100 grid points. This low resolution is sufficient because of the rapid spreading of the wave, in contrast to the fast wave where the spreading is very slow. Figure \ref{fig:srw} shows the exact numerical solution (solid lines) versus the results of computer simulations (markers) at the time $t=3 (t\sub{R}=4)$. Again, there is a good agreement between the solutions everywhere, apart from the vicinity of the leading and trailing fronts. The loss of accuracy near the trailing front is higher due to the higher jumps of the first derivatives there.    

\subsection{Shock waves}

Magnetosonic shock waves present the most challenging type of RMHD solutions for standard unsplit numerical schemes in the high-$\sigma$ regime. The huge variation of the spatial gradients of physical parameters at shocks even with a well-resolved numerical structure yields large numerical errors, and this increases the chance for the computed conserved variables to escape from the physically meaningful domain. The same applies to the splitting scheme.  Moreover, there may be no FFDE shock solution which can be considered as a good first approximation  to an RMHD shock. For example, fast waves of FFDE propagate in all directions with the speed of light, whereas for an RMHD shock on can always find a frame where it is stationary. This makes the perturbation component of the electromagnetic field $(\vv{B}_\1, \vv{E}_\1)$ comparable  to its FFDE component  $(\vv{B}_\0, \vv{E}_\0)$, particularly the electric field.      

We tested numerical shock solutions obtained with our scheme against the exact solutions, obtained by solving numerically the shock equations as described in \citep{MA87}.  Here the results of some of the tests are described. The corresponding solutions of the shock equations are given in table  \ref{tab:swt}.

\begin{table*}
\caption{Parameters of shock wave tests. In all the tests, the left state is the shock upstream state. $\sigma$ is the magnetisation of the upstream state, $M\sub{f}$ and $M\sub{s}$ are respectively the relativistic fast- and slow-magnetosonic Mach numbers of the shock, and $v\sub{sh}$ is the shock speed relative to the grid.} 
 \label{tab:swt}
\begin{center}
\begin{tabular}{ l|l|l|l } 
 \hline
  Test    &   Left state &   Right state  & Other parameters   \T\B\\ 
 \hline
FS5    & $\vv{v}$ =( 0.99968283E+00, 0, 0)   &  $\vv{v}$ = ( 0.99768146E+00, 0.17248747E-01, 0)    &  $v\sub{sh}$=-0.5   \T\\
	  & $\vv{B}$ =( 50, 0.19853866E+04, 0) &$\vv{B}$ =( 50, 0.19886156E+04, 0)& $M\sub{f}=2$ \\
	  & $\vv{E}$ =( 0, 0, -0.19847569E+04) & $\vv{E}$ =( 0 , 0 , -0.19831425E+04)&  $\sigma=10^3$\\
          & $p$=1.0, $\rho$=1.0 & $p$ = 0.44243911E+01, $\rho$ = 0.26176303E+01 &   \B\\
\hline
FS5A  & $\vv{v}$ =( 0, 0, 0)   &  $\vv{v}$ =  ( -0.75954175E+00, 0.16485693E+00, 0)    &  $v\sub{sh}$=-0.99989427E+00   \T\\
	  & $\vv{B}$ =( 50, 50, 0) & $\vv{B}$ =( 50, 0.24230060E+03, 0)& $M\sub{f}=2$ \\
	  & $\vv{E}$ =( 0, 0, 0) & $\vv{E}$ =( 0, 0, 0.19228027E+03) &  $\sigma=10^3$\\
          & $p$=1.0, $\rho$=1.0 & $p$ = 0.44243911E+01, $\rho$ = 0.26176303E+01 &  \B\\
\hline
FS7    & $\vv{v}$ =( 0.57368310E+00, 0, 0)   &  $\vv{v}$ =   ( 0.19727530E+00, 0.34774998E-02, 0)    &  $v\sub{sh}$=0.1   \T\\
	  & $\vv{B}$ =( 0.22803509E-01, 0.27840482E-01,  0) & $\vv{B}$ =( 0.22803509E-01, 0.13638473E+00, 0)& $M\sub{f}=5$ \\
	  & $\vv{E}$ =( 0, 0 ,  -0.15971614E-01) & $\vv{E}$ =(0, 0, -0.26826038E-01) &  $\sigma=10^{-3}$\\
          & $p$=0.01, $\rho$=1.0 & $p$ = 0.28341867E+00, $\rho$ = 0.58282475E+01 &  \B\\
\hline
FS9   & $\vv{v}$ =( 0, 0, 0)   &  $\vv{v}$ =   ( -0.57243160E-01, 0.31963724E-02, 0)    &  $v\sub{sh}$=-0.70530352E-01   \T\\
	  & $\vv{B}$ =(0.70724819E-02, 0.70724819E-02, 0) & $\vv{B}$ =(0.70724819E-02, 0.39243124E-01, 0)& $M\sub{f}=5$ \\
	  & $\vv{E}$ =( 0, 0 , 0) & $\vv{E}$ =(0, 0, 0.22690067E-02) &  $\sigma=10^{-4}$\\
          & $p=10^{-4}$, $\rho$=1.0 & $p$ = 0.34131551E-02, $\rho$ = 0.52994146E+01. &  \B\\
\hline
SS1   & $\vv{v}$ =( 0.19953950E+00, 0, 0)   &  $\vv{v}$ = ( -0.42122856E+00, -0.63382468E+00, 0)    &  $v\sub{sh}$=-0.5   \T\\
	  & $\vv{B}$ =( 50, 0.51026147E+02, 0) & $\vv{B}$ =( 50, 0.50825161E+02, 0)& $M\sub{s}=2.101839785$ \\
	  & $\vv{E}$ =( 0, 0 , -0.10181732E+02) & $\vv{E}$ =(0, 0, -0.10282225E+02) &  $\sigma=10^3$\\
          & $p$=1.0, $\rho$=1.0 & $p$ = 0.14412306E+02, $\rho$ = 0.58792375E+01 &   \B\\
\hline
\end{tabular}
\end{center}
\end{table*}

\subsubsection{FS7. Fast shock in weakly-magnetised plasma} 

We start with the case of fast shock in low-$\sigma$ plasma. This case is selected to demonstrate the very good performance of splitting scheme performance in the low-$\sigma$ regime, even if it was designed specifically with the high-$\sigma$ regime in mind. In addition, this case allows us to illustrate the inner workings of the splitting approach without resorting to sophisticated plotting techniques.

\begin{figure*}
\centering
 \includegraphics[width=0.3\textwidth]{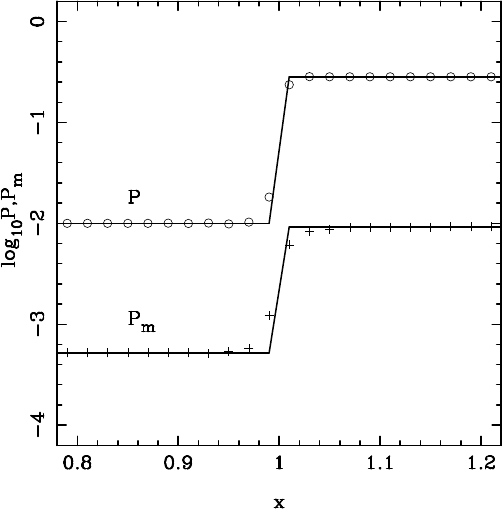}
 \includegraphics[width=0.3\textwidth]{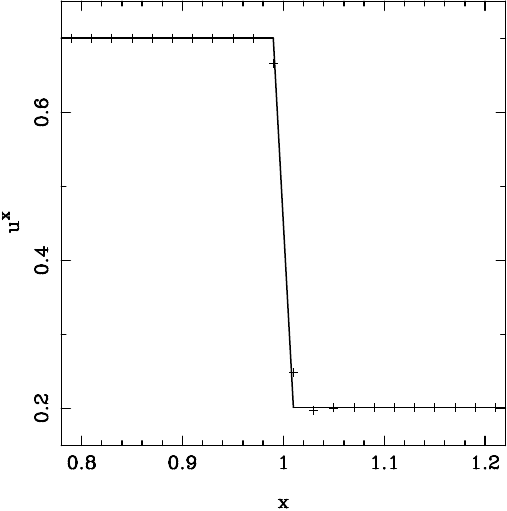}
 \includegraphics[width=0.3\textwidth]{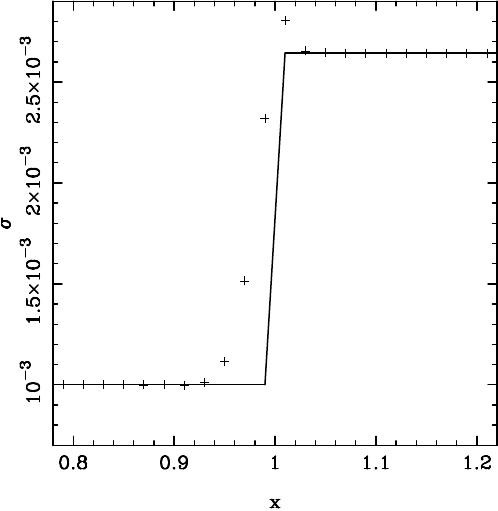}   
 \includegraphics[width=0.3\textwidth]{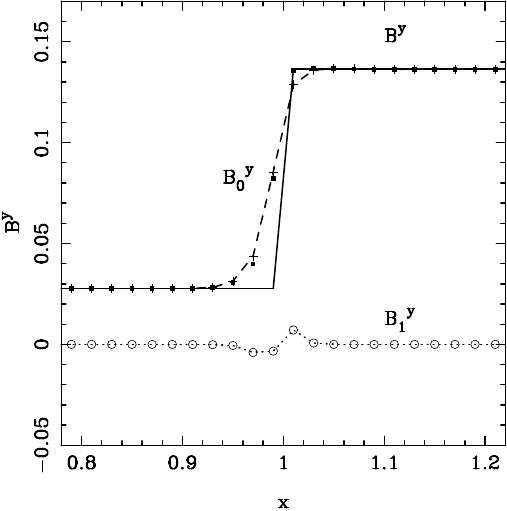} 
 \includegraphics[width=0.3\textwidth]{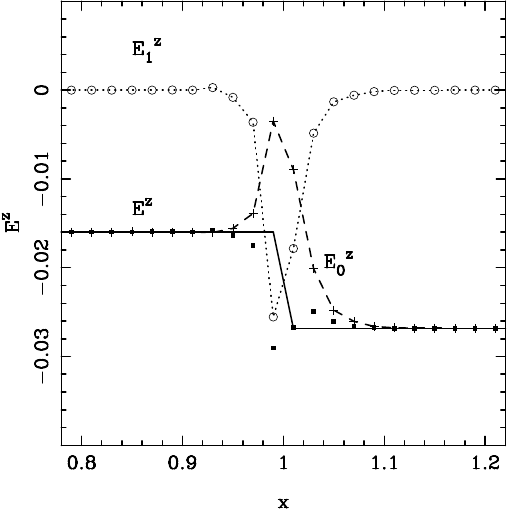}  
 \includegraphics[width=0.3\textwidth]{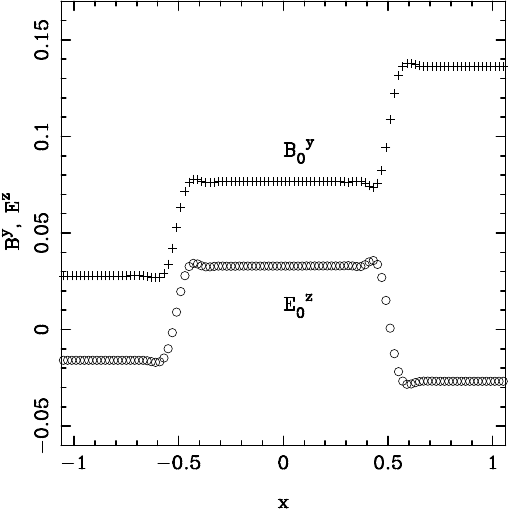}  
 
 \caption{Fast shock FS7 of weak magnetization. $t=10$. The bottom right panel shows the FF electrodynamic solution at $t=0.5$ for the same initial conditions.}  
 \label{fig:fs7}
\end{figure*}

In the upstream (left) state, $p=10^{-2}$, $\rho=1$, and $\sigma=10^{-3}$. The corresponding sound and Alfv\'en speeds are $c_s=0.11$ and $c\sub{A}=0.022$, respectively. In the rest frame of the upstream state, the shock moves in the negative x direction with the fast magnetosonic Mach number $M_f=5$, where 
$$
M_f=\frac{\gamma_s v_s}{\gamma_f v_f}\,,
$$
$v_s$ is the shock speed, $v_f$ is the fast magnetosonic speed along the shock normal, and $\gamma_s$ and $\gamma_f$ are the corresponding Lorentz factors.  
The angle between the shock normal and the magnetic field $\alpha\sub{B}=45^\circ$.   For the test simulations, the shock is setup in the inertial frame where it moves in the positive x direction with the speed $v_s'=0.1$.  The domain is $(-0.5,1.5)$ with 100 grid points. Initially, the shock is located at $x=0$.  Figure \ref{fig:fs7} illustrates the solution at $t=10$, when the shock is expected to reach $x=1$.  In its plots, the solid lines show the exact solution, and the markers show the simulation results obtained with the splitting scheme.

One can see that the shock is captured very well, both in terms of the shock speed  and the jumps of the fluid parameters. The bottom-left panel shows the jump in the total magnetic field $B^y$ and its perturbation components $B^y_\1$, which vanishes in the upstream and downstream uniform states and remains quite low even at the shock front. The bottom-centre panel, shows the total electric field $E^z$, its FFDE component $E^z_\0$, and the perturbation components $E^z_\1$.  The perturbation component vanishes in the upstream and downstream uniform states, where $E^z=E^z_\0$.   However in the shock layer, $E^z_\0$ strongly deviates from $E^z$ and develops a conspicuous upward 'spur'. The perturbation component also has a spur there but in the opposite direction, thus keeping the total electric field $E^z$ close to the exact solution.  The behaviour of $E^z_\0$ is consistent with the pure FFDE solution to the Riemann problem with the same electromagnetic left and right states. The bottom-right panel of Figure \ref{fig:fs7} illustrates this solution at $t=0.5$. In involves two fast waves moving with the speed of light in the opposite directions, and a uniform state in between, where $E^z_\0>0$. The FFDE component of the splitting scheme attempts to evolve the total electromagnetic field in the same direction, but the perturbation component prevents it from getting there.

\subsubsection{FS9. Sub-relativistic fast shock} 

In this case, both the plasma temperature and magnetisation are lower than in FS7, allowing to describe it as a sub-relativistic problem.  The results of this test show that the splitting scheme can be used to simulate such plasmas without significant decrease of accuracy.   This is important as in many astrophysical applications both the ultra-relativistic and sub-relativistic plasmas coexist, e.g. an accretion disk or interstellar gas next to a relativistic jet.

In the upstream (left) state, $p=10^{-4}$, $\rho=1$, $\sigma=10^{-3}$, and the non-relativistic magnetisation parameter $\beta=p/p\sub{m}=2$. The corresponding sound and Alfv\'en speeds are $c_s=0.011$ and $c\sub{A}=0.0071$, respectively. The shock moves through this state in the negative x direction with the fast magnetosonic Mach number $M_f=5$. The angle between the shock normal and the magnetic field $\alpha\sub{B}=45^\circ$.   The test simulations are setup in the rest frame of the upstream state.  In this frame,  the shock speed $v_s=-0.0705$.  The domain is $(-0.35,0.05)$ with 100 grid points. Initially, the shock is located at $x=0$.  The left panel of Figure \ref{fig:othershocks} illustrates the solution at $t=3$, when the shock is expected to reach $x=-0.212$.  In the plot, the solid lines show the prediction based of the shock speed of the exact solution, and the markers show the numerical solution obtained with the splitting scheme.

\subsubsection{FS5. Fast shock in highly-magnetised plasma}

\begin{figure*}
\centering
\includegraphics[width=0.3\textwidth]{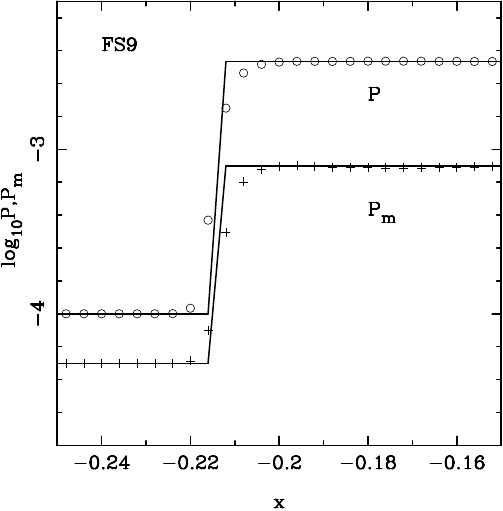} 
 \includegraphics[width=0.3\textwidth]{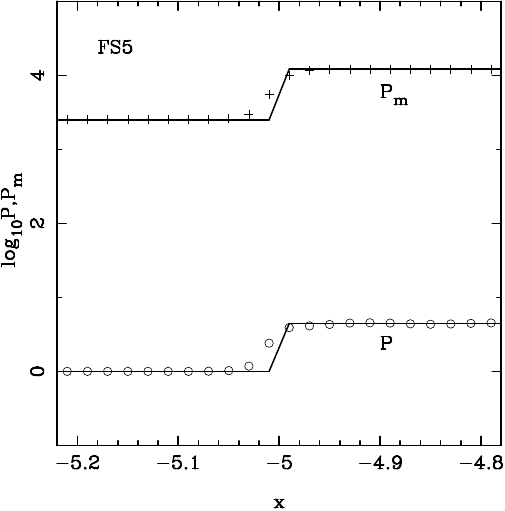}
 \includegraphics[width=0.3\textwidth]{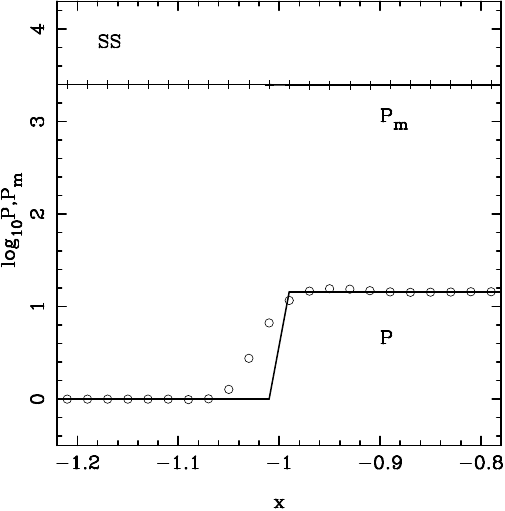} 
 \caption{Left panel: Sub-relativistic shock FS9 at $t=3$.  Middle panel:  Fast shock FS5 of strong magnetisation at $t=10$.
 Right panel: Slow shock SS of strong magnetisation at $t=2$. All these shocks are initially located at $x=0$.}  
 \label{fig:othershocks}
\end{figure*}

This is an example of fast shock in highly-magnetised plasma. In the rest frame of the upstream state, $p=\rho=1$ and $\sigma=10^{3}$.  The shock moves through this state in the negative x direction with the fast magnetosonic Mach number $M_f=2$. The shock speed in this frame is $v_s=-0.99968$, and the angle between the shock normal and the magnetic field $\alpha\sub{B}=45^\circ$.   The test simulations are setup in the frame where the shock speed is $v_s'=-0.5$.  The domain is $(-5.5,0.5)$ with 300 grid points. Initially, the shock is located at $x=0$.  The middle panel of Figure \ref{fig:othershocks} illustrates the solution at $t=10$, when the shock is expected to reach $x=-5.0$.  In the plot, the solid lines show the exact solution, and the markers show the results of computer simulations. Once again both the shock speed and its jumps are well captured by the splitting scheme.  When the energy transfer algorithm is turned off, the errors increase. In particular, the gas pressure is about 20\% lower. The plot also shows a slight shift of the numerical  solution relative to the exact one, implying the possibility of a small error in the shock speed. However, this shift is already seen at $t=2$, where it has the same size. This suggests that the shift is more likely a property of the numerical shock structure.

\subsubsection{SS. Slow shock in highly-magnetised plasma} 

This is an example of slow shock in highly-magnetised plasma. The upstream state is exactly the same as the FS5 example.  The shock moves through this state in the negative x direction with the slow magnetosonic Mach number $M_f=2.1$, the shock speed in this frame is $v_s=-0.63$.  The test simulations consider the flow in the rest frame where the shock speed is $v_s'=-0.5$.   The domain is $(-1.5,0.5)$ with 100 grid points. Initially, the shock is located at $x=0$.  The right panel of Figure \ref{fig:othershocks} illustrates the solution at $t=10$, when the shock is expected to reach $x=-5.0$.  In the plot, the solid lines show the prediction based of the shock speed of the exact solution, and the markers show the numerical solution obtained with the splitting scheme. One can see that this shock is also well captured. The small 'separation' between the curves of the exact and numerical solutions does not increase with time and seems to have the same origin as in the case FS5.

\begin{figure*}
\centering
\includegraphics[width=0.3\textwidth]{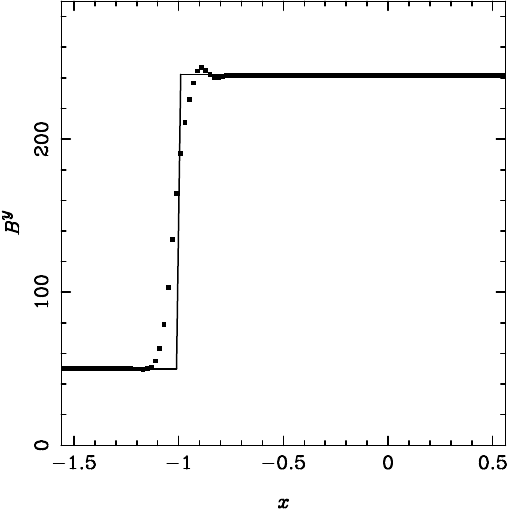} 
 \includegraphics[width=0.3\textwidth]{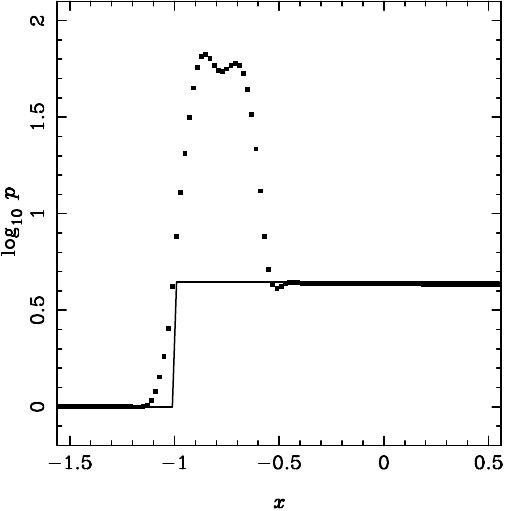}
 \includegraphics[width=0.3\textwidth]{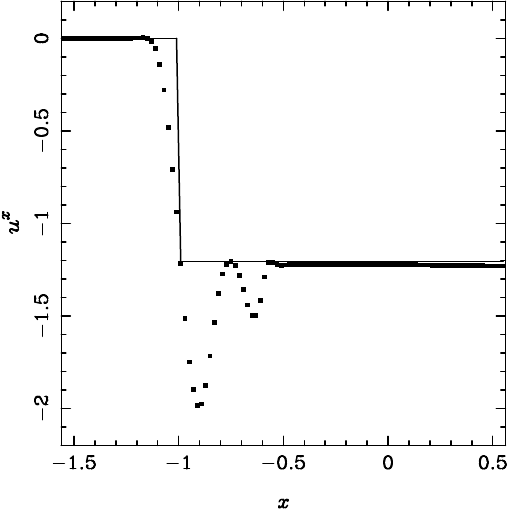} 
 \caption{Fast shock FS5A of strong magnetisation at $t=1$.  The plots the solutions for $B^y$ (left panel), gas pressure $p$ (middle panel) and $u^x$ (right panel).  The solid lines show the exact solutions and the marker show the numerical solution.}  
 \label{fig:fs5a}
\end{figure*}

\subsubsection{FS5A. Fast shock in highly-magnetised plasma}

This is a problematic case where the numerical solution suffers from large computational errors.   
The shock is the same as FS5 but now the simulations are set in the rest frame of its upstream state.  

The results are illustrated by figure \ref{fig:fs5a}. As far as the electromagnetic field is concerned the numerical solution is quite accurate, with the shock speed and jumps across the shock being captured quite well (see the left panel of figure \ref{fig:fs5a}).  The plasma parameters, however, show very large errors. As on can see in the middle panel figure ref{fig:fs5a}, the gas pressure of the numerical solution overshoots the pressure of the exact solution by more than ten times.  

One probable reason for the large errors is the very large shock speed, $v_s=-0.99968$.  At such a speed, the non-linear steepening is extremely slow and hence not as efficient at balancing the magnetic field diffusion due to numerical resistivity as in slower shocks.  As a result, the shock structure keeps spreading out until the numerical diffusion becomes sufficiently reduced. The spreading is accompanied by excessive numerical heating of plasma, which explains the high gas pressure of shocked plasma.  This interpretation is consistent with the fact that the heating is particularly intense at the start of the simulation when the shock is just beginning to develop its numerical structure. Moreover, switching off the energy transfer allows to reduce the amount of numerical heating, which also supports this interpretation. The latter does not cure the problem, however, because the energy transfer is not the only mechanism of plasma heating (see section \ref{sec:heating} ).  Using smooth shock profile in the initial solution does not help much either. 

In addition to the extremely fast motion relative to the grid, the FS5A shock is characterised by much stronger jump of the tangential component of the magnetic field across the shock than in the FS5 shock. If in the FS5 case, $\Delta B^y \approx 3$, in the FS5A it is $\Delta B^y\approx 2\times10^2$, leading to about one hundred times stronger numerical magnetic dissipation.  

\vskip 0.5cm
Summarising the results of our 1D shock wave tests, the splitting method captures strong shocks quite well, especially in the low-$\sigma$ regime. However, in the high-$\sigma$ regime, very fast shocks with large jumps of magnetic field are problematic.

\section{2D test simulations}
\label{sec:2Dtests}

We used some of the 1D tests problems described in section \ref{sec:1Dtests} in setups aligned with the x and y direction to make sure that the results of 1D tests are reproduced with the 2D code. These tests do not reveal anything new and their results are not described in this section, where only the results of inherently 2D problems are presented.  All the 2D simulations are carried out in Cartesian coordinates.

\subsection{Magnetic rope}
\label{sec:mrope}

Lundquist's magnetic rope is a steady-state axisymmetric force-free magnetic configuration, where the magnetic pressure and tension  perfectly balance each other \citep{Lundquist50}.  In our simulations of a stationary rope, the force-free equilibrium is preserved, subject to slow numerical diffusion and magnetic dissipation.  Here we present the results of a more challenging problem, where the rope moves along the x direction with a relativistic  speed.  

In the rest frame of the rope, its magnetic field is given by  

\beq
\begin{cases}
\tilde{B}^x=-B_0 \dfrac{y}{r} J_1\left(\alpha \dfrac{r}{r_0}\right) \,,\\
\tilde{B}^y=B_0 \dfrac{x}{r} J_1\left(\alpha \dfrac{r}{r_0}\right) \,,\\
\tilde{B}^z=B_0  J_0\left(\alpha \dfrac{r}{r_0}\right)\,,
\end{cases} 
\eeq
where $r_0$ is the rope radius. In these equations, $J_n$ are Bessel's functions, $\alpha$ is the first root of $J_1$ and $r=\sqrt{\tilde{x}^2+\tilde{y}^2}$ is the radial distance from the rope axis \citep{Lundquist50}. Outside of the rope, for $r>r_0$, $\vv{B}=(0,0,B_0J_0(\alpha) )$. The gas pressure and density are uniform. 

The initial solution for the rope moving with the speed $v$ in the x direction, is obtained using the Lorentz transformation for the electromagnetic field $\{\vv{E},\vv{B}\}$ and the Lorentz length contraction $x=\tilde{x}/\gamma$. The model parameters of the test simulations are $B_0=100$, $r_0=1$, $p=1$, $\rho=1$, and $v=0.8$. The corresponding magnetisation reaches $\sigma\approx 2000$ in the centre of the rope.  The domain is $(-2,2)\times(-2,2)$ with 200 uniformly spaced grid points in each direction. The periodic boundary conditions are used at both the x and y boundaries.  

\begin{figure*}
\centering
 \includegraphics[width=0.3\textwidth]{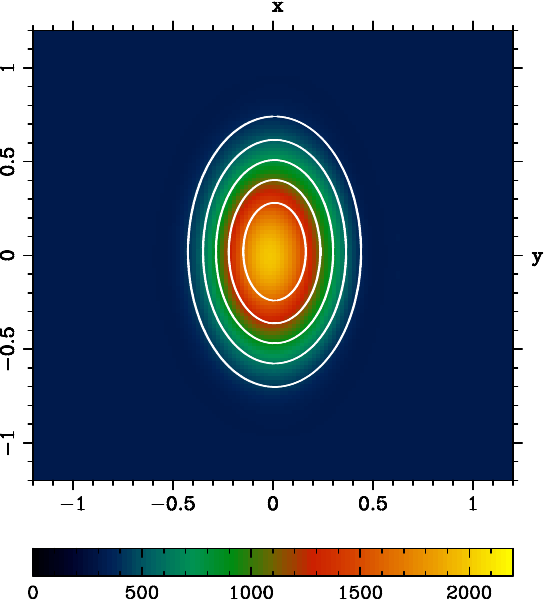}
\includegraphics[width=0.3\textwidth]{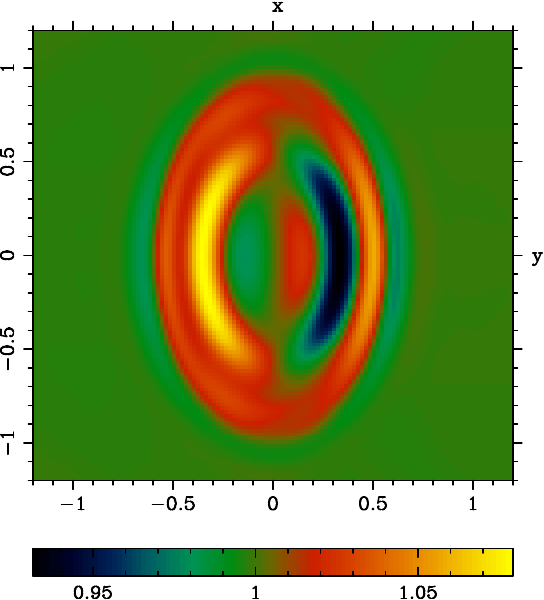} 
 \includegraphics[width=0.3\textwidth]{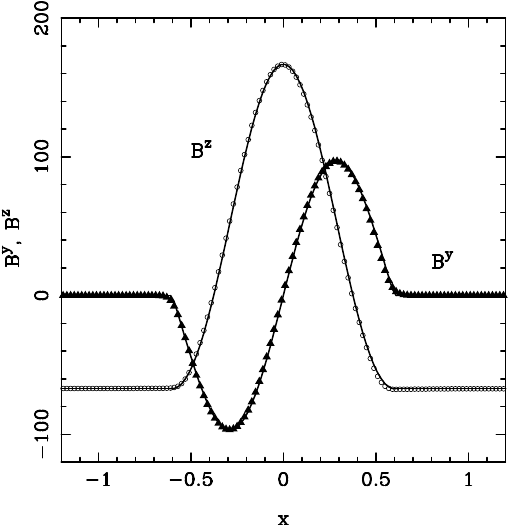} 
 \caption{Magnetic rope. Left panel:  magnetisation parameter $\sigma$ and magnetic field lines. The colour map shows $\sigma$ at $t=5$. Two sets of 5 magnetic field lines, one for $t=0$ and another for $t=5$. They are indistinguishable.   Middle panel: gas pressure $p$ at $t=5$. Right panel: magnetic field along the line   $y=0$. The markers show the numerical solution, and the solid lines show the exact solution at $t=5$.}  
 \label{fig:rope}
\end{figure*}

Figure \ref{fig:rope} compares the numerical solution with the exact solution at $t=5$, by which time the rope has crossed the domain twice and returned to its initial position.  In the left panel, the colour map shows the distribution of $\sigma$ at $t=5$. The plot also includes two sets of contour lines of the magnetic flux function, one set for $t=0$ and another for $t=5$. These are indistinguishable in the plot. The middle panel shows the pressure distribution at $t=5$. Here one can clearly see the numerical errors, which in places reach eight percent. In the right panel, the magnetic field in the numerical solution along the line $y=0$ (markers) is compared to the magnetic field in the exact solution. Here, the errors are hardly visible.

\subsection{Oblique degenerate Alfv\'en wave. Anisotropy of numerical resistivity}
\label{sec:odaw}

Here, we return to the problem of section \ref{sec:daw-1d} and consider the case where the wave vector points at $45^\circ$ to the x axis.  The aim is to evaluate the anisotropy of the numerical resistivity relative to the computational grid.  For such an obliqueness, the solution \eqref{eq:wave-decay}  reads   
\beq
\vv{B}(t)=B_0 \left(-\frac{1}{\sqrt{2}}\cos\phi(x,y,k),  \frac{1}{\sqrt{2}}\cos\phi(x,y,k),\sin\phi(x,y,k) \right) \exp(-\omega t) \,,
\label{eq:wave-decay-2d}
\eeq
where $\phi=(k/\sqrt{2})(x+y)$ and $\omega =\eta k^2$.

\begin{figure}
\centering
\includegraphics[width=0.35\textwidth]{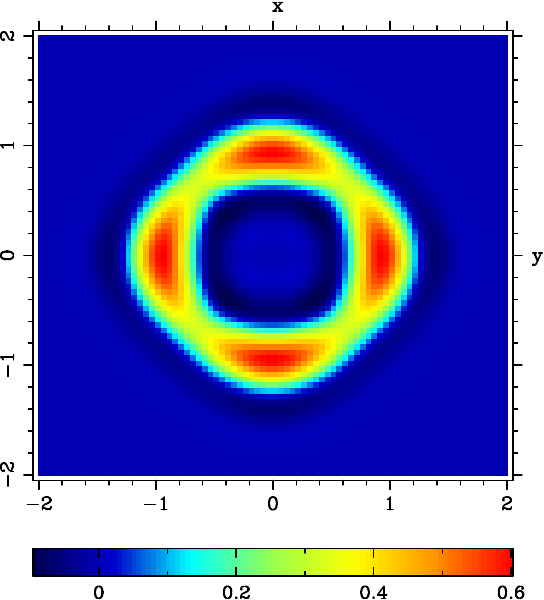} 
 \caption{Anisotropy of numerical resistivity. 
 Entropy distribution in the simulations of stationary magnetic rope at $t=10$.}   
 \label{fig:anis-eta}
\end{figure}
 
 In the test simulations, the domain is $(0,1)\times(0,1)$, with equal resolutions in the x and y directions, and the periodic boundary conditions.  These boundary conditions are satisfied only for the wavenumbers $k_n=2\sqrt{2}\pi n$, $n\in Z$.  The model parameters are the same as in the 1D simulations, $B_0=50$, $\vv{v}=\vv{0}$, $p=1$ and $\rho=1$. Table \ref{tab:AW3} shows the results obtained for the wave with $k=2\sqrt{2}\pi$ and the same resolution as in the 1D test.  Comparing these results with the 1D results for the wave with $k=2\pi$,  given in table \ref{tab:AW2}, one can see that the resistivities are exactly the same. Since $\eta\propto k^2$, this means that for the same wave number the resistivity in the oblique case is smaller by the factor of 2.
   
Clearly, the resistivity must be a smooth periodic function $f(\theta)$ of the angle $\theta$ between the wavevector $\vv{k}$ and the unit vector $\ort{x}$ of the x direction, with the period of $\pi/2$. Moreover, it must be symmetric with respect to the angles $\theta_s=n\pi/4$, $n\in Z$, so that $f(\theta_s+a)=f(\theta_s-a)$. The simplest function satisfying these conditions is
\beq
\eta\sub{num}=\frac{\eta_0}{4}(3+\cos{4\theta}) \,.
\label{eq:num-resist-2D}
\eeq
where $\eta_0$ is given by the equation \eqref{eq:num-resist}.

\begin{table}
\caption{Oblique degenerate Alfv\'en wave simulations.  $n_x=n_y$ is the numerical resolution, $\eta$ is the numerical resistivity,  $A_\eta$ is the normalisation factor in \eqref{eq:num-resist-2D}.} 
\label{tab:AW3}
\begin{center}
\begin{tabular}{ cllll} 
 \hline
 $n_x=n_y$    &    10    &    20  & 40 &  80 \T\B\\ 
 \hline
$2\eta k^2$     &  0.39   &  $0.026$  &   $0.0030$   &  $0.00038$ \T  \\
$\eta\sub{num}$            & $0.50\times10^{-2}$ & $0.33\times10^{-3}$ & $0.38\times10^{-4}$ & $0.48\times10^{-5}$\\
$A_\eta$        &  0.064   & 0.034 &  0.031  &  0.031 \B \\
\hline
\end{tabular}
\end{center}
\end{table}

To explore this issue a little bit further, we inspected the results of the stationary magnetic rope simulations ( see section \ref{sec:mrope}) for the signs of anisotropic resistivity.  The entropy $s=\ln p/\rho^\Gamma$ of the exact solution is uniform. However, the plasma heating associated with the numerical resistivity is expected to yield a non-uniform distribution $s(r,\theta)$, which is periodic in $\theta$ and peaks along the x and y axes.  This is exactly what is observed in these simulations (see figure \ref{fig:anis-eta}.)

\subsection{Cylindrical explosion in uniform magnetic field}
\label{sec:cyl-expl}

This is now a standard test problem for RMHD codes \citep[e.g.][]{ssk-godun99,Liesmmann05,MB06,DZ07}. In the initial solution of this problem, a cylindrical volume filled with plasma of very high pressure and temperature (the result of an explosion) is surrounded by plasma of low pressure and density. To make the problem more interesting, the whole space is threaded with a uniform magnetic field directed perpendicular to the cylinder, which breaks the axial symmetry of the problem.  Although there is no exact analytic solution to this problem, one can compare the results of simulations to the solutions obtained with other numerical methods.    

Following \citet{ssk-godun99}, the density and pressure of the surrounding plasma are set to $\rho\sub{e}=10^{-4}$ and $p\sub{e}=3\times10^{-5}$. The hot cylinder is centred on the z axis and has the radius $r_0=0.9$. Its density and pressure are set  to $\rho_0=10^{-2}$ and  $p_0=1$, respectively. The initial jumps of both the gas pressure and its density are soften with the same $\tanh$-profile
\beq
  f(r) = \frac{1}{2}[(f_0-f\sub{e})\tanh((r-r_0)/\Delta r) + (f_0+f\sub{e})] \,,
 \label{eq:bw}  
\eeq
where $\Delta r=0.03$. The simulation domain is $(-6,6)\times(-6,6)$, with 400 uniformly spaced grid points in each direction. We explored four models with $B_0=0.01$, 0.1, 1.0, and $10^3$.

\begin{figure*}
\centering
 \includegraphics[width=0.28\textwidth]{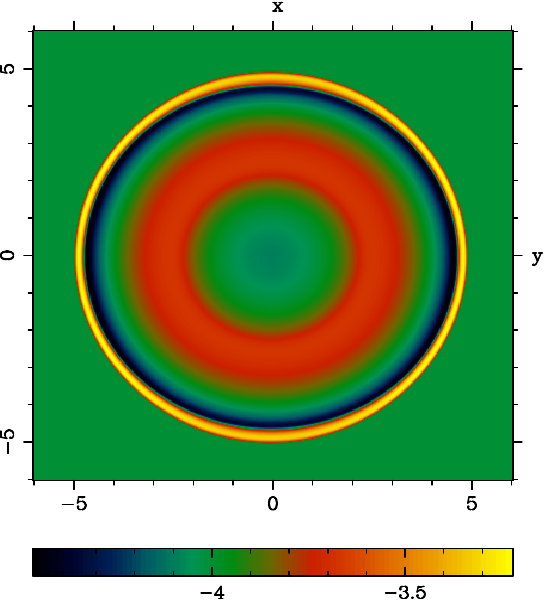}
\includegraphics[width=0.28\textwidth]{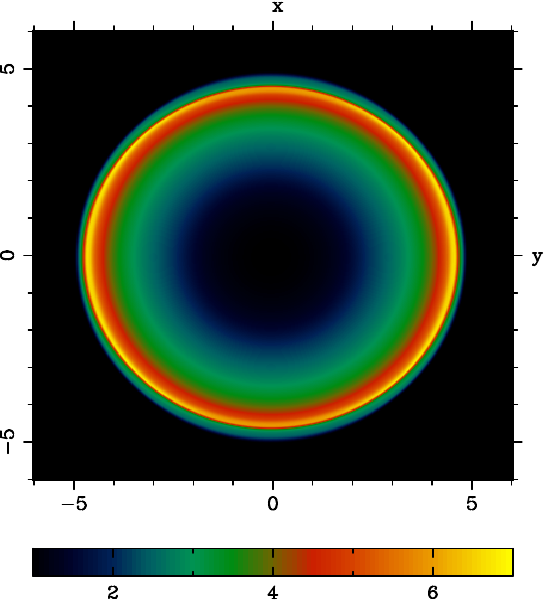} 
 \includegraphics[width=0.28\textwidth]{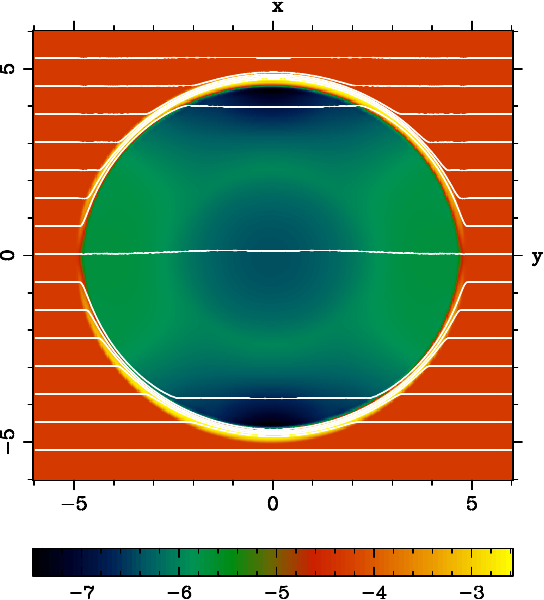} 
 \includegraphics[width=0.28\textwidth]{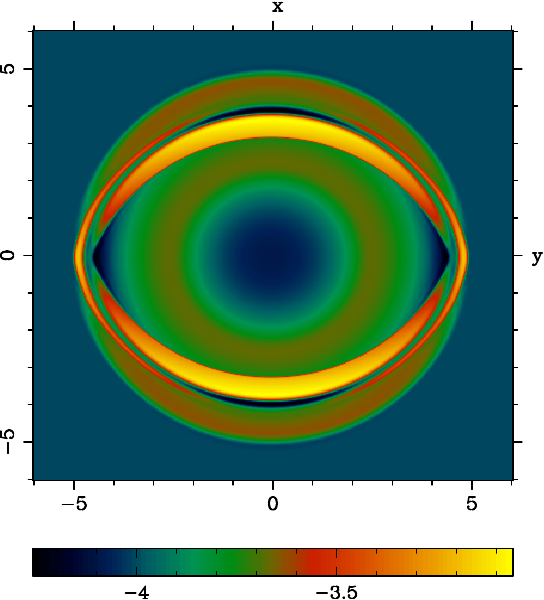}
\includegraphics[width=0.28\textwidth]{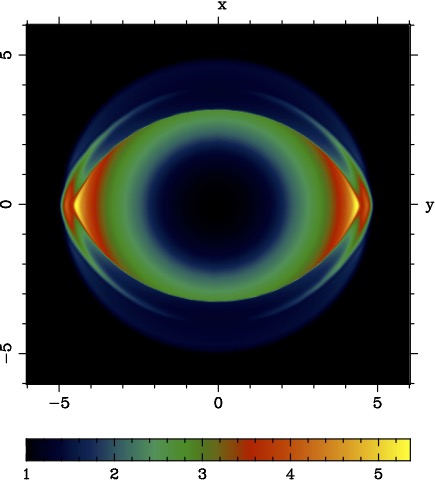} 
 \includegraphics[width=0.28\textwidth]{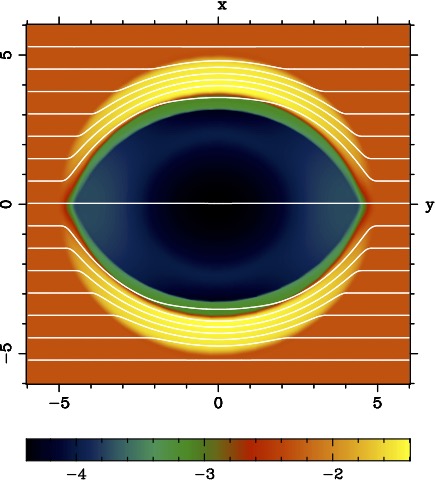} 
\includegraphics[width=0.28\textwidth]{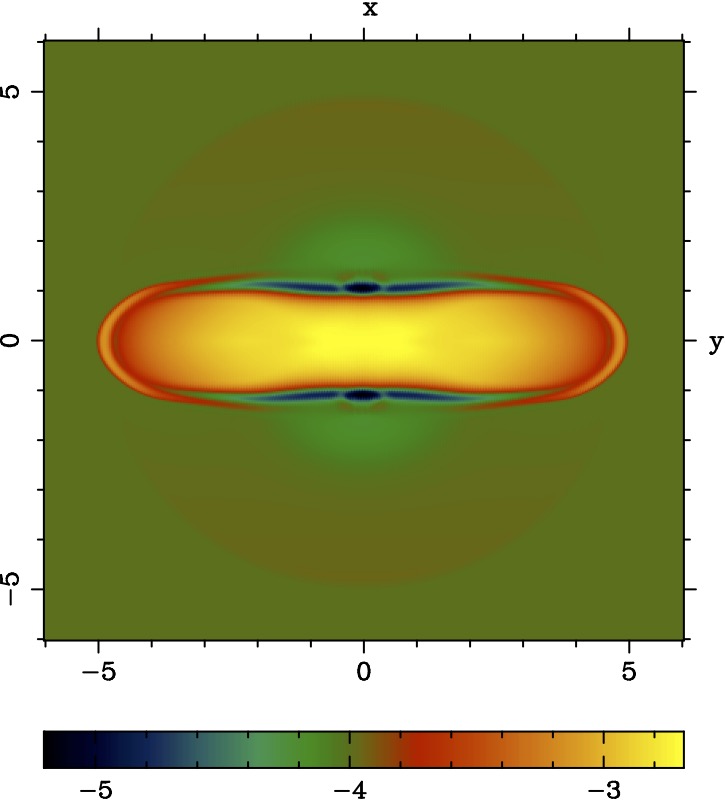}
\includegraphics[width=0.28\textwidth]{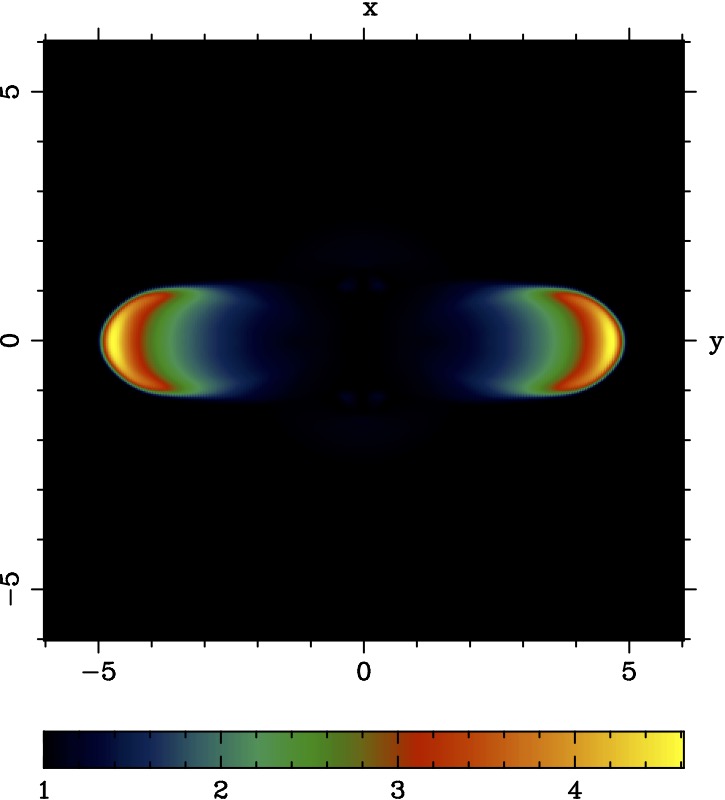} 
 \includegraphics[width=0.28\textwidth]{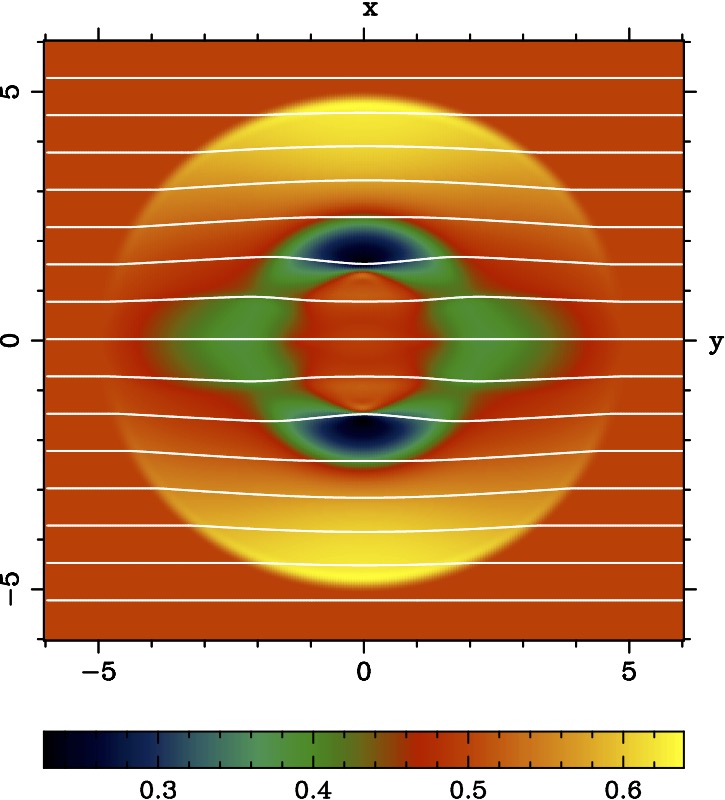} 
  \includegraphics[width=0.28\textwidth]{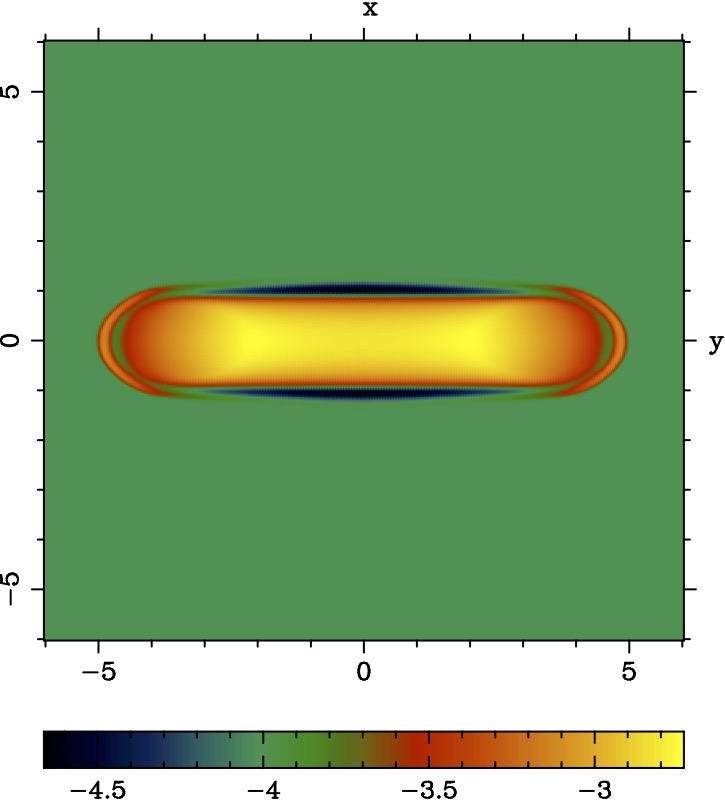}
\includegraphics[width=0.28\textwidth]{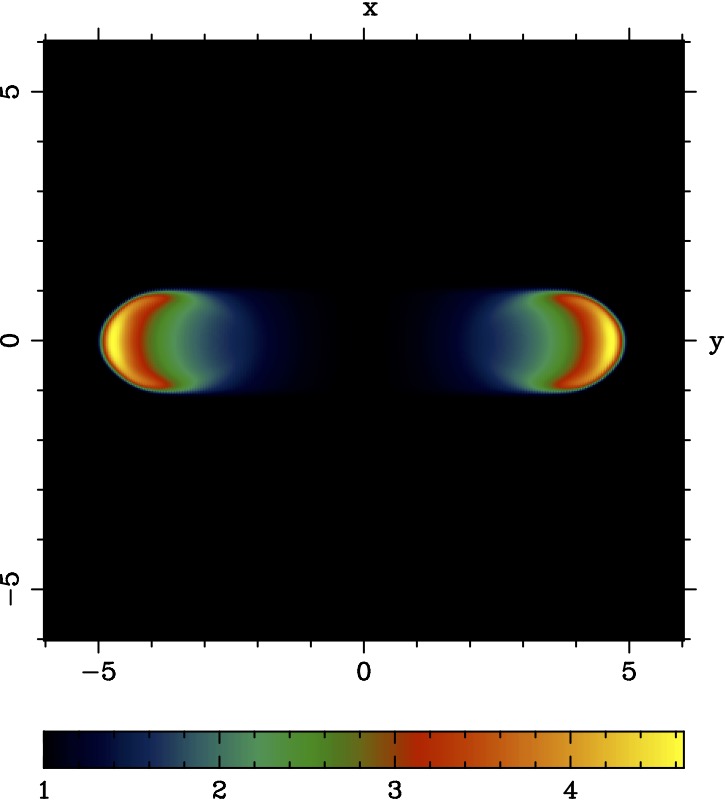} 
 \includegraphics[width=0.28\textwidth]{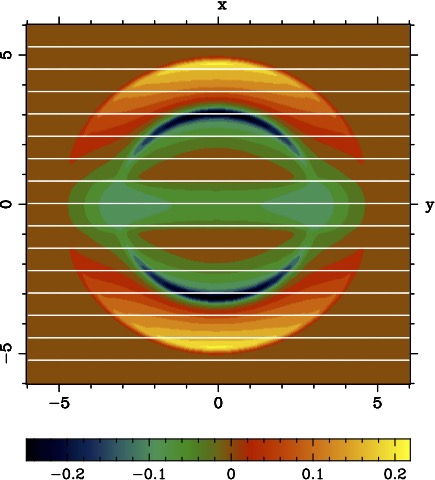} 

 \caption{\bf Cylindrical explosion. From top to bottom, solutions for the models with $B_0=0.01$, 0.1, 1.0 and 1000 at $t=4$. Left panels: $\log_{10} \rho$; Middle panels: Lorentz factor $\gamma$; Right panels:  the magnetic field lines and $\log_{10} p\sub{m}$ for $B_0=0.01$ and 0.1, $p\sub{m}$ for $B_0=1$ and  $\delta p\sub{m}=p\sub{m}-5\times10^5$ for $B_0=1000$. }  
 \label{fig:bw2d-coll}
\end{figure*}

The top row of figure \ref{fig:bw2d-coll} illustrates the solution for the model with the magnetic field strength $B_0=0.01$ at $t=4$. The corresponding magnetisation is $\sigma_0=2.5\times10^{-5}$ inside the cylinder and $\sigma\sub{e}=0.45$ in its surroundings. The magnetic pressure is very low compared to the gas pressure in the cylinder, with $\beta_0=2\times10^4$, and for some time the magnetic field has little influence on the solution.  This is manifested in the central symmetry of the images in this figure.  One can spot the slightly lower flow Lorentz factor and the slightly lower density of the shocked shell along the y axis, indicating that the magnetic field is beginning to have a noticeable effect on the solution at $t=4$.  The second row of figure \ref{fig:bw2d-coll} illustrates the solution for the model with $B_0=0.1$. The corresponding magnetisation is $\sigma_0=2.5\times10^{-3}$ ($\beta_0=2\times10^{2}$) inside the cylinder, and $\sigma\sub{e}=45.0$ in its surrounding.  In this model, the magnetic field is sufficiently strong to have a pronounced effect on the solution, slowing down the flow expansion in the y direction. This leads to the spectacular morphology reminiscent of an eye, which was seen in the test simulations by many other research groups.  The third row  illustrates the solution for the model with $B_0=1$.  The corresponding magnetisation is $\sigma_0=2.5\times10^{-1}$ ($\beta_0=2$) inside the cylinder and $\sigma\sub{e}=4.5\times10^3$ in its surroundings. Now the magnetic is so strong that it prevents the hot plasma from expanding in the y-direction and the explosion proceeds almost entirely along the magnetic field lines. The weak fast shock, however, still has a cylindrical shape, thanks to the fast speed being very close to the speed of light in all directions. One can also see that the magnetic are still a bit distorted by the explosion. The bottom row shows the solution for $B_0=10^3$. The corresponding magnetisation is $\sigma_0=2.5\times10^{5}$ ($\beta_0=2\times 10^{-6}$) inside the cylinder and $\sigma\sub{e}=4.5\times10^9$ in its surroundings. In this case, the distortion of magnetic field lines is so weak that it cannot be seen with a naked eye, and to visualise the fast shock we had to plot not the magnetic pressure $p\sub{m}$ but $p\sub{m}-p\sub{m,0}$, where $p\sub{m,0}=5\times10^5$ is the initial magnetic pressure. For this model, a couple of first time step had to be done with a smaller Courant number,  $C=0.1$. This was    needed for the shock identification subroutine to capture the forward shock before the errors associated with the DER step became too high.  We also run a model with $B_0=10^4$. There we had to use even smaller $C$ and for a larger number of time steps before switching back to the standard $C=0.5$. The results for this model were almost indistinguishable from the results for $B_0=10^3$.   

\begin{figure*}
\centering
\includegraphics[width=0.3\textwidth]{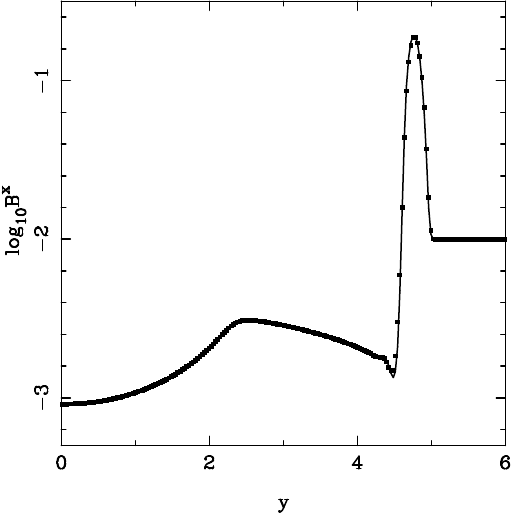} 
\includegraphics[width=0.305\textwidth]{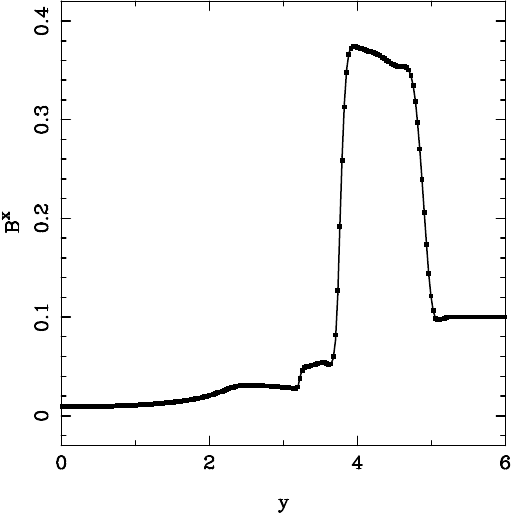} 
 \includegraphics[width=0.3\textwidth]{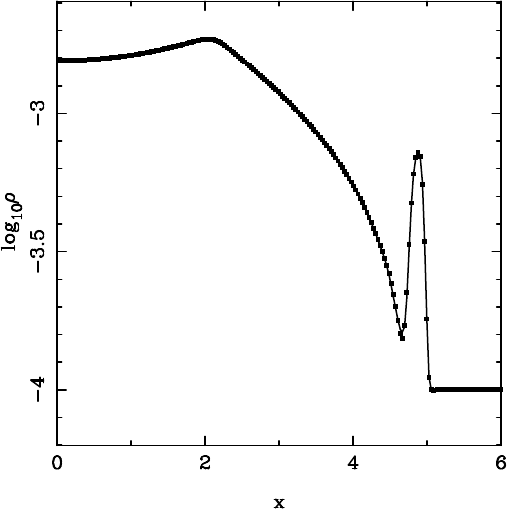}
 \caption{Cylindrical explosion. Splitting approach vs standard RMHD scheme.  Left panel: $B^x$ along the line $x=0$ at $t=4$ solutions for model with $B_0=0.01$.  The solid line shows the solution obtained with the standard approach and the markers show the solution obtained with the splitting approach.  Middle panel: $B^x$ along the line $x=0$ at $t=4$ for the model with $B_0=0.1$. The solid line shows the solution obtained with the standard approach and the markers show the solution obtained with the splitting approach. Right panel: $\log_{10}\rho$ along the line $y=0$ at $t=4$ for the model with $B_0=1000$.  Markers show the 2D solution obtained with the splitting approach and the solid line shows the solution for unmagnetised 1D flow  in the problem with the same initial distribution of flow parameters along the x axis.}  
 \label{fig:comps}
\end{figure*}

In the models with $B_0=0.01$ and 0.1, the magnetisation is sufficiently low to be handled with the standard RMHD codes. The case with $B_0=0.1$ is a particularly popular test.  On visual inspection, the results obtained for this test with the splitting scheme look indistinguishable from those obtained with standard conservative schemes  previously \citep[e.g.][]{ssk-godun99,Liesmmann05,MB06,DZ07}.  For a more detailed comparison, and to compare like with like,  we run this model in the standard RMHD mode of our code (see section \ref{sec:NumSplitting}).  The results are illustrated in the middle panel of figure \ref{fig:comps}  which shows the distributions of $B^x$ along the line $x=0$, with the line corresponding to the solution obtained in the standard mode and markers to the solution obtained in the splitting mode.  They are so close that one may think that both the line and the markers show the same solution.  The same applies to other parameters.   We did the same comparison for the model with $B_0=0.01$. This case is interesting, because the magnetic field is very weak and far from being in a force-free configuration. Given the fact that the splitting approach involves advancing the electromagnetic component using the FFDE  approximation, one could anticipate large errors in $\vv{B}$. However, this is not the case, as illustrated in the left panel of figure \ref{fig:comps}. The solution obtained using the splitting approach is still almost indistinguishable from solution obtained with the standard approach. 

The model with $B_0=1$ seems to be at the border line or already beyond the capabilities of the standard approach. Although one of us presented results for this model in the past  \citep{ssk-godun99}, which actually look quite similar to what is shown in figure \ref{fig:bw2d-coll}, they were unable to reproduce this result later on request, indicating some unusual undocumented tweaking of the code.   The simulations in the standard mode of the current code also crashed.      

The model with $B_0=1000$ is much more extreme than the one with $B_0=1$, and the simulations in the standard mode expectedly failed.  However, because the magnetic field of this model remains highly uniform, a comparison with the solution of a different kind suggests itself. Since the flow is basically one-dimensional, one may check it against the Cartesian 1D hydrodynamic ($B=0$) solution with the same initial distribution of pressure and density. This 1D HD solution is particularly close to the 2D solution along the $y=0$. The symmetry of the 2D problem implies that $y=0$ is a magnetic field line, and hence both the magnetic and electric forces along it vanish.  Thus, the flow along this line is driven solely by the gas pressure force.   The two solutions are compared in the right panel of figure \ref{fig:comps}, showing the distribution of $\rho$ along the line $y=0$. As expected, they look almost indistinguishable from each other.

Following \citet{DZ07},  we used the $B_0=0.1$ model on the $200\times200$ grid to benchmark the performance of our code. It was compiled with {\it gfortran} using the -O optimisation option, which does not allow automatic parallelisation, and was run on a single core of the Apple M2 3.49GHz processor. It took 24 cpu seconds (134 timesteps) to reach $t=4$. In the standard RMHD mode, the code was only about 20\% faster.  At $t=4$, the variable conversion takes about  36\%  of the computational time.

\subsection{Tearing instability of Harris current sheet} 
\label{sec:tihcs}

In this test, the initial solution describes a Harris current sheet, with the  $\vv{B}=(B^x,0,0)$, where 
\beq
B^x=B_0 \tanh\frac{y}{a} \,,
\eeq
and the gas pressure
\beq
p= p_0+ \frac{B_0^2}{2} \left(1-\tanh^2\frac{y}{a}\right) \,,
\eeq
where $a$ is the half-thickness of the current sheet, and $p_0$ and $B_0$ are the asymptotic values (as $y\to\pm\infty$) of the $B^x$ and $p$ respectively. In addition, $\rho=\rho_0$ and $\vv{v}=\vv{0}$. The computational domain is $(-1,1)\times(-1,1)$ with 400 uniformly-spaced grid points in the both directions, periodic boundary conditions in the x direction and zero-gradient boundary conditions in the y directions. The zero-gradient boundary conditions result in artefacts near the y boundaries, which become noticeable in log-scale plots towards the end of the simulations. However they remain at sufficiently low amplitude and do not influence the sheet dynamics.      
 
The parameters used in the simulations are $p_0=\rho_0=1$, $B_0=500$ and $a=0.01$. The corresponding asymptotic value of plasma magnetisation $\sigma_0=5\times10^4$. The selection of the very small value for $a$ is determined by the intention of setting as thin current sheet as allowed by the numerical resistivity. The value of numerical resistivity in the current sheet can be estimated using equation \eqref{eq:num-resist}. The corresponding length scale, as determined by equation \eqref{eq:lengthscale}, 
$$
\cL=\frac{a}{\sqrt{2}} \cosh\frac{x}{a} \,,
$$
now depends on the location. At $x=a$, $\cL\approx 0.009$, and with $A_\eta=0.034$, equation \eqref{eq:num-resist} yields $\eta\sub{num}\approx 10^{-4}$.  
The corresponding resistive time-scale  $\tau_\eta = a^2/\eta\approx 1$, whereas the Alfv\'en time-scale based on the half-length $L=1$ of the current sheet, $\tau\sub{A} = L/c=1$. Given that $\tau_\eta\propto a^4$, even a moderately smaller value of $a$ would result in rapid thickening of the sheet.    

\subsubsection{Linear phase}  

The equilibrium is perturbed by introducing the vertical component of magnetic field 
\beq
  B^y = \sum_{j=1}^{20} A_j\sin\left(\frac{\pi j}{L} x+ 2\pi r_j\right) \,,
  \label{eq:ti-perturb}
\eeq
where $A_j=10^{-5}B_0$, and $0<r_j<1$ is a random number. 

\begin{figure*}
\centering
\includegraphics[width=0.3\textwidth]{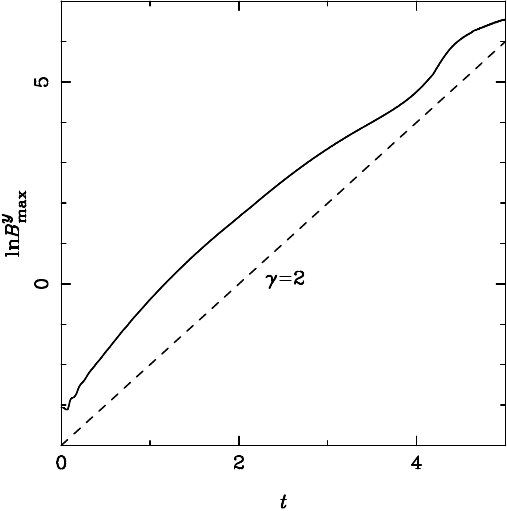}
\includegraphics[width=0.3\textwidth]{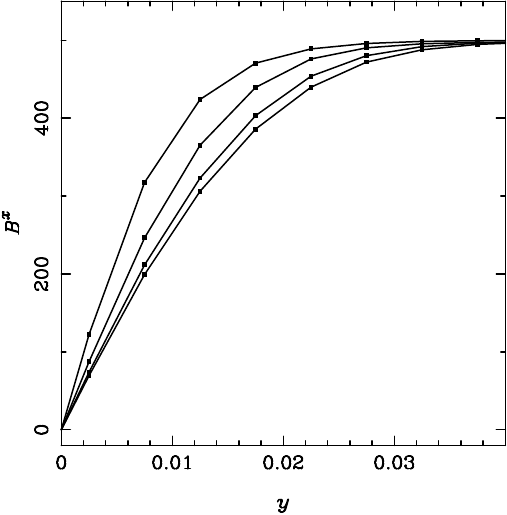} 
\includegraphics[width=0.3\textwidth]{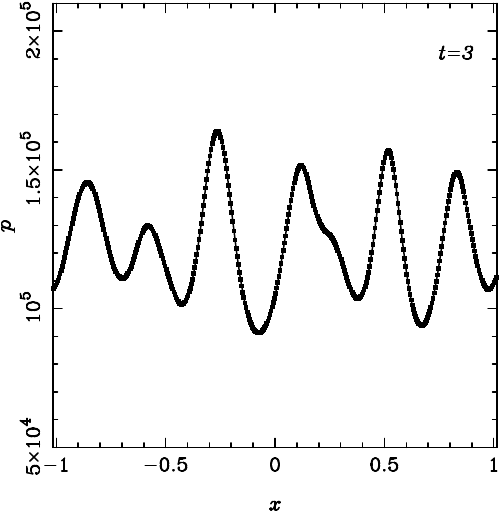} 
 \caption{Tearing instability of Harris current sheet. {\it Left panel}: Maximal value of $B^y$ over the whole domain as a function of time during the linear phase. The dashed line shows the exponential function $\propto e^{2t}$ for comparison. {\it Middle panel}: Diffusive spreading of the sheet during the linear phase. The lines show $B^x$ along the line $x=0$ at $t=0,0.5,1.5$ and 2.5 (from the narrowest to the widest of the profiles respectively). {\it Right panel}:  Gas pressure in the middle of the current sheet ($y=0$) near the end of the linear phase, at $t=3$.}   
 \label{fig:ti-lin}
\end{figure*}

\begin{figure*}
\centering
\includegraphics[width=0.3\textwidth]{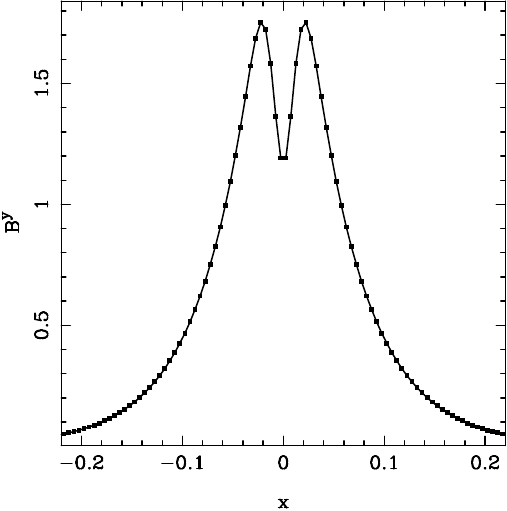} 
\includegraphics[width=0.3\textwidth]{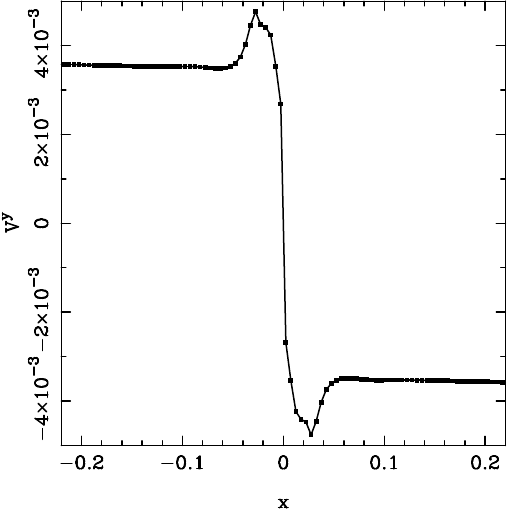} 
\includegraphics[width=0.3\textwidth]{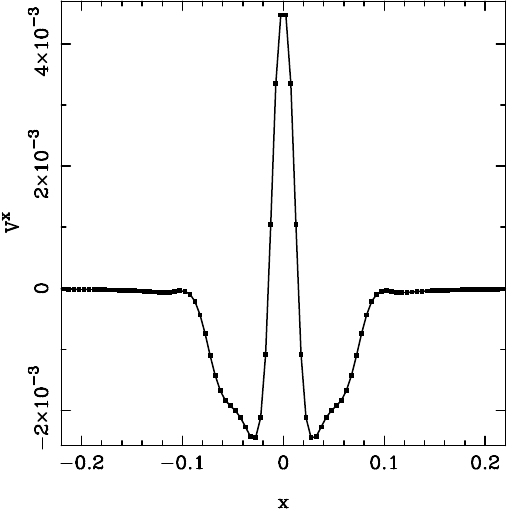} 
 \caption{Tearing instability of Harris current sheet. Numerical ''eigenmodes''  in the case of single perturbation with $k=6\pi$.}    
 \label{fig:ti-mode}
\end{figure*}

The right panel of figure \ref{fig:ti-lin} shows the function $B^y_{max}(t)$ obtained in the simulations. Using the expected exponential growth of a single eigenmode $B^y_{max}(t) \propto e^{\omega t}$, we find $\omega\approx 2.7$ for $0<t<1$,   $\omega\approx 2.0$ for $1<t<2$, and $\omega\approx 1.7$ for $2<t<3$. The variation could be related to the thickening of the current sheet from $a=0.01$ at $t=0$ to $a\approx0.013$ at $t=0.5$,  $a\approx0.015$ at $t=1.5$,  and $a\approx0.016$ at $t=2.5$ (see the middle panel of figure \ref{fig:ti-lin}). According to the theory of tearing instability, the maximum growth rate occurs for the mode with the wavenumber $k\sub{m}$, given by the equation
\beq
k\sub{m} a \approx 1.4\, {S^*}^{-1/4}  \,,
\label{eq:fastest-k}
\eeq
and it has the value
\beq
\omega\sub{m} \tau\sub{A}^* \approx 0.63 \,{S^*}^{-1/2} \,,
\label{eq:highest-om}
\eeq
where 
\beq
S^* = \frac{a c\sub{A}}{\eta} \,,
\eeq 
is the Lundquist number,  and 
\beq
\tau\sub{A}^*=a/c\sub{A} \,,
\eeq
is the Alfv\'en time-scale of the current sheet based on the sheet thickness \citep{Furth63}.  Since the number of plasmoids emerging in the simulations (see the right panel of figure \ref{fig:ti-lin}) is $n_p=6$ the fastest growing mode in the simulations has $\lambda=0.33$ and $k\sub{m}=6\pi$, which  is inside the range set by the perturbation (see equation \ref{eq:ti-perturb}). Hence one may use the above equations to estimate $S^*$ due to the numerical resistivity, assuming domination of the fastest mode. Substituting the measured values of $\omega\sub{m}$ and $a$ into equation \eqref{eq:highest-om} yields $320<S^*<480$, where the lower limit corresponds to the data for $0<t<1$ and the upper limit for $2<t<3$. For $0<t<1$, the corresponding resistivity  is $\eta\approx 4\times 10^{-5}$, which is only 2.5 times lower than the initial numerical resistivity estimated via \eqref{eq:num-resist}.   The corresponding Lundquist number based on the half-length of the current sheet 
$$
S=\frac{L c\sub{A}}{\eta} \approx 3\times10^4 \,.
$$ 
Next, one can use equation \eqref{eq:fastest-k} to check if the value of $S^*$ based on the growth rate is consistent with the number of emerged plasmoids. Substituting the values of $S^*$ and $a$ into equation \eqref{eq:fastest-k} yields $0.25 < \lambda\sub{m}< 0.36$, where again where the lower limit corresponds to the data for $0<t<1$ and the upper limit for $<2t<3$.  Somewhat surprisingly, the observed value $\lambda=0.33$ fits perfectly this theoretical prediction.

For further comparison with the results of analytical and numerical studies of the tearing instability in the framework of resistive MHD,  we also run a model with a single sinusoidal perturbation $B^y=10^{-4}\sin{6\pi x}$.  Figure \ref{fig:ti-mode} illustrates the profiles of the key flow parameters across the current sheet along the line $x=1/12$, where $B^y$ is maximum. These vary very little during the linear phase and have almost the same shape along all other lines $x=$ const. So, one may call them numerical ''eigenmodes''.  Qualitatively, they are similar to the eigenmodes found in the resistive RMHD simulations by \citet{DelZanna16}, though there are some differences too.  For example, the central dip in the profile of $B^y$ in not as deep, the increase of $|v^y|$ prior to vanishing at $x=0$ is not as strong, and the central peak of $v_x$ is surrounded by a broad depression not seen in the resistive data.       

Overall, given the fact that the numerical resistivity is more complex than the uniform scalar resistivity used in the theory of tearing instability,  the agreement between this theory and the results of our simulations is quite remarkable.

\begin{figure*}
\centering
\includegraphics[width=0.45\textwidth]{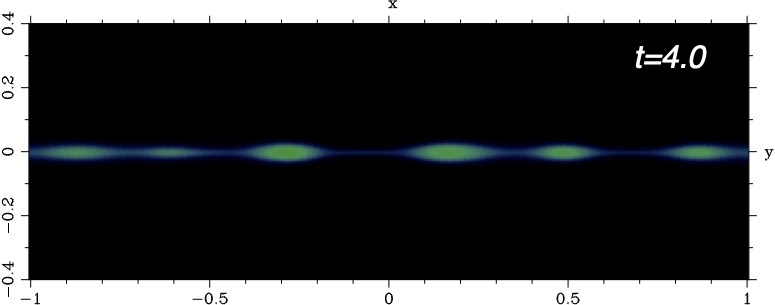}
\includegraphics[width=0.45\textwidth]{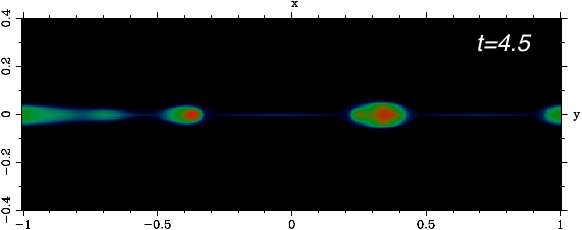} 
\includegraphics[width=0.45\textwidth]{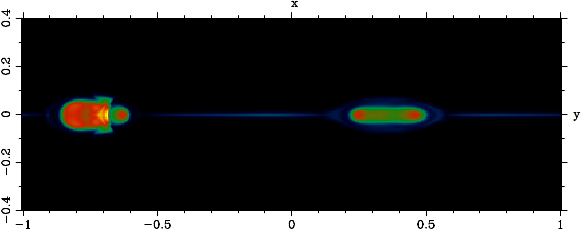} 
\includegraphics[width=0.45\textwidth]{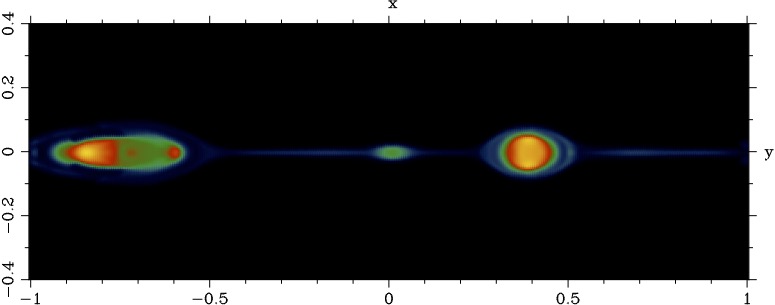}
\includegraphics[width=0.45\textwidth]{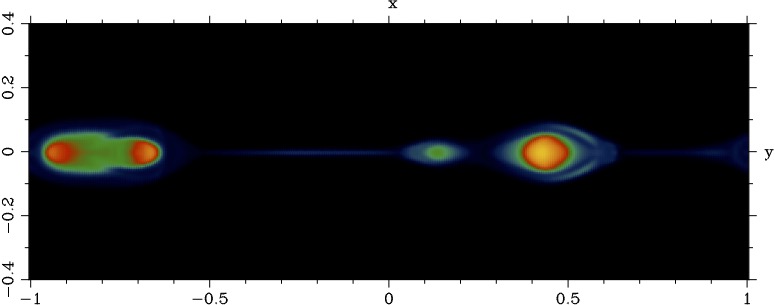} 
\includegraphics[width=0.45\textwidth]{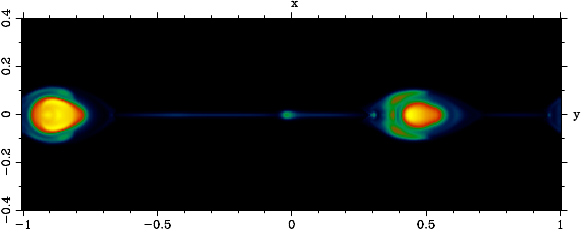} 
\includegraphics[width=0.45\textwidth]{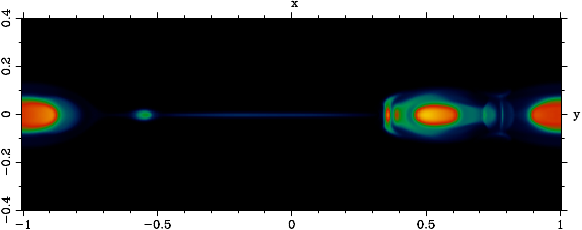} 
\includegraphics[width=0.45\textwidth]{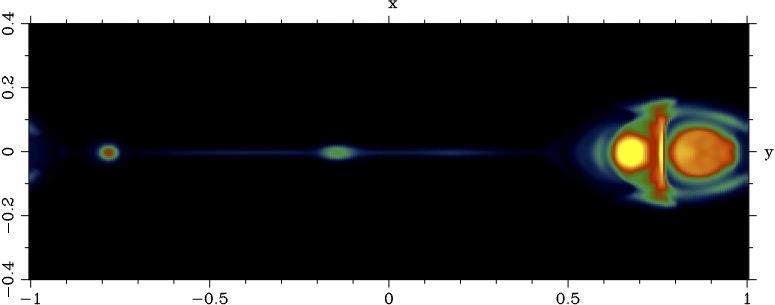} 
\includegraphics[width=0.45\textwidth]{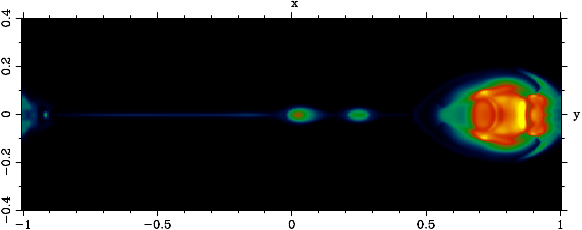} 
\includegraphics[width=0.45\textwidth]{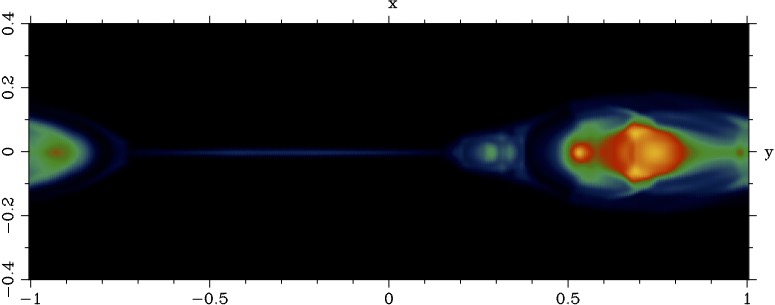} 
\includegraphics[width=0.3\textwidth]{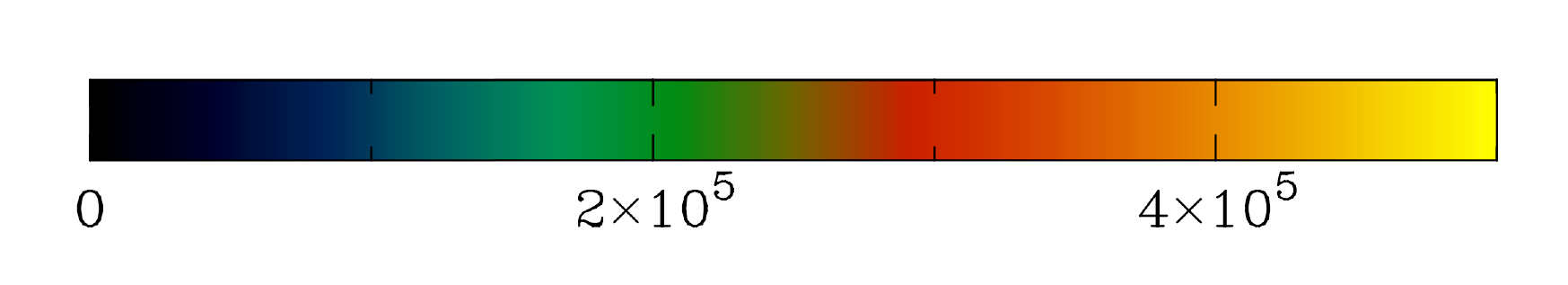} 
 \caption{Non-linear phase of the tearing instability. Gas pressure at $t=4,4.5,\dots,8.5$ (from left to right and from top to bottom).}  
 \label{fig:ti-nlin}
\end{figure*}

\subsubsection{Nonlinear phase} 

Once the multiple plasmoids developed in the current sheet, its subsequent evolution proceeds in the plasmoid-dominated regime.  Smaller plasmoids merge to form larger ones, the sections of the current sheet between them lengthen and suffer secondary tearing instability. Secondary plasmoids emerge and merge with the larger plasmoids or other secondary plasmoids (see figure \ref{fig:ti-nlin}), trying to establish a hierarchy of scales \citep{uzd-10}.  The plasma of the current sheet gets heated up to very high temperature, typically $\zeta=kT/mc^2=10^5$. This is consistent with the magnetic energy per particle $\zeta_{B}=B^2/8\pi nmc^2=1.25\times 10^5$ in the external plasma.  In places, the Lorentz factor of the flow in the current sheet reaches $\gamma=3$, and the collisions of the fast moving plasma with plasmoids drive shock waves.

\begin{figure*}
\centering
\includegraphics[width=0.33\textwidth]{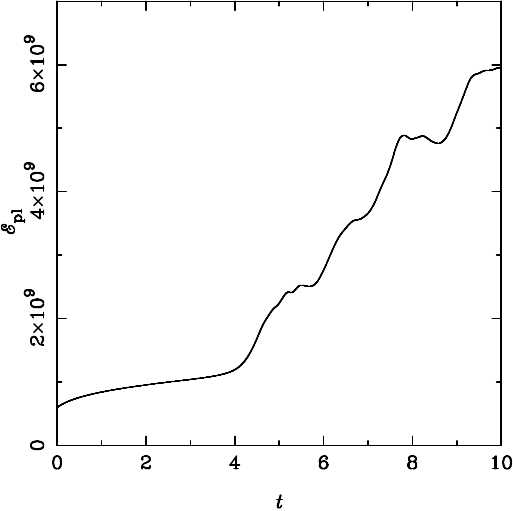}
\includegraphics[width=0.315\textwidth]{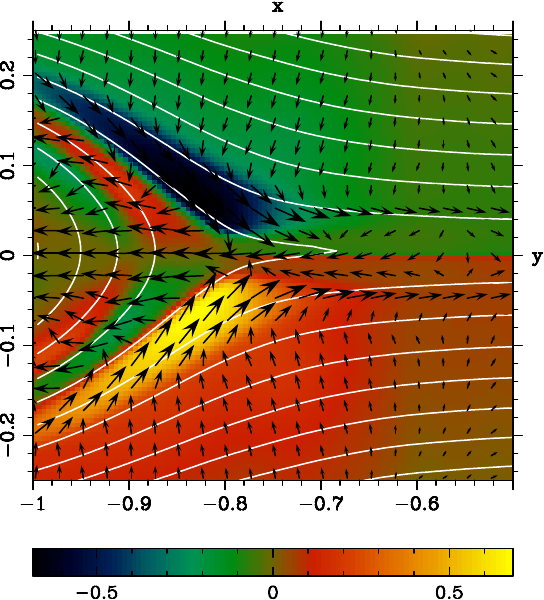}
\includegraphics[width=0.33\textwidth]{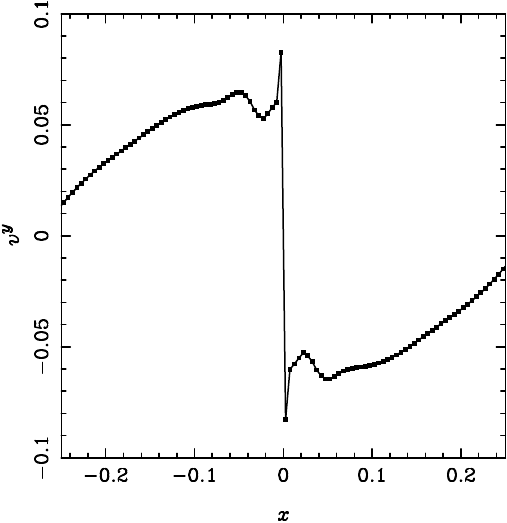}
\caption{Non-linear phase of the tearing instability. Left panel: Total plasma energy in the domain as function of time. Middle panel: The colour image shows the $v^y$ component of velocity near the large plasmoid at $t=9$. The contours show the magnetic field lines and the arrows are the velocity vectors.  Right panel: The inflow velocity of magnetic field along the line $x=-0.56$, where the velocity field shown in the middle panel indicates an x-point in the current sheet.   }
\label{fig:ti-rr}
\end{figure*}

Given the efficient heating of plasma  in the current sheet, the global reconnection rate can be derived from the rate of increase of the total plasma energy in the computational domain. This energy is dominated by the thermal energy of plasma in the current sheet.  The left panel of figure \ref{fig:ti-rr} shows the total plasma energy $\cE\sub{pl}(t)$ computed via equation 
\beq
\cE\sub{pl} =\sum_{i=1}^{n_x} \sum_{j=1}^{n_y} \cE\sub{pl}^{i,j} \,,
\label{eq:total-energy}
\eeq
where the cell volume factor is ignored. Up to $t=4$ its increase is associated with the resistive spreading of the current sheet, and thereafter with the magnetic reconnection. The total increase of the plasma energy for $4\le t \le 10$  is $\Delta{\cE}\sub{pl}=0.475\times 10^{10}$. The total initial electromagnetic energy in the domain $\cE\sub{em}=0.197\times10^{11}$.  Ignoring the residual magnetic energy of plasmoids, 
\beq 
\Delta{\cE}\sub{pl} = \frac{\cE\sub{em}}{L} \langle v_r\rangle \Delta t \,,
\eeq
where $\langle v_r\rangle$ is the average speed of the electromagnetic energy inflow. For the above measurements, this equation yields $\langle v_r\rangle = 0.04$.  

The middle panel of figure \ref{fig:ti-rr} shows the solution at $t=9$ around the x-point near the largest plasmoid of the current sheet at this stage. Based on the velocity field, the x-point is located at $(x,y) \approx (-0.56,0)$. The right panel of this figure shows $v^y(y)$ along the line $x=-0.56$. One can see that the plasma (and the magnetic field) flows towards the x-point with the speed $\approx 0.05$, in agreement with the above estimate of the global reconnection rate.  This reconnection rate is only slightly below the 'universal' maximal reconnection rate $R\approx0.1$ found in resistive MHD, Hall-MHD and particle-in-cell (PIC) simulations, and in the observations of Earth and Solar magnetospheres \citep[see the references in ][]{Liu17}.  

This is another test problem where the DER step had to be switched off in order to avoid conversion failures at shocks. The same applies to the remaining tests described further down.

\subsection{ABC grid of magnetic ropes}
\label{sec:abc}

The double-periodic 2D ABC configuration of magnetic ropes is interesting because it is unstable and involves developing of current sheets at the non-linear phase of the instability via collapse of x-points \citep[e.g.][]{Parker83, East15,LSKP-17b}.  Its magnetic field is force-free with 
\beq
\begin{cases}
B^x=-B_0 \sin(k y)\,,\\
B^y=B_0 \sin(k x)\,,\\
B^z=B_0 (\cos(k x)+\cos(k y))\,.
\end{cases}
\eeq
The ropes with $B^z>0$ are located at $(x_i,y_j)=(2\pi/k)(i,j)$, the ropes with $B^z<0$ at $(x_i,y_j)=(\pi/k)(1+2i,1+2j)$, and the x-points (out-of-plane x-lines) at $(x_i,y_j)=(\pi/k)(2i+1,2j)$ and  $(x_i,y_j)=(\pi/k)(2i,2j+1)$, where $i,j\in Z$.  

In the test simulations,  $B_0=100$, $k=2/\pi$, and $p=\rho=1$.  The magnetisation varies from $\sigma=0$ at the x-points to $\sigma=2\times10^3$ in the centre of the magnetic ropes (islands). The domain is $(-1,1)\times(-1,1)$ with 400 uniformly spaced grid points in each direction, and periodic boundary conditions.  The initial equilibrium is perturbed by imposing the velocity field 
\beq
\vv{v}(x,y)=\frac{v_0}{\sqrt{2}}\left(-\cos\frac{k}{2}(x+y), \cos\frac{k}{2}(x+y), 0\right) \,,
\label{eq:abc-pert}
\eeq
with $v_0=0.01$. Such a perturbation is expected to trigger the shear-type mode of the instability \citep{LSKP-17b}.

\begin{figure*}
\centering
\includegraphics[width=0.3\textwidth]{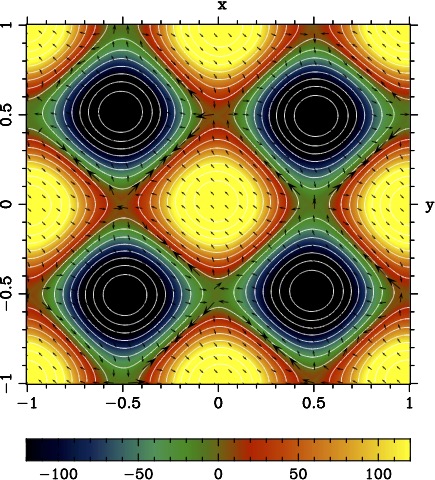}
\includegraphics[width=0.3\textwidth]{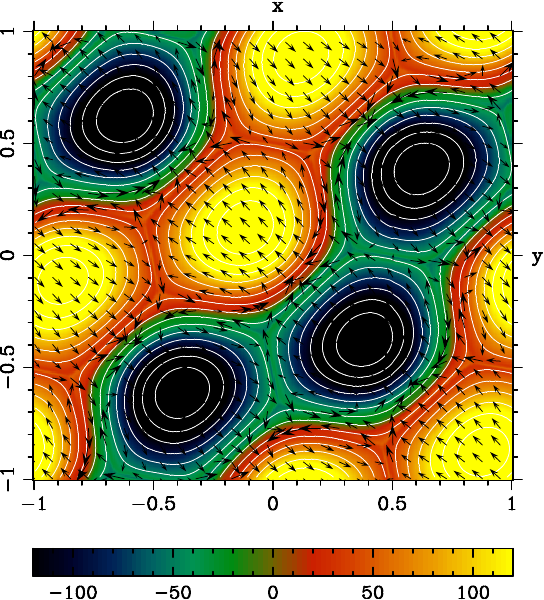} 
\includegraphics[width=0.3\textwidth]{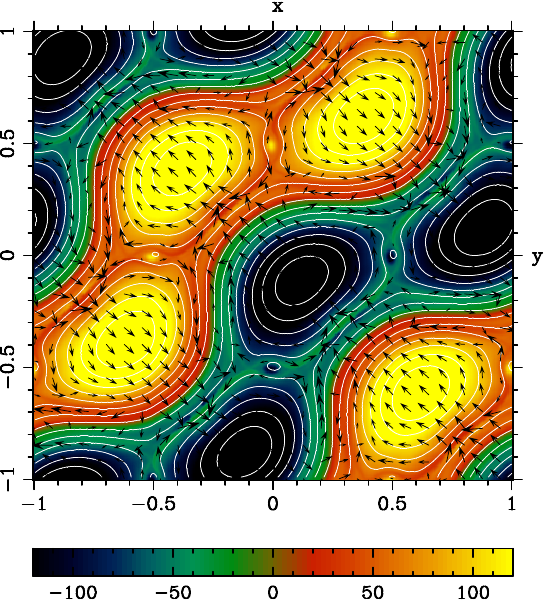} 
\includegraphics[width=0.3\textwidth]{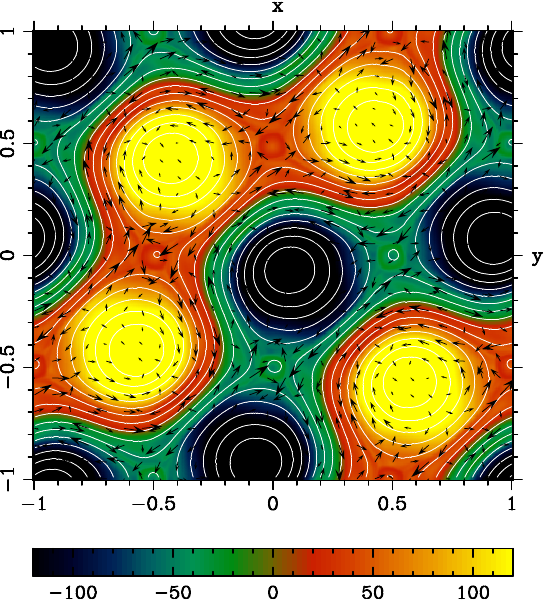}
\includegraphics[width=0.3\textwidth]{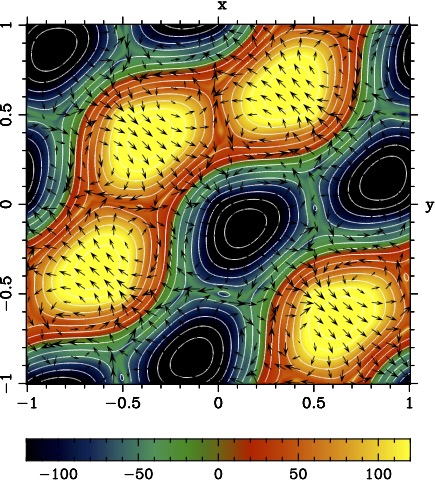} 
\includegraphics[width=0.3\textwidth]{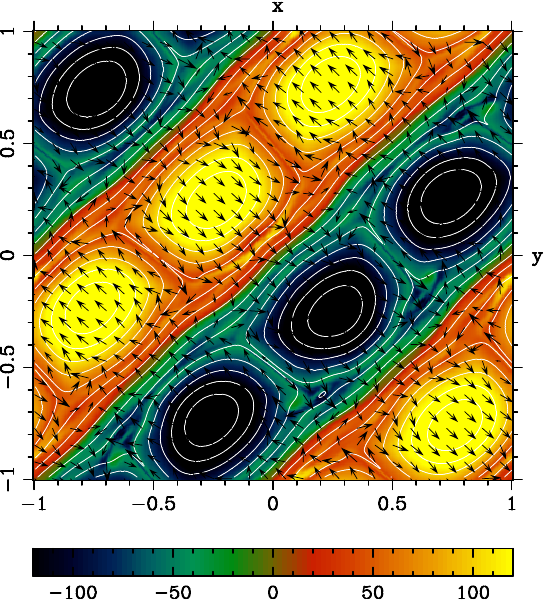} 
 \caption{ABC grid.  The colour map shows $B^z$, the contours show the magnetic field lines, and the arrows show the velocity field $\vv{v}$.  From left to right, $t=$1.0, 2.0, 3.0 in the top row, and  $t=$3.5, 4.0, 4.5 in the bottom row.}  
 \label{fig:abc1}
\end{figure*}

The global dynamics of the ABC grid is illustrated in figure \ref{fig:abc1}. Initially, the speed of global motion set by the perturbation \eqref{eq:abc-pert} increases, reaching the maximum value of $v\approx 0.35$ at about $t=2.5$. At this point,  the ropes of the same polarity (the same sign of $B_z$) form a linear chain running at the angle of $45^\circ$  to the $x$ axis, for the first time. The high value of the speed shows that the initial perturbation may be considered as small. At around $t=3.5$ there is a turning point, when the ropes start moving in the opposite direction. The subsequent global motion is a decaying oscillation about the state with the $45^\circ$-alignment. In the ideal model, this state is a stable equilibrium \citep{LSKP-17b}. 

\begin{figure*}
\centering
\includegraphics[width=0.23\textwidth]{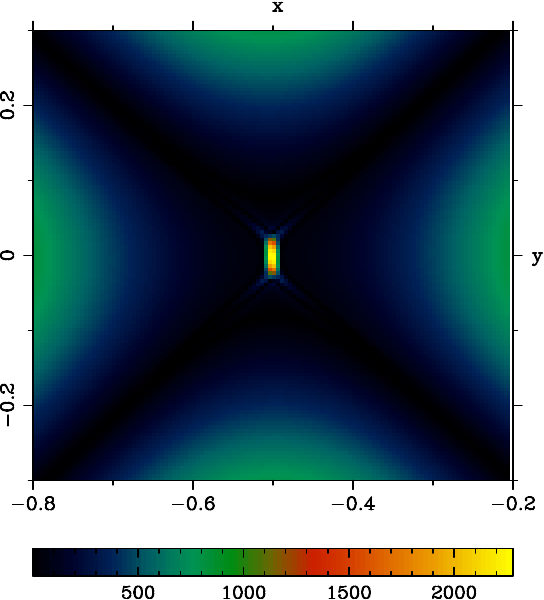}
\includegraphics[width=0.23\textwidth]{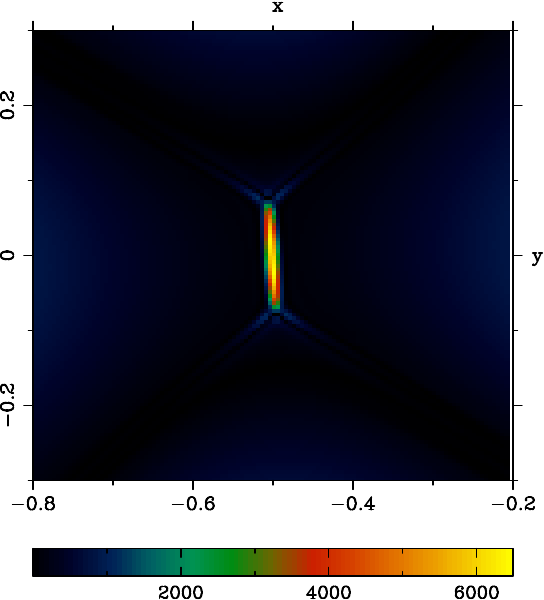}
\includegraphics[width=0.23\textwidth]{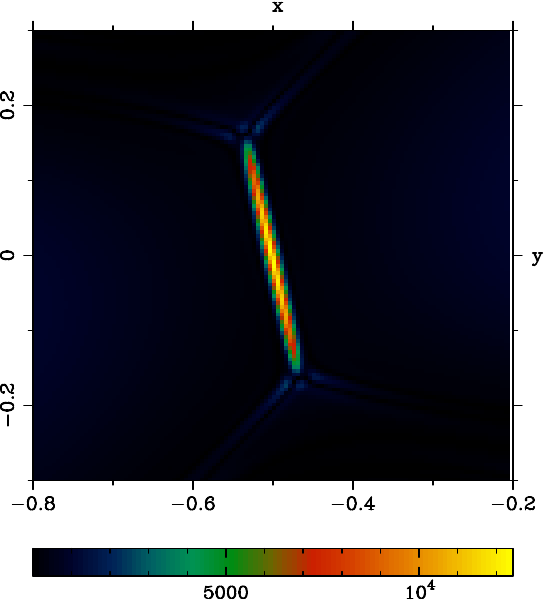} 
\includegraphics[width=0.23\textwidth]{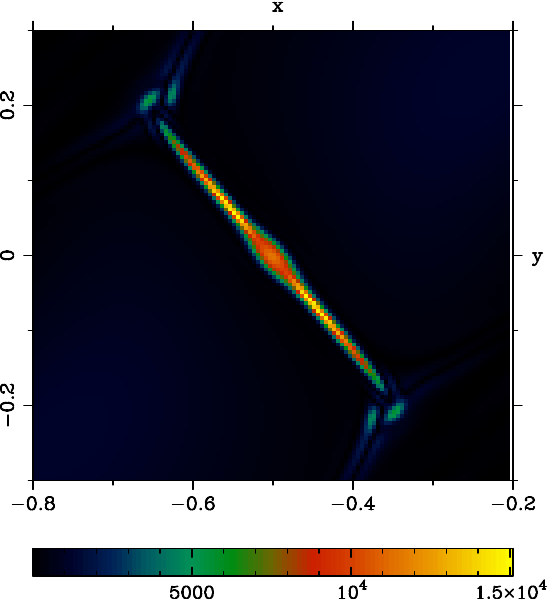} 
\includegraphics[width=0.23\textwidth]{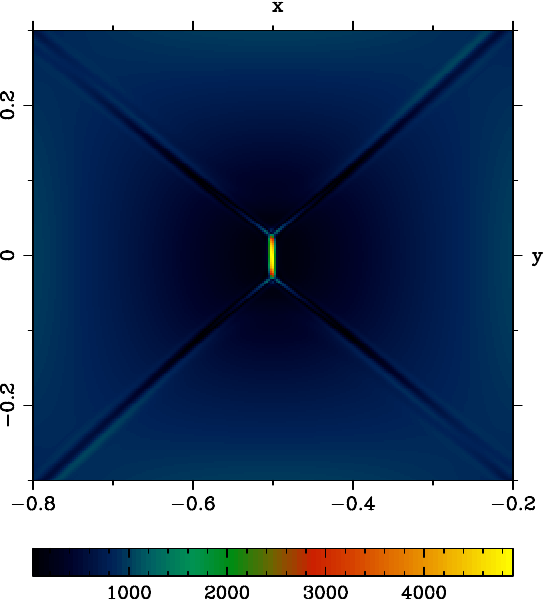}
\includegraphics[width=0.23\textwidth]{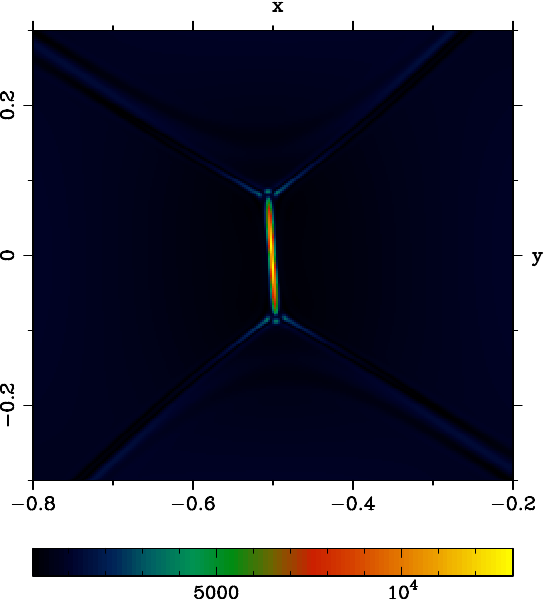}
\includegraphics[width=0.23\textwidth]{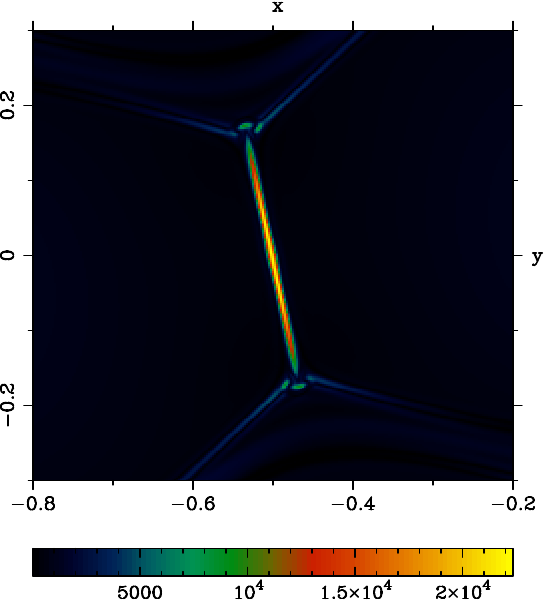} 
\includegraphics[width=0.23\textwidth]{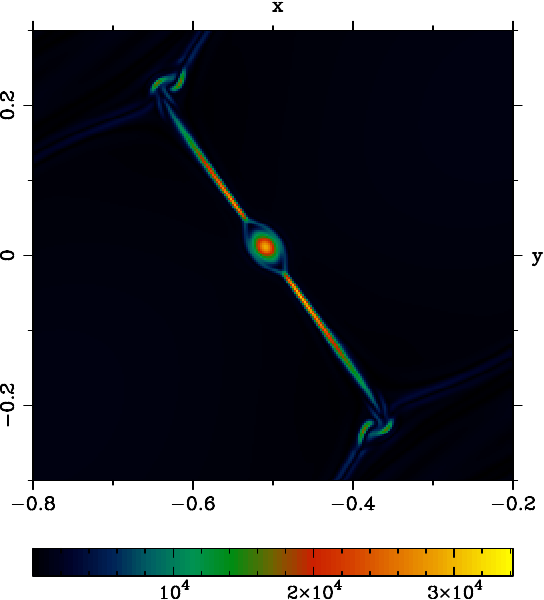} 
 \caption{ABC grid. From left to right,  the total electric current density for the current sheet emerging near the point $(x,y)=(-0.5,0)$ at $t=1.0$, 1.5, 2.0, and 2.5 respectively. The bottom row shows the solution obtained with the doubled resolution.}
 \label{fig:abc3}
\end{figure*}

On approach to the oblique alignment, the x-points collapse into current sheets separating ropes of the same polarity (see the top-middle panel of figure  \ref{fig:abc1}).  These current sheets appear to suffer the tearing instability, and very soon a single plasmoid emerges in the middle of each sheet (the top-right panel of figure  \ref{fig:abc1}). Figure \ref{fig:abc3} zooms into the current sheet located around the point $(x,y)=(-0.5,0)$.   As one can see, the current sheet is not yet developed at $t=1.0$.  At  $t=1.5$,  it appears as a vertical linear structure, whose length is approximately $3.5$ times shorter than its ultimate length.  At $t=2.0$, its length increases approximately by a factor of two and its orientation in space changes, reflecting the relative motion of the flux ropes.  At $t=2.5$, the current sheet is inclined at about $45^\circ$ to the y axis, and in the middle of it there is a bulge visible with a naked eye. Thus, the plasmoid had only time $\Delta t\approx1$ to grow from perturbation.  

This current sheet is as thin as the initial current sheet in the tearing instability simulations described in section \ref{sec:tihcs}, both in terms of the number of cells, approximately four, and in terms of the linear size, $a\approx 0.013$.  Hence, based on the numerical resistivity Lundquist number is also approximately the same, $S^*\approx 300$. The total length of the current sheet is $2L\approx 0.45$, leading to the aspect ratio $a/L\approx 0.045$.   The aspect ratio of Sweet-Parker's equilibrium current sheet, $(a/L)\sub{SP} \simeq S^{-1/2}=1/S^*$ \citep{SP-57,SP-58}, corresponding to the same value of $S^*$ is much smaller, $(a/L)\sub{SP}\approx 0.003$. Taking into account the reduction of the numerical resistivity for current sheets inclined at $45^\circ$ by the factor of two would make this estimate even lower.  Therefore one may ignore the flow inside the current sheet and apply the results of \citet{Furth63} on the growth of the tearing instability \citep[c.f.][]{DelZanna16}.   Equation \eqref{eq:fastest-k} then gives the wavelength of the fastest growing mode  $\lambda\sub{m}\approx 0.25$, which is consistent with the fact that only one plasmoid emerges in this current sheet. 

\citet{Pucci14} argued that current sheets are unable to reach the Sweet-Parker equilibrium because they get fragmented by the tearing instability before reaching $(a/L)\sub{SP}$. They also proposed the current sheets scaling  $(a/L)\sub{PV}= S^{-1/3}={S^*}^{-1/2}$, where $(a/L)\sub{PV}$ can be interpreted as the lower limit on the aspect ratio of non-fragmented current sheets. For $S^*=300$, this yields $(a/L)\sub{PV}\approx0.057$, which is similar to the ultimate aspect ratio of the current sheets in our ABC simulations. The Pucci-Velli scaling can be supported with a simple causality argument. The minimum time required to form a current of the half-length $L$ is its Alfv\'en time  $\tau\sub{A}\simeq L/c\sub{A}$. The e-folding time $\tau\sub{m}=1/\omega\sub{m}$ of the fastest growing unstable mode cannot be much shorter than $\tau\sub{A}$, as otherwise the current sheet fragments already during its formation.  Hence for the longest non-fragmented current sheet, $\omega\sub{m}\tau\sub{A}\sim1$.  Equation \eqref{eq:highest-om} for the growth rate of the fastest mode can be conveniently written as
\beq
\frac{a}{L} \approx \fracp{0.63}{\omega\sub{m} \tau\sub{A}}^{2/3} S^{-1/3} \,,
\label{eq:omega-ideal}
\eeq  
which yields the Pucci-Velli scaling when $\omega\sub{m}\tau\sub{A}\sim1$. 

For $S^*=300$ and $a/L=0.045$, equation \eqref{eq:omega-ideal} yields $\omega\sub{m} \tau\sub{A}\approx 0.81$, consistent with the current sheets just braking the fragmentation threshold.  Given the observed $\tau\sub{A}\approx 0.22$, the estimated value of $\omega\sub{m}$ yields  $\tau\sub{m}\approx 0.27$. Because the plasmoids emerge on the timescale $\Delta t \approx 1$, this implies  that their amplitude could grow only by the factor $\approx\exp(1/0.27)\approx 40$. This is a small growth compared to what is normally achieved in the numerical studies of instabilities, which start with very small perturbations.  However, the current sheets in the ABC simulations are highly dynamic from the start and hence not expected to be near to such an almost perfect balance at any point in their evolution. 


We also run this problem at the doubled resolution, and found a very similar evolution, especially at the early phase. In particular, the plasmoids emerge on the same time scale (see the  bottom row of figure  \ref{fig:abc3}). There is still only one plasmoid per current sheet, but the secondary current sheets have approximately the same aspect ratio as the primary current sheet at the lower resolution, suggesting that secondary plasmoids may emerge when the resolution is increased furthermore.  

The PIC simulations of this problem \citep{LSKP-17b} show a similar dynamics, but with some quantitative differences. In these simulations, the ABC grid has the same linear scales, and the Alvf\'en speed is also very close to the speed of light. Hence, no time rescaling is required.   The initial plateau phase in the PIC simulations continues up to $t=4$, not $t=1.5$ like in our simulations.  However, this difference is attributable to the amplitude and nature of the initial perturbation of the ABC grid and simply requires us to shift the timing of the PIC simulations back by about $\Delta t=2.5$ for comparison with our results.  With this shift applied, by $t=10$ the total electromagnetic energy in the PIC simulations is reduced by about 40\%, compared to the 18\% found here.  This implies an approximately twice as fast reconnection rate in the PIC simulations compared to ours. Moreover, by this time the initial periodic structure of the ABC grid is erased, with the ropes of single polarity merged into larger structures (see figure 8 in \citet{LSKP-17b}), whereas in our simulations the individual ropes are still identifiable.  This is also consistent with the higher reconnection rate of the PIC simulations. According to the figure 8 in \citet{LSKP-17b}, the plasmoids are not seen at $t=1.5$ and 2.5, but fully formed at $t=3.5$. Thus, they emerge on approximately the same time scale as in our simulations. This suggests that the timing is dictated by the macroscopic dynamics of the system rather than by the details of the microphysics. The number of plasmoids is also about one per current sheet. (However in other PIC runs, which yields thicker current sheets, the plasmoids do not emerge at all. See the discussion around equation \ref{eq:LarmorR}. )       

\subsection{Magnetic field errors in the 2D simulations}
\label{sec:Berrors}

In our implementation of the splitting approach we used the GLM method to keep the magnetic field near, but not exactly in, the divergence-free state.  The deviation from the divergence-free state originates due to the truncation errors in the numerical integration of the Faraday equation and this allows us to estimate the errors in the magnetic field.  This can be used to assess the potential impact of such errors on the conversion failures when the same problems are attempted in the standard mode of our code. 

 According to the analysis of section \ref{sec:IRMHD}  (see equations  \ref{eq:Berr-para} and \ref{eq:Berr-perp}), in order not to cause the conversion failures, the relative error in magnetic field must satisfy the condition
$$
\left|\frac{\delta B}{B}\right|\lesssim \frac{\gamma^2}{\sigma}\,.
$$
To apply this result, we first estimate the relative error in the magnetic field as 
$$
\fracp{\delta B}{B}_{ij} \simeq \left((B^i_{i+1,j}-B^i_{i-1,j})+(B^j_{i,j+1}-B^j_{i,j-1})\right)/||\vv{B}_{i,j}|| \,,
$$
and then compute the error parameter 
$$
\err=\left|\frac{\delta B}{B}\right|\frac{\sigma}{\gamma^2}\,.
$$ 
When $\err \gtrsim 1$, the error is sufficiently large to result in an unphysical state and hence cause a failure of the variable conversion. When  $\err < 1$, the error is below the safety limit. Since the analysis leading to these expectations is not comprehensive but confined to simple special cases, some caution needs to be exercised here.

\begin{figure*}
\centering
\includegraphics[width=0.35\textwidth]{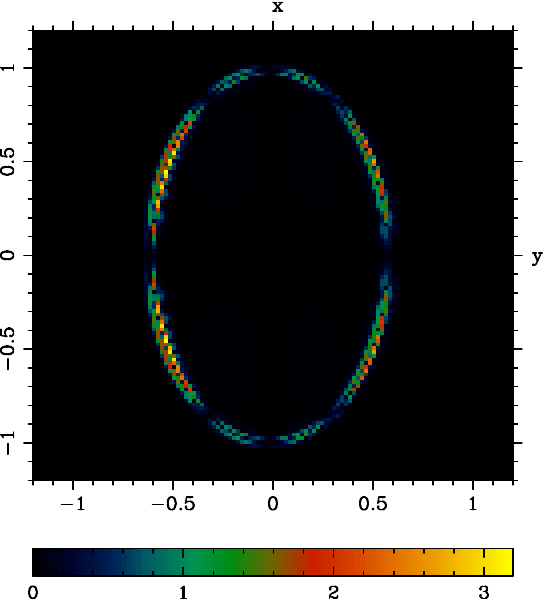}
\includegraphics[width=0.35\textwidth]{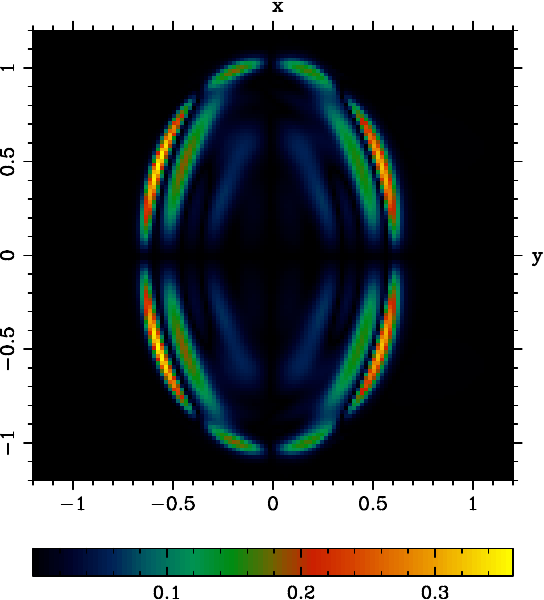}
\includegraphics[width=0.35\textwidth]{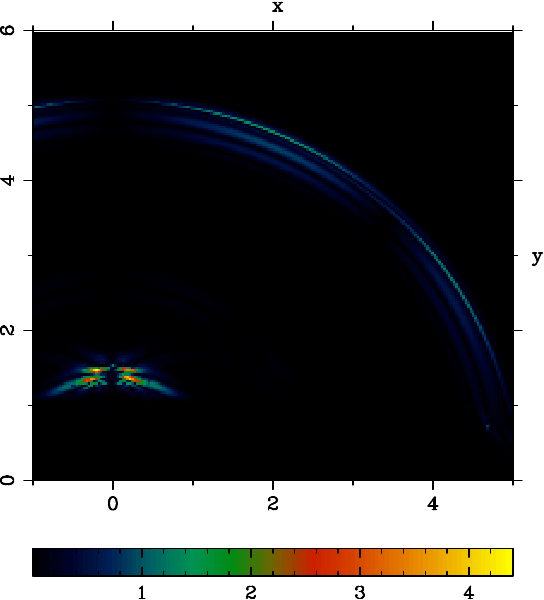}
\includegraphics[width=0.35\textwidth]{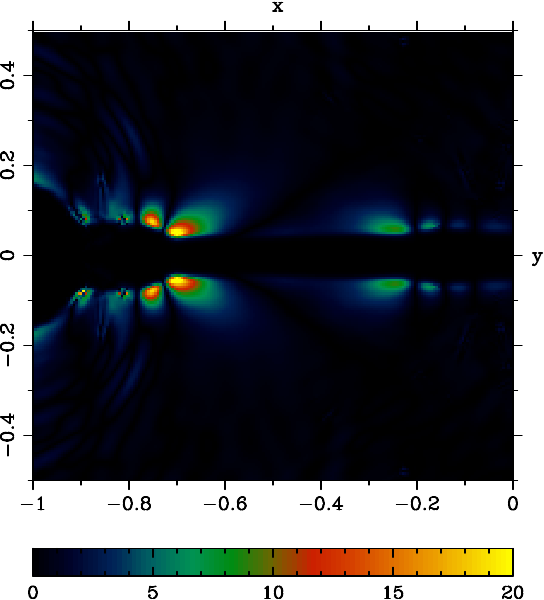} 
 \caption{ \bf The error parameter $\err$. Top-left panel: The magnetic rope test with $B_0=100$ at $t=0.02$. Top-right panel: The magnetic rope test with $B_0=100$ at $t=5$. Bottom-left panel: The cylindrical explosion test with $B_0=1$ at $t=4$. The high residuals colocate with the transient shock waves associated with the oscillations in the direction normal to the magnetic field.     Bottom-right panel: The tearing instability test with $B_0=500$ at $t=7.5$. }
 \label{fig:divb}
\end{figure*}

Figure  \ref{fig:divb} shows the error parameter for the magnetic rope, cylindrical explosion and  tearing instability tests.  At the start of the magnetic rope test, $\err$ reaches the values significantly exceeding unity near the rope surface (see the top-left panel of figure \ref{fig:divb}).  Based on these values, we anticipated the simulations run in the standard mode of the code to crash at the very start, and they did even for the Courant number as small as $\mbox{C}=0.01$. By $t=5$, the maximum value of $\err$ reduces to  $\approx 0.36$, suggesting that it may be possible to continue the simulations from this point in the standard mode. This was indeed the case, but only with the Courant number reduced down to $C=0.4$. 

The bottom-left panel of figure \ref{fig:divb}) shows $\err$ for the explosion test with $B_0=1$ at $t=4$.  Even at this time it remains about unity at the external fast shock, and reaches $\err\approx 4$ in the vicinity of the point $(x,y)=(0,1.3)$,  where reverberation of the magnetic field appears to have triggered a secondary shock wave.  As already stated in sec.\ref{sec:cyl-expl}, our attempt to run this test in the standard mode of the code has failed.  For the test with $B_0=0.1$, the error parameter is much lower, $\err\lesssim 0.3$, reaching the maximum at the external shock.  These low values are consistent with the fact that for this value of $B_0$ the test can be run with the standard code from the start to the finish.     

In the tearing instability test, $\err$ is small at the linear phase, but then strongly increases in the non-linear phase. As one can see in the bottom-right panel of figure \ref{fig:divb}, the error parameter can reach the values $\err\approx 20$ in the vicinity of the current sheet. Although ${\delta B}/{B}$ itself is higher inside the current sheet, where the magnetic field is highly distorted, leading to a stronger departure from the divergence-free state, the magnetisation $\sigma$ is significantly lower there, due to the presence of very hot plasma. In the outskirts of the current sheet, $\sigma$ is very high and the magnetic field is still strongly distorted by the plasmoids.  Our attempts to continue the simulations from this point in the standard mode have failed.

 Overall, the data support the conclusion that the errors in magnetic field are responsible for failures of standard schemes in the high-$\sigma$ regime. In principle, an increase of the accuracy in the numerical integration of the Faraday equation may help to extend the applicability of standard schemes.  In this regard, schemes utilising the CT  approach may be more robust as they eliminate the errors associated with the $\vdiv{B}=0$ constraint. Other errors, however, will remain and may still be too high.  Increasing the order of scheme's accuracy could help too. This is expected to be very effective in regions with smooth magnetic field, but not at discontinuities.

\section{Summary and discussion}
\label{sec:summary}

The main goal of this study was to find a new approach to numerical RMHD in the high-magnetisation regime, where the standard conservative schemes turned out to be highly unreliable.  Its direction was motivated by the understanding that the most attractive feature of such schemes, the conservation of total energy-momentum, is also the main reason for their failures in the high-$\sigma$ regime. For such a high magnetisation, the energy-momentum tensor is dominated by the electromagnetic field, and even relatively small errors emerging in the numerical integration of the Faraday equation can render the set of conserved variables unphysical.   This understanding invited us to search for a way of breaking the strict link between the energy-momentums of plasma and electromagnetic field imposed by the total energy-momentum conservation and enforced in the standard conservative schemes. Moreover, in this regime the electromagnetic field is largely force-free, and its evolution is well approximated by the equations of FFDE, which can be considered as a singular limit of RMHD when $\sigma\to\infty$. This invited us to study the potential of the perturbation approach, where the electromagnetic field is evolved mostly as force-free, and the plasma introduces only a small perturbation to the FFDE solution, with $\sigma^{-1}$ playing the role of a small parameter.   However, the standard asymptotic expansion approach is complicated, with higher order terms needed for accuracy in the case of moderate $\sigma$.  Moreover, it is not suitable for $\sigma\leq 1$, significantly limiting the area of application. 

Instead, we opted for a generalisation of the approach proposed by \citet{Tanaka94}, where the perturbation is governed the RMHD equations with the energy-momentum tensor modified via explicit subtraction of the energy-momentum tensor of the force-free field. In contrast to \citet{Tanaka94}, where  the strong force-free background field is stationary, in our approach it is dynamic.  So, we enlarge the system of differential equations, which is now composed of two linked subsystems: the FFDE system for the electromagnetic field, and the perturbation system for the plasma. The latter has the same number of equations as the original RMHD. These subsystems are linked via the interaction terms in the perturbation system and the perfect conductivity condition.   This approach delivers a numerical scheme which can be applied in both the high- and low-$\sigma$ regimes.          

The equations of the enlarged system are integrated simultaneously, and at the end of each time step the electromagnetic field of the FFDE system and its perturbation are recombined.  Thus, the final result is a splitting scheme, which is similar in spirit to operator-splitting schemes, but different in form. Like in the operator-splitting method, we separate processes of different nature and do this to bypass the stiffness of differential equations. However, if the operator-splitting method is focused on the stiffness arising due to the very different timescales associated with the involved differential operators (processes), our splitting scheme deals with the stiffness arising due to the significant difference in the magnitude of contributions to the conserved quantities from components of different nature. If the operator-splitting method involves successive integration of  simplified versions of differential equations, where some of the operators are dropped, we solve the whole system of equations simultaneously. This simplifies development of higher-order schemes.            

Both the subsystems of split RMHD can be written as conservation laws, and hence can be numerically integrated using the standard methods developed for such laws. We adopted the 3rd-order WENO approach similar to that of \citet{DZ07}, with some modifications. In particular, 1) we developed a new 3rd-order WENO interpolation, which allows rapid transition to the 3rd-order scaling of computational errors at low resolution and does not result in a loss of accuracy at turning points; 2) the code required a new variable conversion algorithm; 3)  we used the GLM method  \citep{Dedner02} to keep the magnetic field nearly divergence free; 4) we developed a simple algorithm to locate strong shocks in order to switch off the DER step of  \citet{DZ07} at their locations. The latter is needed to suppress the spurious oscillations capable  of causing conversion failures at high-$\sigma$ shocks.           

Only the momentum density is used for the variable conversion of the FFDE subsystem. As a result, the energy of the FFDE subsystem and hence the total energy are not conserved. This break of conservation is at the centre of our splitting method. One can compute the difference between the energy density of the FFDE subsystem based on the energy conservation law and the one based on the updated $\vv{E}_\0$ and $\vv{B}_\0$, and transfer it to the perturbation subsystem, thus enforcing the total energy conservation. Such energy exchange between plasma and electromagnetic field implicitly occurs in standard conservative schemes, where it facilitates plasma heating in current sheets.  However, it is also responsible for their failures in the high-$\sigma$ regime.  Hence in the splitting approach, the energy transfer must be conditional, filtering out the cases where this may lead to a crash.  A simple analysis shows that the energy transfer is safe when the transferred energy is positive.  In this case, the energy transfer amounts to plasma heating via the numerical dissipation of the electromagnetic energy.  When the positivity is the only condition,  plasma is also heated by weak waves generated in active regions.   By setting positive lower limits on the transferred energy,  this low-level heating can be suppressed.   In the splitting approach, there is another, now uncontrollable mechanism of plasma heating, which involves the interaction terms of the perturbation equations. This mechanism accounts for about 50\% of heating in current sheets.                     

The 1D and 2D test simulations of continuous hyperbolic waves and associated shock waves have shown that the splitting method remains robust and accurate when applied to problems with very high $\sigma$.  This is particularly true for continuous waves.  Shock waves are more problematic, and in some cases the code can fail to deliver accurate values for the plasma parameters.  Our test results suggest that this occurs when the tangential component of magnetic field experiences large jumps across the shock, leading to excessive plasma heating via the uncontrollable numerical dissipation of electromagnetic energy. As a result, the shock fails to develop monotonic structure. Although such shocks do exist,  they are unlikely to be common.  For example,  in the 2D simulations of explosions in strong uniform magnetic field, the variation of the tangential magnetic field is much smaller.  

The splitting approach delivers accurate solutions not only for high-$\sigma$ problems, but also for problems with low magnetisation, as illustrated by the shock tests FS7 and FS9, where the magnetisation of the upstream state is only $\sigma=10^{-3}$, and by the blast wave simulations with $\sigma_0\sim10^{-5}$.  Moreover, as the magnetisation decreases, the shock solutions become progressively more accurate. In fact, for unmagnetised plasma, the splitting scheme reduces to the standard conservative scheme for relativistic hydrodynamics. For sub-relativistic problems, the splitting approach also performs very well, as demonstrated by the FS9 test where the sound speed $c_s\approx 0.01$ and the Alfv\'en speed $c\sub{A}\approx 0.007$.  Thus, the splitting approach can be applied to many complex astrophysical problem involving states with vastly different parameters, like active galactic nuclei, where the low-$\sigma$ accretion disk coexists with the high-$\sigma$ magnetosphere of the central black hole.  

Our test simulations of problems involving current sheets have demonstrated that the splitting approach can capture the active phenomena of plasma astrophysics involving fast magnetic reconnection. The fast reconnection plays an important part in many astrophysical phenomena, resulting in explosive dynamics, plasma heating, and acceleration of nonthermal particles responsible for high-energy emission. The latter is particularly important for high-$\sigma$ relativistic plasmas, where PIC simulations of collisionless shocks revealed their low efficiency in particle acceleration \citep{ss-09,ss-11}.   The reconnection events are preceded by the formation of current sheets, which can emerge spontaneously in quasi-static configurations, or forced by plasma motion in highly-dynamic conditions \citep[e.g.][]{PP22}.  

The detailed structure and evolution of current sheets depends on the microphysics responsible for the deviation from the magnetic flux freezing approximation of ideal MHD. Interestingly, most numerical  methods for ideal MHD also break the flux freezing because of the truncation errors of numerical algorithms. This phenomena is called the numerical resistivity.  Although magnetic reconnection has been seen in ideal MHD simulations \citep[see e.g.][for more recent examples]{Lait05,Ripperga22,FFG23,BMBM24}, this has been treated with a great deal of scepticism.   However, in the plasmoid-dominated regime the overall dynamics of current sheets and the reconnection rate do not seem to be that sensitive to the incorporated microphysics \citep[e.g.][]{Liu17,PP22}.  This is even more so in the theory of turbulent reconnection, where the reconnection rate does not depend on the microphysics altogether \citep{LV99,Laz20}.  This motivated us to include problems involving current sheets in the suite of test simulations.  
   
We started by studying the properties of numerical resistivity in our scheme, using as a guide the ansatz of \citet{Remb17}. The 1D simulations of degenerate Alfv\'en waves  (section \ref{sec:daw-1d}) are in agreement with the simple prescription for the numerical resistivity \eqref{eq:num-resist} based on the value of the rounding error.  They confirm the dependence of numerical resistivity on the scheme's order of accuracy, numerical resolution, and the characteristic length scale of the magnetic field variation $\cL$. Since the equation  \eqref{eq:num-resist} states $\eta\sub{num}\propto \cL^{-2}$, the numerical resistivity is similar to the so-called anomalous resistivity, with $\eta\propto j^2$, used in resistive MHD simulations to achieve fast magnetic reconnection \citep[e.g.][]{YS94,SPC19,FNC23}.   In our 2D simulations, the corresponding magnetic Reynolds number varies from $\mbox{Re}\sub{m} \sim 10^2$ for current sheets which are only few cells wide, to $\mbox{Re}\sub{m} \sim 10^8$ on the domain scale. Thus, the numerical resistivity has little effect on the large-scale dynamics but very important in 'paper-thin' current sheets.  As expected, the numerical resistivity is anisotropic. Our initial investigation of this issue suggests that it is highest when magnetic field is aligned with the grid lines and reduces by the factor of two when the magnetic field is at the angle of $45^\circ$ to the grid lines.    

In section \ref{sec:tihcs},  we described the simulations of the tearing instability for the case of a very long and thin, only few grid cells across, Harris current sheet aligned with the computational grid. Quite remarkably, the results of these simulations are in good agreement with the key conclusions of the basic theory of this instability developed within the framework of Newtonian resistive MHD with constant scalar resistivity \citep{Furth63} (Although the theory of the tearing instability was developed in the Newtonian framework, the relativistic results are basically identical \citep{ssk-tearing,DelZanna16}.). In particular, the wavelength of the fastest growing mode and its growth rate agree with the theoretical values obtained with the numerical resistivity (in place of the actual $\eta$) assuming $\cL\approx a$, where $a$ is its half-width of the current sheet. This result is somewhat surprising, as the theory predicts the existence of a narrow resistive (tearing) sublayer (boundary layer) in the middle of the current sheet. The thickness of this sublayer is 
\beq
     a\sub{sub} \approx 1.5\, {S^*}^{-1/4} 
\eeq        
\citep{Furth63}. For the consistent with the simulations value $S^*\approx400$, the corresponding $a\sub{sub}\approx 0.3a=3\times10^{-3}$, whereas the cell size $\Delta y=5\times 10^{-3}$, and hence the sublayer is not resolved.  In fact, it is collapsed into a discontinuity (see the right panel of figure \ref{fig:ti-rr}).  On the other hand, it has been claimed that many properties of reconnection are largely determined by  the ideal MHD dynamics outside of the sublayer and only weakly depend on its microphysics \citep{Liu17,PP22}. This is especially clear in the case of forced reconnection, where the reconnection rate is set by the externally-determined rate of plasma inflow into the current sheet.   At the nonlinear phase of our ideal MHD simulations, the dynamics of the current sheet is also very similar to what in seen in resistive MHD and PIC simulations, including the development of primary plasmoids, their merger, the emergence of secondary plasmoids in secondary current sheets etc. \citep[e.g.][]{Bhatta09, DelZanna16,PS18}. The estimated global reconnection rate is about $0.04$. 

The simulations of the unstable ABC grid of magnetic ropes (section \ref{sec:abc}) allowed us to study the case where the current sheets are not present in the initial solution, but develop as a result of the x-point collapse. These current sheets produced solitary plasmoids on the timescale which is only few times longer than their ultimate Alfv\'en time scale.  These results are in agreement with the conclusion reached by \citep{Pucci14} that current sheets become fragmented by the tearing instability well before they reach the Sweet-Parker equilibrium, thus making  studies of Sweet-Parker current sheets a matter of purely academic interest.      

It is quite interesting that the PIC simulations of the ABC problem for electron-positron plasma \citep{LSKP-17b} yield very similar results in terms of the timescale of the current sheet fragmentation, the number of emerging plasmoids, and the reconnection rate.  As noted in  \citet{LSKP-17b}, the half-thickness of the collisional current sheets emerging in the PIC simulations is set by the Larmor radius of the plasma particles heated in the sheet, $a\sim r\sub{L,h}$. For relativistic plasma, this is approximately 
\beq
   r\sub{L,h} = \sigma_0 \gamma\sub{t,0} r\sub{L,0} \,,
   \label{eq:LarmorR}
\eeq
 where $\sigma_0=B_0^2/4\pi w_0$ is the magnetisation of the inflowing plasma, $\gamma\sub{t,0}$ is the thermal Lorentz factor of its particles, and $r\sub{L,0}=m\sub{e} c^2/e B_0$.  They have also found that the emergence of plasmoids depends on the parameter $r\sub{L,h}/D$, where $D$ is the wavelength of the ABC grid. Namely, they begin to emerge when $D/r\sub{L,h}>126$. Since the half-length of the current sheets $L \approx D/3$, this can be written as
$$
       \frac{r\sub{L,h}}{L} < 0.02 \,. 
$$     
Thus, even the fragmentation threshold  is similar to what is found in our ideal RMHD simulations.      

The results of our study of current sheets suggest that in principle the fast reconnection events can be captured in simulations even with ideal RMHD and MHD codes.  Although the development of plasmoids and explosive reconnection has already been reported in the ideal RMHD simulations of neutron-star magnetospheres \citep{Bucc06} and black hole accretion \citep{Ripperga22},  our study seems to be the first one where the plasmoid-dominated regime of magnetic reconnection is studied more or less systematically (a more advanced study is under way), and an agreement with the resistive MHD theory is found.  This warrants a closer look at the numerical resistivity and its properties in different numerical schemes. It is quite possible that its properties are close to those of the proper resistivity only in some schemes and drastically different in others. For example,  \citep{Remb17} found negative resistivity for their scheme.  It is possible that the peculiarities of the splitting approach play a role too.  Especially the fact that in the ideal FFDE approximation current sheets collapse into discontinuities, with the corresponding reconnection rate approaching the speed of light.           

Our results show that for the thinnest current sheets allowed by the code, only few cells wide, the current sheets should be at least $\sim$100 cells long for the tearing instability to trigger fast reconnection on the Alfv\'en timescale.  Very long current sheet are know to exist in stellar magnetospheres, including the high-$\sigma$ magnetospheres of black holes and neutron stars. However in other astrophysical problems, current sheets may be much smaller compared to the dynamical scales of interest. For example, the size of reconnection sites responsible for the gamma-ray flares in the Crab nebula is only about one light day, whereas the size of the nebula is about 10 light years.  For such problems, code's ability to efficiently resolve small thin structures becomes paramount.           
          
Somewhat paradoxically, the ideal MHD codes might end up being more suitable than the resistive codes for large-scale problems of astrophysical interest \citep[c.f. ][for the simulations of MHD turbulence]{DelZanna24}.  First, the actual resistivity of resistive codes has to be much higher than the numerical one to make its introduction meaningful. This would make current sheets significantly thicker and hence they would have to be much longer to allow fast reconnection.  Second,  uniform scalar resistivity will have strong effect on the magnetic field, and hence the plasma dynamics, outside current sheets, leading to much lower magnetic Reynolds numbers on the large scales compared to what it would be with an ideal code \citep{Mattia23,Mattia24}.  In principle, this can be mitigated with anomalous resistivity, which depends of the strength of the electric current. Finally, the resistive codes are great for verifying the analytical results of resistive MHD and exploring their nonlinear regime, but since the astrophysical plasma is mostly collisionless, the actual  benefits of the resistive model in astrophysics are not that obvious.  

For RMHD, the fact that the numerical resistivity is not Lorentz-invariant is likely to be an issue for the simulations involving fast relativistic flows. As can be seen in \eqref{eq:relat-resist}, for such flows the resistivity reduces like $\gamma^{-1}$, whereas the numerical resistivity does not. One relevant example of such flows is the striped pulsar wind, where the time-dilation effect may prevent the reconnection of stripes till the wind passes through its termination shock \citep{LK01}. However,  direct numerical simulations of such wind in the pulsar frame are extremely challenging and require significant simplification anyway.

Over the last decade, the kinetic approach based on the particle-in-cell (PIC) method was successfully applied to numerical simulations of pulsar and black hole magnetospheres \citep[e.g.][]{PS14,Parfrey19,Crinquand20,SCC24}. This approach has no difficulty in dealing with highly magnetised plasma but suffers from the scale-separation issue.  PIC simulations must resolve the microphysics scales, which severely limits the accessible macroscopic scale and makes the method computationally expensive.  Although the most recent studies show that the macroscopic size of some astrophysical problems can be scaled down towards the microscopic scales, without the large-scale dynamics being "contaminated" by the microphysics, in general the issue is here to stay.  One approach to mitigating this issue is the use of hybrid schemes, where PIC computations are limited in extent and carried out only where they are unavoidable, for example to compute the nonthermal radiation \citep[e.g.][]{SCC24}.  Another option is not to use PIC simulations directly altogether, but to incorporate the PIC predictions on particle acceleration and non-thermal emission at the sub-grid level of fluid simulations. This requires accurate treatment of plasma in the high-$\sigma$ regime, including the value of $\sigma$ itself, and this is where the splitting approach to numerical RMHD promises to be most useful.          

\section{Conclusions}
\label{sec:conclusions}

In this work, we developed a novel numerical method for integrating RMHD equations, which allows to extend the applicability domain into the regime of extremely high magnetisation (high-$\sigma$) typical to the magnetospheres of neutron stars and black holes, and expected in the magnetised relativistic outflows from them as well.  The method is based on splitting the RMHD equations into interacting (linked) subsystems, one governing the electromagnetic field, and another governing the motion of plasma. The splitting  breaks the stiffness of RMHD equations in the high-$\sigma$ regime, where the total energy-momentum tensor is largely dominated by the electromagnetic field.  The method sacrifices the total energy-momentum conservation of standard conservative schemes for RMHD, and this does not allow the small numerical errors in magnetic field to result in catastrophic errors for the plasma parameters.   Both the subsystems have the form of conservation laws, which allows to combine the splitting method with various numerical methods developed for such laws. In the current code, we applied the 3rd-order accurate WENO approach.     

The suitability of the splitting method to high-$\sigma$ problems has been confirmed by a variety of 1D and 2D test simulations presented in this paper. Moreover, the code remains accurate for low-$\sigma$ problems, including the unmagnetised regime ($\sigma=0$), and the sub-relativistic problems. Thus, the splitting method can be used for numerical simulations of complex astrophysical phenomena, which involve components with vastly different physical parameters, with no need for development of hybrid codes.   

Given the importance of fast magnetic reconnection in high-energy astrophysics,  particular attention has been paid to determining the numerical resistivity of the code and to test problems involving long and thin current sheets.  
Studying the numerical decay of periodic degenerate Alfv\'en waves, we verified and calibrated a simple model of numerical resistivity, and found it to be similar to the anomalous resistivity. In the 2D simulations of the tearing instability in a long Harris current sheet, we found the results to be in good agreement with the basic theory by \citet{Furth63} when the resistivity proper is replaced with the numerical resistivity. At the nonlinear phase, the simulations exhibited the typical properties of the fast magnetic reconnection in the plasmoid-dominated regime.  The 2D simulations of the ABC grid of magnetic ropes allowed us to study the dynamics of current sheets emerging via x-point collapse. These current sheets became fragmented by tearing instability on Alfv\'enic timescale before they could reach the aspect ratio of the Sweet-Parker sheets, in agreement with   the analytical results by \citet{Pucci14}.  These results suggest that ideal RMHD codes, at least those based on the splitting method, may be applicable to problems involving fast magnetic reconnection.


\section*{Data Availability}     

The data underlying this article will be shared on reasonable request to the corresponding author.

\section*{Acknowledgments}
We thanks the anonymous referee for their extensive and constructive report which led to noticeable improvements on the original manuscript. David Phillips acknowledges support from STFC in the form of research studentship.     

\bibliographystyle{mnras}
\bibliography{../BibFiles/plasma,../BibFiles/komissarov,../BibFiles/numerics,../BibFiles/bholes,../BibFiles/hea,../BibFiles/astro,../BibFiles/jets,../BibFiles/pwn}

\appendix

\section{3rd order WENO interpolation}
\label{sec:app-weno}

Below,  only the interpolation in the $x$ direction is considered, and all other spatial indices are dropped for brevity. In the other directions, the procedure is the same. 

\subsection{Modified 2nd order TVD weights}

Consider a 3-point stencil $S=\{x_{i-1},x_i,x_{i+1}\}$ and its two sub-stencils $S_-=\{x_{i-1},x_i\}$ and $S_+=\{x_i,x_{i+1}\}$. Each of the sub-stencils yields a linear polynomial for interpolation to the $i$th cell interfaces $x_{i+1/2}=x_i+\Delta x/2$ and $x_{i-1/2}=x_i-\Delta x/2$ on a uniform grid,
\beq
   P_-(x)=u_i+\frac{(u_i-u_{i-1})}{\Delta x}(x-x_i) \,,
   \label{eq:lin-pol-m}
\eeq 
and
\beq
   P_+(x)=u_i+\frac{(u_{i+1}-u_{i})}{\Delta x}(x-x_i) \,.
      \label{eq:lin-pol-p}
\eeq 
Any linear combination of these interpolants ensures 2nd order spatial accuracy in smooth regions of numerical solution. \citet{F-91} used a TVD slope limiter which is equivalent\footnote{\citet{F-91} also use the polynomial $P_0(x)=u_i$, for the case where  $\beta_+\beta_-\le 0$. } using following linear combination of the polynomials $P_\pm$  
\beq
P(x)=w_- P_-(x)+w_+ P_+(x)\,,
\label{eq:comb-poly}
\eeq
where
\beq
    w_-=\frac{\beta_+}{\beta_+ + \beta_-}\,,\quad w_+=\frac{\beta_-}{\beta_+ + \beta_-}\,,
    \label{eq:orig-weights}
\eeq
are the weights and 
\beq
\beta_-=(u_i-u_{i-1})^2\,,\quad \beta_+=(u_{i+1}-u_{i})^2
\label{eq:orig-betas}
\eeq
are the 'roughness' indicators. Incidentally, these indicators are the same as in \citet{JS96} for a 3rd-oder WENO interpolation. The weights \eqref{eq:orig-weights} satisfy the constraint 
\beq
w_-+w_+=1 \,.
\label{eq:constraint2}
\eeq
It is clear that not the absolute values of $\beta_+$ and $\beta_-$ but their ratio determines the weights:
\begin{itemize}
\item $w_-,w_+\to 1/2$ as $\beta_-/\beta_+\to 1$\,;
\item $w_-\to 1$ and $w_+\to 0$ as $\beta_-/\beta_+\to 0$\,; 
\item $w_-\to 0$ and $w_+\to 1$ as $\beta_+/\beta_-\to 0$\,.
\end{itemize}
This combination favours the interpolant with smaller gradient, thus reducing oscillations at regions with rapid variation of the numerical solution, such as shock waves. For example, suppose  that $u_{i+1}=u_i$, like in the upstream state of a shock, whereas $u_{i-1}\ne u_i$ is a point of numerical shock structure.  Then $\beta_+=w_-=0$ and $P(x_i)=P_+(x)=u_i$. 

Interestingly, these weights treat  critical points of smooth solutions almost on the same footing as shocks. To illustrate this,  suppose that a local maxima is located exactly between $x_i$ and $x_{i+1}$, so that $u_{i+1}=u_i$. Then, like in the shock example, $\beta_+ = w_-= 0$ and $P(x)= P_+(x)= u_i$.  
Generalising, any weights based on the ratios of the roughness indicators do not differentiate  between shocks and critical points. This applies to the WENO weights proposed by \citet{JS96}, which results in a loss of accuracy in the vicinity of critical points.

To remove this confusion, we propose the modified smoothness indicators 
\beq
\beta_\pm=(u_i-u_{i\pm1})^2 +  U^2 \fracp{\Delta x}{L}^2 +\epsilon\,, 
\label{eq:beta-m}
\eeq
where  
\beq
U=\max(|u_{i+1}|,|u_{i}|,|u_{i-1}|) \,,
\label{eq:U}
\eeq
is the maximal magnitude of $u$ on the stencil, $L\gg \Delta x$ is the minimal characteristic length scale of what can be considered as a computationally smooth solution, and $\epsilon$ is a small number, introduced to avoid division by zero when $u_i=u_{i-1}=u_{i+1}=0$. Hence,   

\begin{itemize}

\item In smooth regions away from local extrema, 
$$
(u_{i \pm 1}-u_i)^2 \approx \left(\pder{u}{x}\right)_i^2 \Delta x^2  
\leq U^2 \fracp{\Delta x}{L}^2\,.
$$
Hence, $\beta_-/\beta_+\approx 1$ and  $w_\pm\approx 1/2$, like in the original TVD scheme.

\item At strong shocks, either 
$$
(u_{i+1}-u_i)^2 \approx U^2 \gg \frac{U^2}{L^2} \Delta x^2\,,
$$
or
$$
(u_{i-1}-u_i)^2 \approx U^2 \gg \frac{U^2}{L^2} \Delta x^2  \,,
$$
or the both of them. In any of the cases, the new terms introduced in \eqref{eq:beta-m} have a little impact on $w_\pm$. 

\item Near the critical of points of smooth solutions, 
$$
(u_{i\pm1}-u_i)^2 \approx \fracp{\partial^2 u}{\partial x^2}_{i}^2 \Delta x^4  
\approx U^2 \fracp{\Delta x}{L}^4 \ll U^2 \fracp{\Delta x}{L}^2\,.
$$
Hence, $\beta_-/\beta_+\approx 1$ and $w_\pm\approx 1/2$, like at any other point of smooth solutions.
 
\end{itemize}

As to the value of $L$, it is reasonable to use $L=n_{sm}\Delta x$, with $5\lesssim n_{sm}\lesssim10$, leading to the final expression for the modified weights

\beq
\beta_-=(u_i-u_{i-1})^2 + \frac{U^2}{n_{sm}^2}  +\epsilon\,, 
\label{eq:beta-mf}
\eeq
\beq
\beta_+=(u_{i+1}-u_{i})^2 + \frac{U^2}{n_{sm}^2}  +\epsilon\,. 
\label{eq:beta-pf}
\eeq
For the test simulations described in this paper, we set $n_s=10$ and $\epsilon=10^{-25}$.  

\subsection{3rd-order WENO weights}

3rd-order WENO interpolation utilises the fact that the linear interpolation \eqref{eq:comb-poly} 
yields the same value  at $x=x_{i+1/2}$ as the quadratic interpolation based on the all three points of the stencil $S$ if $w_- =1/4$ and $w_+=3/4$, and the same value at $x=x_{i-1/2}$ if $w_- =3/4$ and $w_+=1/4$. Thus, two linear interpolants of the form \eqref{eq:comb-poly}, one per each interface of the cell,  can be used to achieve 3rd-order accurate interpolation to the both interfaces.  $\gamma\sub{a}=1/4$ and $\gamma\sub{b}=3/4$ are known as the ideal or linear weights. 
We denote the interpolant used  for the interpolation to the $x_{i-1/2}$ interface of $i$-th cell as
\beq
P^l(x)= w_-^l P_-(x) + w_+^l P_+(x)\,,
\eeq
 and the interpolant used  for the interpolation to the $x_{i+1/2}$ as
\beq
P^r(x)= w_-^r P_-(x) + w_+^r P_+(x)\,.
\eeq
Their weights satisfy exactly the same constraint as before 
\beq
     w_-^l+w_+^l=1\,, \quad w_-^r+w_+^r=1 \,.
     \label{eq:constraints3}
\eeq  
One may put $w_+^r=w_-^l=\gamma\sub{b}$ and $w_+^l=w_-^r=\gamma\sub{a}$,  but this will lead to violent oscillations at shocks. Instead, WENO weights are nonlinear, reducing to the ideal weights only on very smooth solutions.  At shocks, the linear interpolant with lower gradient should dominate.  Since the 2nd-order TVD interpolation, described earlier, is also based on the three-point stencil $S$, has exactly the same form as the 3rd-order WENO interpolants,  and already has the required behaviour at shocks,  a mapping of the TVD weights, which is closed to the identity mapping at shocks but yields ideal weights on smooth solutions, suggests itself.      

\begin{figure*}
\centering
 \includegraphics[width=0.3\textwidth]{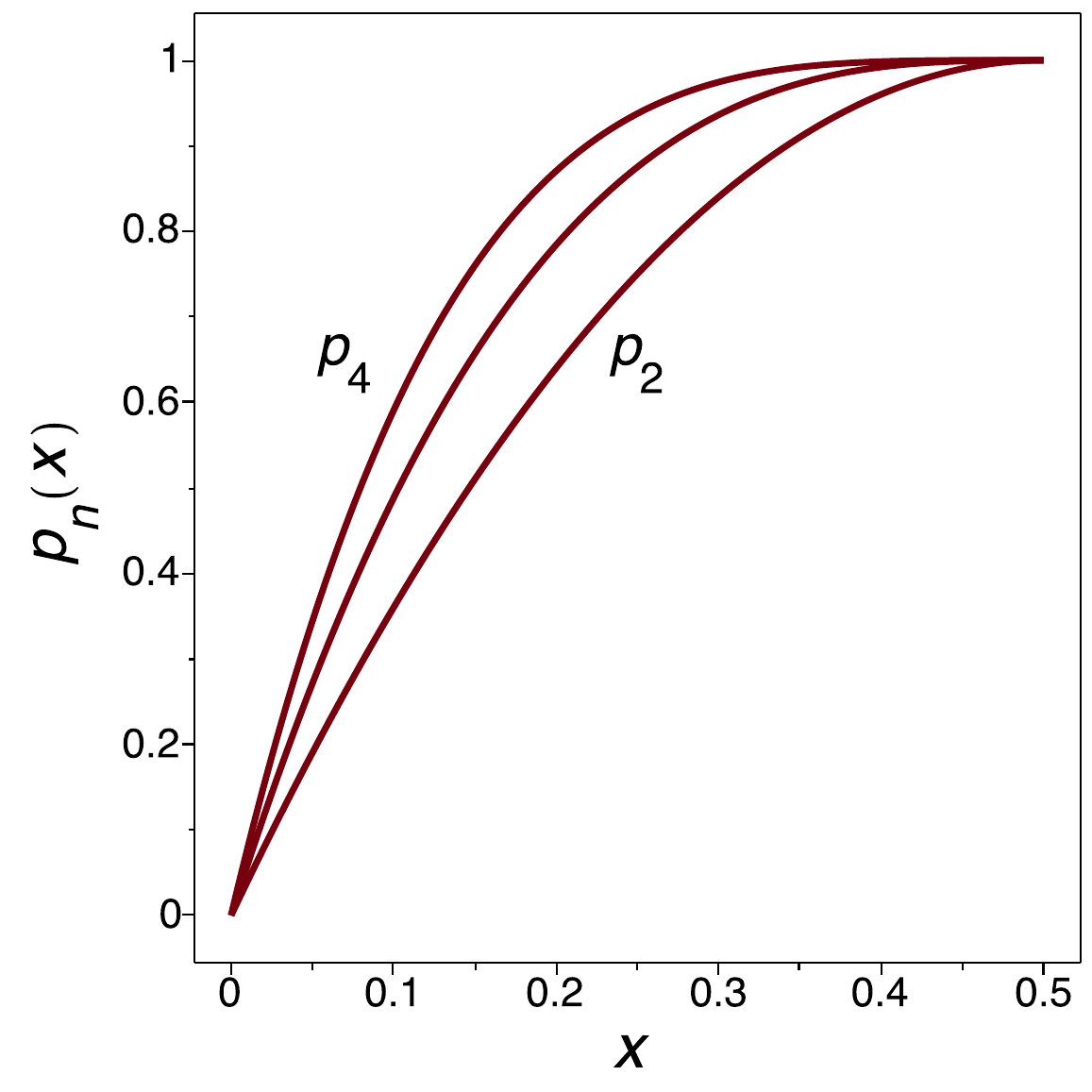}
 \includegraphics[width=0.3\textwidth]{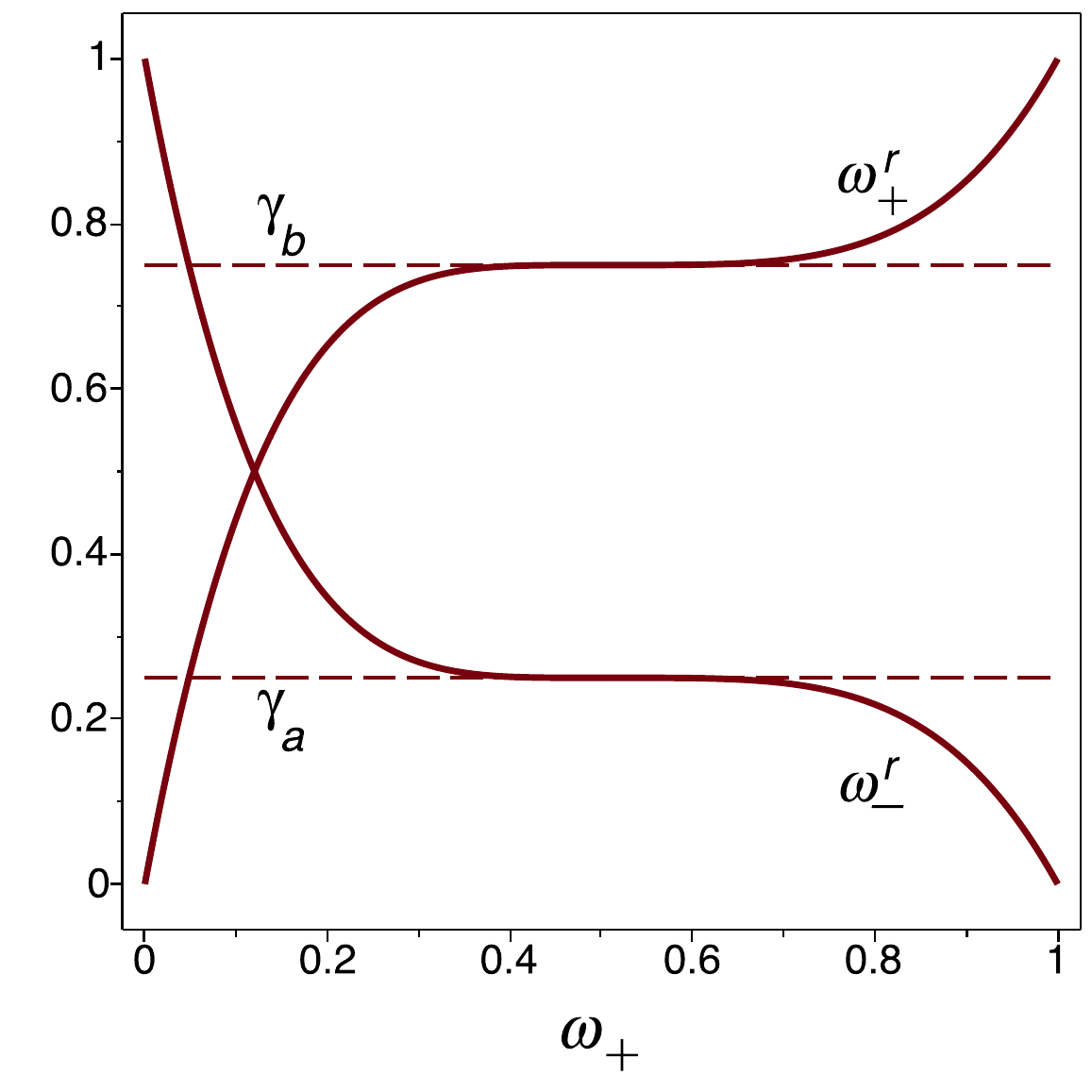}
\caption{Left panel: Mapping polynomials $p_n(x)$. Right panel: Non-linear WENO weights for $P^r(x)$ obtained with $\alpha(x) =p_4(x)$.}
\label{fig:weights}
\end{figure*}

So, we look for the mapping  $w_+ \to \{w_+^l,w_+^r\}$ such that 

 \beq
w_+^l \to \gamma\sub{a},\quad w_+^r \to \gamma\sub{b} \etext{as} w_+\to 0.5 \,,  
\eeq

\beq
 w_+^r, w_+^l   \to w_+ \etext{as} w_+\to 0 \etext{or} w_+\to 1  \,.
\eeq 
It also makes sense to require the functions $w_+^l(w_+)$  and $w_+^r(w_+)$ to be monotonic. Hence, if 
\beq
   w_+^r=\gamma\sub{b} \alpha(w_+) \,,
   \label{eq:map-p}
\eeq   
then $\alpha(x)$, $x\in[0,1]$,  must be a monotonic function of $x$ satisfying the conditions   
\beq
   \alpha(0)=0\,, \quad \alpha (0.5) =1\,,\quad  \alpha(1)=1/\gamma\sub{b} \,.
    \label{eq:cond0}
\eeq
In addition, it is desirable to have a reasonably wide region near x = 0.5 where $\alpha(x)$ remains close to 1. Hence, one may also require a number of its low-order derivatives to vanish at x = 0.5. 
For $x\in[0,0.5]$ these conditions are satisfied by the polynomials 
\beq
p_n(x)=1-(1-2x)^n\,, 
\eeq
where $n\ge2$.  The first three examples of such polynomials are shown in the left panel of figure \ref{fig:weights}.

To determine $\alpha(x)$ for $x\in[0.5,1]$, we require the function $w_-^r(w_-)$ to
be the same as  $w_+^r(w_+)$, apart from $\gamma\sub{b}$ replaced by $\gamma\sub{a}$, and write 
\beq
w_-^r=\gamma\sub{a} \alpha(w_-) \,.
\label{eq:map-m}
\eeq
Given the constraints   (\ref{eq:constraint2}) and (\ref{eq:constraints3}), one can write this equation as
\[
  w_+^r = 1-\gamma\sub{a} \alpha(1-w_+) \,.
\] 
This allows us to fully specify $w_+^r(w_+)$ and $w_-^r(w_+)$, 
\begin{align}
w_+^r(w_+) & = 
     \begin{cases}
       		\gamma\sub{b} p_n(w_+),  & 0\leq w_+ \leq 0.5 \,,\\
     		1-\gamma\sub{a} p_n(1-w_+), & 0.5 < w_+ \leq 1 \,,
      \end{cases} \\
w_-^r(w_+) & = 1-  w_+^r(w_+) \,.
\end{align}
Similarly, one finds 
\begin{align}
w_+^l(w_+) & = 
     \begin{cases}
       		\gamma\sub{a} p_n(w_+),  & 0\leq w_+ \leq 0.5 \,,\\
     		1-\gamma\sub{b} p_n(1-w_+), & 0.5 < w_+ \leq 1 \,,
      \end{cases} \\
w_-^l(w_+) & = 1-  w_+^l(w_+) \,.
\end{align}
Figure \ref{fig:weights} shows the nonlinear weights based on $p_4(x)$.

\subsection{Downgrading to 2rd-order TVD interpolation at strong shocks}

Strong shocks in high-$\sigma$ regime may still exhibit residual numerical oscillations of the flow parameters.    To remove them completely, one can switch to the 2nd-order TVD interpolation in the safety zone around such shocks (see Section \ref{sec:DER}).

\section{Variables conversion}
\label{app:vc}

The conserved variables of the perturbation system are mass density
\beq
  D= \rho\gamma \,,
\label{D}
\eeq
energy density
\beq
\cE_\1=\cE-\cE_\0=\vv{E}_\0\cdot\vv{E}_\1+\vv{B}_\0\!\cdot\!\vv{B}_\1+\frac{E_\1^2+B_\1^2}{2} +w\gamma^2-p\,,
\label{Q1}
\eeq
where 
\beq
\cE=\frac{E^2+B^2}{2} +w\gamma^2-p\,,
\eeq
\beq
\cE_\0=\frac{E_\0^2+B_\0^2}{2}\,,
\eeq
momentum density
\beq
\vv{S}_\1=\vv{S}-\vv{S}_\0=\vv{E}_\0\!\times\!\vv{B}_\1+\vv{E}_\1\!\times\!\vv{B}_\0+\vv{E}_\1\!\times\!\vv{B}_\1 +w\gamma^2\vv{v}\,,
\eeq
where
\beq
\vv{S}=\vpr{E}{B} +w\gamma^2\vv{v}\,,
\eeq
\beq
\vv{S}_\0=\vv{E}_\0\!\times\!\vv{B}_\0 \,.
\eeq
In addition, we have the perfect conductivity condition is
\beq
\vv{E}=-\vpr{v}{B} \,,
\label{PC1}
\eeq
which can also be written as 
\beq
\vv{E}_\1=-\vv{E}_\0-\vpr{v}{B} \,.
\label{PC2}
\eeq
and the polytropic equation of state
\beq
 w = (\rho +\kappa p) \,,
 \label{eos}
\eeq
where $\kappa=\Gamma/(\Gamma-1)$.

Eq.\eqref{PC1} leads to 
$$
\vpr{E}{B}=\vv{B}\!\times\!(\vpr{v}{B})=B^2\vv{v}-(\spr{v}{B})\vv{B}
$$
and hence 
\beq
\vv{S}=(B^2+W)\vv{v} -(\spr{B}{v})\vv{B} \,,
\label{S}
\eeq
where $W=w\gamma^2$. From the last equation it follows that 

\beq
(\spr{B}{v}) = \frac{(\spr{S}{B})}{W} \,.
\label{Bv}
\eeq
Substituting this back in \eqref{S}, we obtain
\beq
 \vv{v}= \frac{\vv{S}+((\spr{S}{B})/W)\vv{B}}{B^2+W} \,.
 \label{v}
\eeq
This equation shows that $\vv{v}$ depends solely on the unknown $W$.  From this result it follows that 
\beq
S^2=(B^2+W)^2v^2 - (2W+B^2) \frac{(\spr{S}{B})^2}{W^2}\,.
\label{S2}
\eeq
Thus, we have an equation for only two unknowns, $W$ and $v^2$.  However, this equation is not immediately suitable for the high magnetisation case as it involves terms of the order $B^4$, that results in large computational errors for the hydrodynamic variables.  As we show later, these terms cancel out.  

Next, we use the perfect conductivity condition \eqref{PC2} to eliminate $\vv{E}_\1$ from the expression \eqref{Q1} for $\cE_\1$. To this end, we first find that 

$$
\vv{E}_\0\!\cdot\!\vv{E}_\1= -E_\0^2-\vv{E}_\0\!\cdot\!(\vpr{v}{B}) \,,
$$
and
$$
E_\1^2=E_\0^2+2\vv{E}_\0\!\cdot\!(\vpr{v}{B})+||\vpr{v}{B}||^2\,,
$$
and hence 
$$
\vv{E}_\0\!\cdot\!\vv{E}_\1+\frac{E_\1^2}{2}=-\frac{E_\0^2}{2}+||\vpr{v}{B}||^2\,. 
$$
This can be reduced further using  

\begin{align*}
||\vpr{v}{B}||^2  &=(\vpr{v}{B} )\cdot(\vpr{v}{B})\\
	&=\vv{v}\cdot(\vv{B}\!\times\!({\vpr{v}{B}})) \\
	&=\vv{v}\cdot(\vv{v} B^2 -\vv{B}(\spr{v}{B}))\\
	&=v^2 B^2 - (\spr{v}{B})^2 \,.
\end{align*}
Substituting the last two results into \eqref{Q1} we obtain 
\beq
\cE_\1=\frac{1}{2}(B^2v^2-(\spr{v}{B})^2)+W-p 
      -\frac{E_\0^2}{2}+\frac{B_\1^2}{2}+\vv{B}_\0\!\cdot\!\vv{B}_\1 \,.
\label{Q1a}
\eeq
The last three terms of the right-hand side are already known. To reflect this, we introduce 
\beq
\bar{\cE}_1=\cE_\1 +\frac{E_\0^2}{2}-\frac{B_\1^2}{2}-\vv{B}_\0\!\cdot\!\vv{B}_\1 \,,
\eeq 
and write \eqref{Q1a} as

\beq
\bar{\cE}_1=\frac{1}{2}B^2v^2+W-p -\frac{1}{2} \frac{(\spr{S}{B})^2}{W^2}\,,
\label{Q1b}
\eeq
where we have also applied \eqref{Bv}.  This equation contains the unknowns $v^2$,  $W$ and $p$. Using EOS \eqref{eos} and equation \eqref{D}, we find that 
\beq
p=\frac{1}{\kappa}(W(1-v^2)-D(1-v^2)^{1/2}) \,,
\eeq
which allows to eliminate $p$ from \eqref{Q1b} and obtain the cubic equation 

\beq
a_3(v^2) W^3+a_2(v^2) W^2 + a_0 =0 \,,
\label{EQ1}
\eeq
where
\beq
a_3=1-\frac{1-v^2}{\kappa}\,,
\eeq
\beq
a_2=\frac{1}{2}B^2v^2-\bar{\cE}_1+D\frac{(1-v^2)^{1/2}}{\kappa} \,,
\eeq
\beq
a_0=-\frac{1}{2}(\spr{S}{B})^2 \,.
\eeq
Thus we have obtained two equations, \eqref{S2} and \eqref{EQ1}, for the unknowns $W$ and $v^2$. This system is to be solved numerically.

Obviously, one can further reduce the system to just one equation, either for $v^2$ or $W$.  
Following the reasonable argument of Del Zanna (2007), it is preferable to eliminate W by solving the cubic equation (12) analytically. This allows us to control the condition $0\le v^2<1$ during the numerical iterations of the Newton method (or its secant version) for the resultant equation. 

The fully expanded expression for the coefficient $a_2$ is 
$$
   a_2= \frac{1}{2}(B^2v^2 - E_\0^2)-\cE_\1+\frac{B_\1^2}{2}+\vv{B}_\0\!\cdot\!\vv{B}_\1+D\frac{(1-v^2)^{1/2}}{\kappa} \,.
$$
The first two terms of this expression constitute the difference between $B^2v^2/2$ and $E_\0^2$. These non-negative terms can be very large and their difference can be a source of large error in computations of $a_2$ in the case of high magnetisation. 

Introducing the drift velocity of force-free approximation
$$
\vv{v}_\0=\frac{\vv{E}_\0\!\times\!\vv{B}_\0}{B_\0^2} \,.
$$
one can write
$$
B^2 v^2-E_\0^2=B^2v^2-B_\0^2 v_\0^2=B_\0^2(v^2-v_\0^2)+(B_\1^2+\vv{B}_\0\!\cdot\!\vv{B}_\1)v^2\,,
$$ 
and hence 

\begin{align}
\nonumber
a_2&=\frac{1}{2}(B_\0^2(v^2-v_\0^2)+(B_\1^2+\vv{B}_\0\!\cdot\!\vv{B}_\1)v^2) - \\
\nonumber
 &-\cE_\1 +\frac{B_\1^2}{2}+\vv{B}_\0\!\cdot\!\vv{B}_\1+D\frac{(1-v^2)^{1/2}}{\kappa} \,.
 \end{align}

Computations of the term $\spr{S}{B}$ may also involve subtraction of large numbers and hence results in large errors. This can be avoided if we note that $\vv{S}_\0\!\cdot\!\vv{B}_\0=0$ and 
write
$$
\spr{S}{B}=\vv{S}_\0\!\cdot\!\vv{B}_\1+\vv{S}_\1\!\cdot\!\vv{B} \,.
$$

Substituting $(\spr{S}{B})^2/W^2$ from eq.\eqref{Q1b} into eq.\eqref{S2} and cancelling out terms of the order $B^4$ results in 

\begin{align}
\nonumber
W^2v^2&+4\bar{\cE}_1 W +4(p-W)\left(W+\frac{B^2}{2}\right)=S_\1^2 +2\vv{S}_\1\!\cdot\!\vv{S}_\0-\\
& - 2\tilde{\cE}_1 B^2 - B_\0^2 v_\0^2 (B_\1^2+2\vv{B}_\0\!\cdot\!\vv{B}_\1)\,,
\label{EQ2}
\end{align}
where
$$
\tilde{\cE}_1=\cE_\1-\frac{B_\1^2}{2}-\vv{B}_\0\!\cdot\!\vv{B}_\1 \,.
$$

\end{document}